\documentclass[12pt,preprint]{aastex}

\newcommand       \Angstrom     {\,{\rm \AA}}

\shorttitle{SDI Survey}
\shortauthors{Biller et al.}

\begin{document}

\title{An Imaging Survey for Extrasolar Planets around 45 Close, Young Stars 
with SDI at the VLT and MMT$^1$}

\author{Beth A. Biller$^1$, Laird M. Close$^1$, Elena Masciadri$^{2}$, 
Eric Nielsen$^1$, Rainer Lenzen$^{3}$, Wolfgang Brandner$^{3}$, Donald McCarthy$^1$, Markus Hartung$^4$, Stephan Kellner$^5$, Eric Mamajek$^6$, 
Thomas Henning$^3$, Douglas Miller$^1$, Matthew Kenworthy$^1$, and 
Craig Kulesa$^1$}

\email{bbiller@as.arizona.edu}

\affil{$^1$ Steward Observatory, University of Arizona, Tucson, AZ 85721}
\affil{$^2$ Observatorio Astrofisico di Arcetri, L.go E. Fermi 5, 50125 Florence, Italy}
\affil{$^3$ Max-Planck-Institut f\"ur Astronomie, K\"onigstuhl 17, 69117 
Heidelberg, Germany}
\affil{$^4$ European Southern Observatory, Alonso de Cordova 3107, 
Santiago 19, Chile}
\affil{$^5$ W.M. Keck Observatory, 65-1120 Mamalahoa Hwy., Kamuela, HI 96743}
\affil{$^6$ Harvard-Smithsonian Center for Astrophysics, 60 Garden St., Cambridge, MA 02138}

\begin{abstract}
We present the results of a survey of 45 young ($\la$250 Myr),
close ($\la$50 pc) stars with the Simultaneous Differential 
Imager (SDI) implemented at the VLT and the MMT for the direct detection 
of extrasolar planets.\footnote[1]{based on observations made with 
the MMT and the ESO VLT at Paranal Observatory under programme ID's 
074.C-0548, 074.C-0549, and 076.C-0094}  As part of the survey, 
we observed 54 objects total,
of which 45 were close, young stars, 2 were somewhat more 
distant ($<$150 pc), very young ($\leq$10 Myr) stars,
3 were stars with known radial
velocity planets, and 4 were older, very nearby ($\leq$20 pc) 
solar analogues.  Our SDI devices use a double Wollaston prism and 
a quad filter to take images simultaneously
at 3 wavelengths surrounding the 1.62 $\mu$m methane absorption 
bandhead found in 
the spectrum of cool brown dwarfs and gas giant planets.  By performing 
a difference of adaptive optics corrected 
images in these filters, speckle noise from
the primary star can be significantly attenuated, resulting in photon 
(and flat-field) noise limited 
data.  In our VLT data, we achieved H band contrasts
$\ga$ 10 mag (5$\sigma$) at a separation of 0.5" from
the primary star on 45$\%$ of our targets
and H band contrasts of $\ga$ 9 mag at a separation of 0.5'' 
on 80$\%$ of our targets.    
With this degree of attenuation, we should be able to 
image (5$\sigma$ detection) a 5 M$_{Jup}$ planet 15 AU from 
a 70 Myr K1 star at 15 pc or a 5 M$_{Jup}$ planet at 2 AU from 
a 12 Myr M star at 10 pc.  Our 45 southern targets were observed with the 
VLT while 11 of our northern targets were observed with the MMT (2 objects
were observed at both telescopes).
We believe that our SDI images are the highest contrast astronomical 
images ever made from ground or space for methane rich companions
$\leq$1'' from their star.  We detected
no tentative candidates with S/N $>$ 2 $\sigma$ which behaved consistently
like a real object.  Followup observations
were conducted on 8 $<$2$\sigma$ candidates 
(with separations of 3 - 15.5 AU and 
masses of 2-10 M$_{Jup}$, had they been real) -- 
none of which were detected at a second epoch.  In the course of our survey, 
we also discovered 5 new close stellar binary systems with measured 
separations of 0.14'' to 0.26''.  For the best 
20 of our survey stars, we attained 50$\%$ 5$\sigma$ 
completeness for 6-10 M$_{Jup}$ planets at semi-major
axes of 20-40 AU.  Thus, our completeness levels are sufficient to 
significantly test theoretical planet distributions.  From our survey null
result, we can rule out (at the 98$\%$ confidence/2.0$\sigma$ level) 
a model planet population using a planet distribution where
N({\it a}) $\propto$ constant out to a distance of 45 AU (further 
model assumptions discussed within).  
\end{abstract}


\section{Introduction}

While over 200 extrasolar planets have been 
detected\footnote[2]{http://exoplanet.eu/catalog.php, maintained by 
Jean Schneider} over the 
last 11 years (mostly via the radial velocity technique), 
very few extrasolar planet candidates have been imaged 
directly~\citep[for~instance,~2MASS~1207b ($\sim$8$\pm$3 M$_{Jup}$),
Oph 1622B ($\sim$13$\pm$5 M$_{Jup}$), and CHXR 73 B ($\sim$12.5$\pm$8 M$_{Jup}$)][]{cha05,clo07a,luh06,bra06}.  
The few candidates
discovered of ``planetary mass'' $<$ 13 M$_{Jup}$ 
are companions to brown dwarfs and possess properties more similar to 
young brown dwarfs (separations $>$ 50 AU; surface gravity g $\ga$ 0.3) 
than to giant extrasolar planets orbiting sun-like stars.  Based on their large
($>$50 AU) separations, these objects appear to 
have formed via a fragmentation process, more similar to brown dwarfs.  Hence, to date
no true images of extrasolar planets have been obtained.

Theoretically, a large telescope (D $>$ 6 meters)
plus an adaptive optics (AO) system should be able to reach the photon-noise limit
at 1\arcsec~separations from the star 
with an hour of exposure time and thus attain the very 
high ($>$10$^5$) contrasts
necessary to image a young extrasolar giant planet.  
Thus, numerous adaptive optics surveys to directly detect extrasolar 
planets have been
completed \citep[for~instance,][]{kai03,mas05}.  These surveys have yielded
interesting contrast limits but no true extrasolar giant planet candidates.

The difficulty in directly imaging extrasolar giant planets
can be attributed to the unfortunate fact that bright quasi-static speckles
(also known as super speckles) 
caused by slowly evolving instrumental aberrations remain in adaptive 
optics images even after adaptive optics correction
\citep[see~for~example][]{rac99}.  These super speckles 
evolve stochastically on relatively long (minute) timescales and also 
vary somewhat chromatically, producing correlated speckle noise which
is very difficult to calibrate and remove \citep[][]{rac99}.  
For photon-noise limited 
data, the signal to noise S/N increases as t$^{0.5}$, where t is the exposure 
time.  Approximately speaking, 
for speckle-noise limited data, the S/N does not increase with time
past a specific speckle-noise floor (limiting AO contrasts often to 
$\sim$10$^3$ at 0.5'', Racine et al. 1999; Masciadri et al. 2005).
More exactly, S/N does continue to increase with time, but as the speckle
noise in successive frames becomes correlated, the N gain becomes 
considerably slower.  Effectively independent exposures then
have durations of many minutes rather than a small fraction of a second
(Racine et al. 1999).   
This correlated speckle noise is considerably above the 
photon noise limit and makes planet detection very difficult.  Interestingly, 
space telescopes such as HST also suffer from limiting correlated 
speckle noise due to temperature variations which induce changes in the PSF 
\citep[known~as~``breathing'',][]{sch03}.

Many observatories, including Gemini, Subaru, and the VLT, 
are currently building dedicated planet-finding AO/coronagraph 
cameras in order to 
overcome this speckle noise floor \citep[][]{doh06,mac06,tam06}.  
A number of instrumental 
speckle-attenuation methods have been proposed, such as 
spectral differential imaging \citep[][]{rac99,mar00,mar02,mar05},
azimuthal differential imaging \citep[][]{mar06}, 
integral field spectroscopy \citep[][]{spa02,ber06,tha07},
precise wavelength control methods such as those developed at the High
Contrast Imaging Testbed \citep[][]{tra04}, 
focal plane wavefront sensing \citep[][]{cod05,ken06}, 
and nulling interferometry \citep[][]{liu06}.

The Simultaneous Differential Imagers at the VLT and MMT, built and commisioned
by our team \citep[][]{len04,len05,clo05a}, 
utilizes a spectral differential speckle-attenuation 
technique \citep[pioneered~by][]{rac99,mar00,mar02,mar05}.  
It exploits a methane absorption feature at 1.62 $\mu$m (see 
Fig.~\ref{fig:SDIFILT}) which
is robustly observed in substellar objects with spectral type later than
T3.5 \citep[][]{geb02,bur01}.  
SDI utilizes specialized hardware to image simultaneously
inside and outside this methane feature with custom 
25 nm filters (see Fig.~\ref{fig:SDIFILT}).  Since the super-speckles are
coherent with the starlight and both starlight and speckles 
have a flat spectrum (see Fig.~\ref{fig:SDIFILT}) 
in this narrow wavelength band ($\delta\lambda$ / $\lambda$ $\simeq$1.6$\%$), 
subtracting the ``on'' and ``off'' methane 
absorption images removes the starlight and its speckles, while preserving 
light from any substellar methane companion to the star.

We have completed a 54 star survey with the SDI device at the VLT and MMT.
Survey stars were chosen primarily according to proximity to the Sun ($\la$50 pc)
and youth ($\la$300 Myr, typically $<$100 Myr).  We observed 47 young ($\la$250 
Myr) stars, 3 nearby stars with known RV planets, and 4 very close ($\la$20 
pc) older solar analogues.  We obtained
contrasts of $\Delta$H$>$10 mag (5 $\sigma$) at 0.5$\arcsec$ for 45$\%$ of 
target objects at the VLT and contrasts of $\Delta$H$>$9 mag (5 $\sigma$) at 
0.5$\arcsec$ for 80$\%$ of our targets.  The VLT 
SDI device is fully commissioned 
and available to the community and the MMT SDI device is a PI instrument
with the ARIES camera.  In contrast, the 
dedicated planet-finding instruments such as Sphere and GPI 
\citep[][]{doh06,mac06} being built at 
the VLT and Gemini will not see first light for several years.  
Thus, as a precursor to planet surveys with these dedicated planet finding 
cameras, the results from the SDI devices are especially timely and relevant,
particularly to inform the large Gemini NICI survey starting in 
2007 \citep[][]{liu05}.

\section{The Simultaneous Differential Imagers at the VLT and MMT}

The VLT Simultaneous Differential Imager (henceforth SDI) was built 
at the University of Arizona by L. Close 
and installed in a special f/40 camera relay for the VLT AO camera 
CONICA built by R. Lenzen at the Max Planck Institute for Astronomy, 
Heidelberg.  These were both installed at the VLT in August 2003.  
The MMT SDI was also
built at the University of Arizona.  In February 2004, it was 
installed in the ARIES f/30 camera built by D. McCarthy.  Both devices are 
available to the observing communities of their respective telescopes.

\subsection{Hardware Considerations}

The SDI device consists of a custom double Wollaston, which splits
the incoming AO beam into 4 identical beams (utilizing calcite
birefringence to minimize non-common path error -- adding only $\la$10 
nm rms of differential non-common path errors per the first 
few Zernikes modes -- Lenzen et al. 2004a).  
Each beam then passes through a narrowband 
filter with a central wavelength either on or off methane absorption.
Three different filters were used; all filters were placed in 
different quadrants on the same substrate.
SDI filters for the VLT and MMT were manufactured by Barr Associates. 
Filter wavelengths were chosen on and off 
the methane absorption feature at 1.62 $\mu$m and were spaced closely (every 
0.025 $\mu$m) in order to limit residuals due to speckle and calcite 
chromatism.  We used four filters F1, F2, F3a, and F3b with 
measured cold central 
wavelengths F1$\tbond$1.575 $\mu$m, F2$\tbond$1.600 $\mu$m, 
and F3a$\tbond$F3b$\tbond$1.625 $\mu$m.  
The filters are approximately 0.025 $\mu$m in bandwidth (1.6$\%$).  The 
SDI filter transmission curves 
overlaid on a theoretical young planet spectrum (private 
communication, D. Sudarsky) are presented in Fig.~\ref{fig:SDIFILT}.   

\subsection{Discoveries with the SDI Cameras}

The SDI device has 
already produced a number of important scientific results: the discovery 
of the important calibrator object AB Dor C \citep{clo05b}
which is the tightest (0.16'') 
low mass (0.090$\pm$0.05 M$_{\odot}$, $\sim$100$\times$ fainter)
companion detected by direct imaging, the most detailed 
methane surface maps of Titan from the pre-Cassini era \citep{har04}, 
the discovery of $\epsilon\,$Ind Ba and Bb, the 
nearest binary brown dwarf \citep{mcc04}, the discovery of 
SCR 1845-6357B, a very close (3.85 pc) T6 brown dwarf \citep{bil06b}, 
and evidence 
of orbital motion for Gl 86B, the first known 
white dwarf companion to an exoplanet host star \citep{mug05}.  In fact, the 
SDI device discovered all known brown dwarfs within 5 pc of the Sun.  It
has also set the best upper limit on the luminosity of the older ($\sim$1 Gyr)
extrasolar planet around $\epsilon\,$Eri.  

\subsection{Observational Techniques and Data Reduction}

\begin{figure}
   \includegraphics[width=\columnwidth]{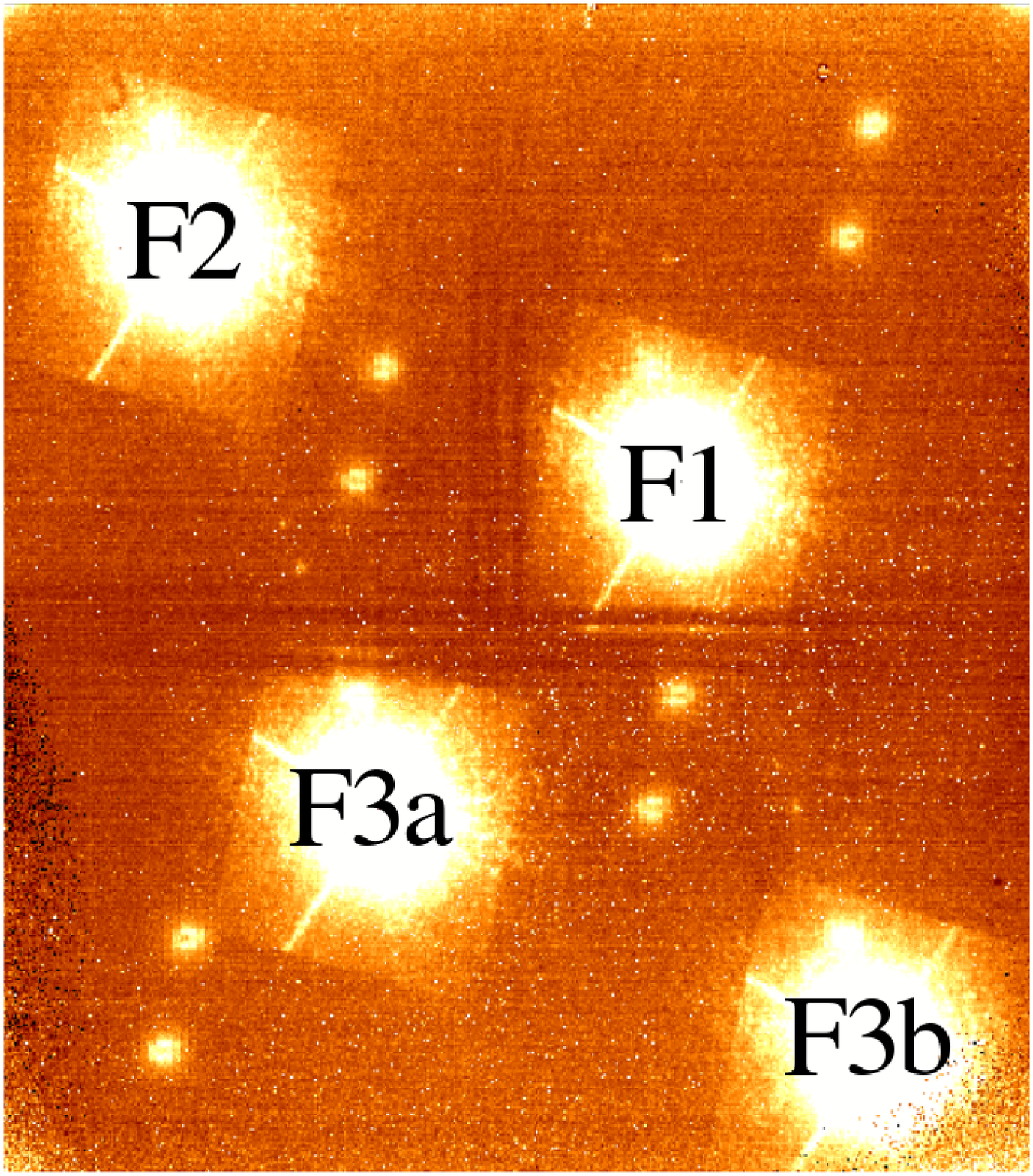} 
   \caption[Raw VLT SDI data]
   { \label{fig:SDIRAW} 
Two minutes of raw SDI data from NACO SDI's 
1024$\times$1024 Aladdin array 
in the VLT CONICA AO camera (Lenzen et al. 2004).  A number of 
electronic ghosts are apparent outside the four square filter apertures
(each aperture is rotated by 30$^{\circ}$);
indeed, filter apertures were specifically selected to exclude these ghosts.
Note that this is an image of the original Alladin array; the current SDI 
array has far fewer bad pixels. 
}
   \end{figure}

To ensure the highest possible signal to noise ratio and to maximize SDI 
speckle attenuation, a complex data 
acquisition procedure was followed for each star.  
For each object observed, we saturated the inner $\sim$0.1'' of the star, thus 
providing a wide dynamic range and contrast down into the halo.  Base 
exposure times (DIT) range from 0.3 to 20 s (typically this was 
$>$ 2s to allow Fowler sampling at the VLT), 
depending on the H magnitude of the
observed star.  A number of exposures (NDIT) with the base exposure time are
then coadded in hardware  
to produce a standard $\sim$2 minute long base datum.
An example raw datum is presented in Fig.~\ref{fig:SDIRAW}
\footnote[3]{As with all our survey data, this was taken with the original SDI 
double Wollaston prism.  In February 2007, 
the original prism was replaced with a next generation prism which is 
cut in such a way that each subimage now subtends a whole 
quadrant of the detector chip.  The new prism is also fabricated from YV04,
a material which produces smaller chromatic errors at 1.6$\mu$m 
than the original calcite.}. 

Base datum are then taken at a grid of dither positions
(4$\times$0.5'' spacings with the MMT, 5$\times$0.5'' spacings with the VLT).
This dither pattern is then repeated at typically 2 telescope ``roll angles''
(where a ``roll angle'' refers to a different
 field derotator position / position angle (henceforth PA) settings).  
A subtraction of data taken at different roll angles further attenuates
super-speckle residuals (since the weak residual speckles after SDI subtraction
are instrumental features in the SDI optics which do
not shift with a change in roll angle) while producing a 
very important signature ``jump'' 
in position for any physical companion (since a physical companion 
will appear to shift by the roll angle difference between datasets).
For a space telescope such as Hubble (where the entire telescope can 
be rolled), a companion detected at the 5$\sigma$ level in two different 
roll angles 
would be detected at the 7$\sigma$ level (a S/N gain of $\sim\sqrt{2}$)
across the entire dataset
(assuming roughly Gaussian statistics).  
This method is somewhat less effective with ground based telescopes where
field rotation is provided by the field derotator rather than rolling the 
entire telescope (thus, super speckles from the telescope optics can 
appear to rotate by the roll angle as well).   
Nonetheless, observing at two roll angles provides us with two independent
detections of a substellar companion at different 
locations on the detector, thus allowing us to rule out a 
``false positive'' detection at an extremely high level of confidence --
indeed, the only 3 faint companions ($\epsilon\,$Ind Bb, SCR 1845-6357B, and
AB Dor C) ever detected with $\geq$5$\sigma$ using SDI in more than one
roll angle have {\it all proven to be real}.
A typical observing block at the VLT then consists of the following series of :
1) $\sim$10 minute long dither pattern taken with a roll angle of 0 degrees.
2) $\sim$10 minute long dither pattern taken with a roll angle of 33 degrees.
3) $\sim$10 minute long dither pattern taken with a roll angle of 33 degrees.
4) $\sim$10 minute long dither pattern taken with a roll angle of 0 degrees.
A custom template was developed at the VLT to automate this process in each 
OB.

Each base datum was reduced using a custom IDL pipeline
(described in detail in Biller et al. (2006a) and Biller et al. (2006c)).  
This pipeline performs sky-subtraction, 
flat-fielding, and bad pixel removal, extracts a square aperture 
around each separate filter image, scales the platescale of each 
filter image so that the speckles in each filter
fall at the same radii despite chromatic differences, scales the 
flux in each image to remove any quantum efficiency differences
between the images, and filters out very low ($>$15 pixels) 
spatial frequencies by unsharp masking each image.
Each filter image is then initially 
aligned to a reference image to within 0.25 pixels 
using a custom shift and subtract algorithm (Biller et al. (2006a,c)).  
One master reference image is used for 
each $\sim$40 minute long dataset. 
After each of the filter images has been aligned to the reference image, we
calculate 2 differences 
which are sensitive to substellar companions 
of spectral types T (T$_{eff}$ $<$ 1200 K) and ``Y'' (T$_{eff}$ $<$ 600 K).
The first is optimal for T spectral types:

\begin{equation}
Difference1 = F1(1.575~\mu m) - F3a(1.625~\mu m) 
\end{equation}

The second is optimal for Y spectral types:

\begin{equation} 
Difference2 = F2(1.6~\mu m) - F3a(1.625~\mu m) 
\end{equation}


An additional alignment is performed before the SDI subtraction; using the F1
image as our reference image, we align images F1 and F3a to within 0.05
pixels.  A similar alignment is performed with images F2 and F3a, using
the F2 image as the reference image.  

These differences are also  
somewhat sensitive to hotter substellar companions (L and early T spectral 
types), due to the fact that 
the platescale in each filter image has been scaled to a reference 
platescale to align the Airy patterns in each image.  A real object 
(as opposed to a speckle) will not
scale with the Airy pattern and thus, after scaling, will appear at a slightly
different radius in each filter image.  Subtracting images in different 
filters will then produce a characteristic dark-light radial pattern for a 
real object.  This effect obviously scales with radius -- at the VLT, 
an object at 0.5'' will be offset by less than 1 pixel between filters, while
an object at 1.5'' will be offset by $\sim$3 pixels, producing a very 
noticeable pattern.  Thus, the SDI subtractions have a  
limited sensitivity to bright L and early T companions.  We note that AB Dor 
C ($\Delta$H $\sim$ 5 mag) was detected at 0.15'' (February 2004, 
Close et al. 2005) and 0.2'' (September 2004, Nielsen et al. 2005) 
separations from 
AB Dor A even though AB Dor C has no methane absorption features (as is expected from its M5.5 spectral type, Close et al. 2007b.)  

We additionally calculate one further non-differenced combination sensitive to
M, L, and early T companions:

\begin{equation}
Broadband = F1(1.575 \mu m) + F2(1.6 \mu m) + F3(1.625 \mu m)
\end{equation}

After each datum is pipelined the data are further processed in IRAF.  
For each $\sim$10 minute long 
dither pattern, all three combinations described above and the four 
reduced filter images are median combined.  Each 10 minute dataset is
then differenced with the following 10 minute dataset (taken at a different
position angle).  All roll-angle differenced images for each target object
observation are then median combined to produce the final data product.

\begin{figure}
\includegraphics[width=\columnwidth]{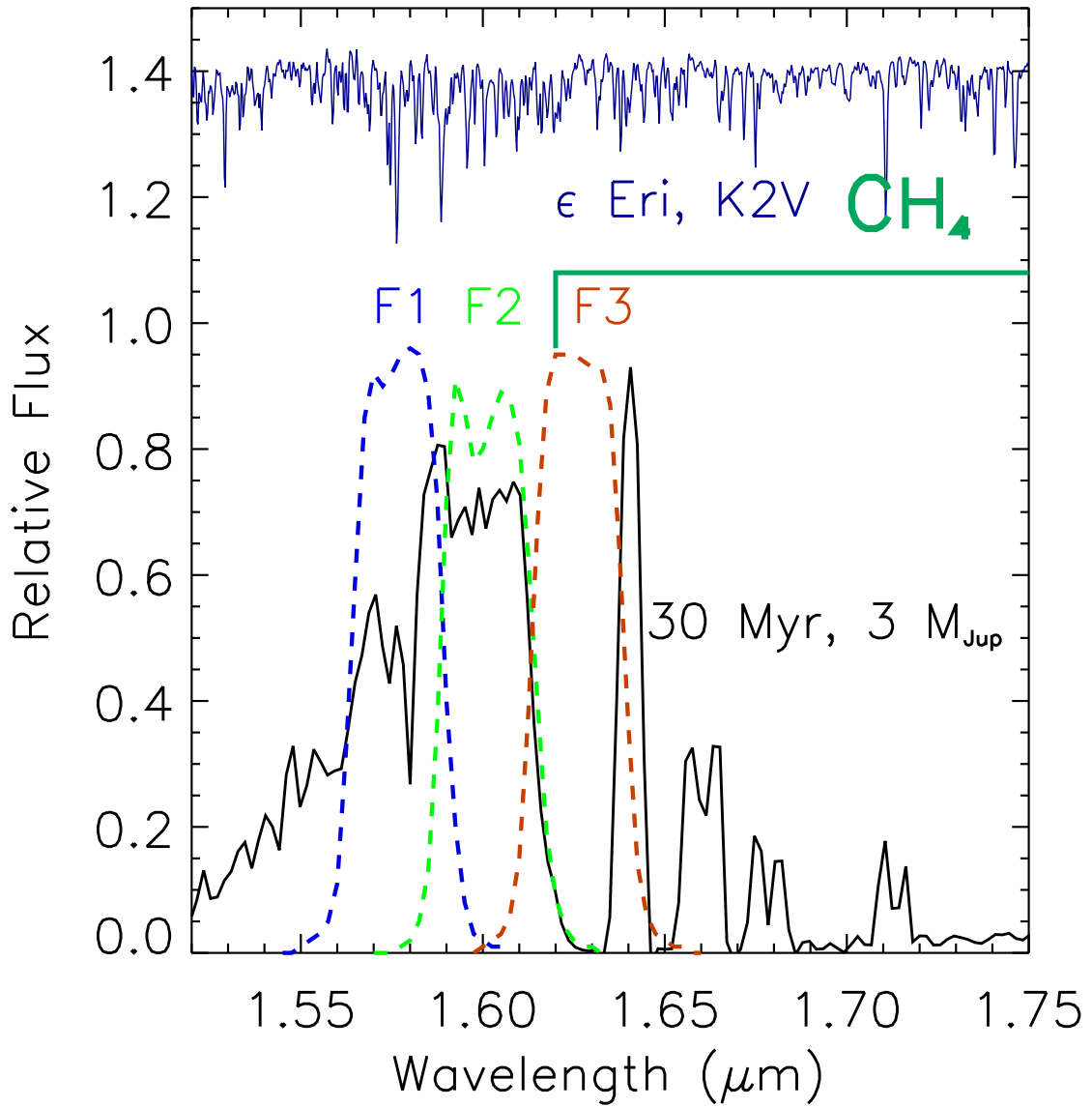}
\caption[SDI filters]
{\label{fig:SDIFILT} SDI filter transmission curves overlaid on the 
theoretical spectrum (private communication, D. Sudarsky) 
of a young extrasolar planet (30 Myr, 3 M$_{Jup}$).
Filters 1 and 2 sample off the 1.62 $\mu$m 
CH$_4$ absorption feature, while filter 3 samples 
within the absorption feature.  In contrast, the spectrum of the K2V 
star $\epsilon\,$Eri (Meyer et al. 1998) 
is flat across the whole wavelength band.
Subtracting images taken in filters ``on'' and ``off'' the methane 
absorption feature will remove
the star and speckle noise (which is coherent with the starlight) 
while preserving any light from giant planet companions. (Details of 
the complex SDI data pipeline are provided in Section 2.3.)}
\end{figure}

\begin{figure}
   \begin{center}
   \begin{tabular}{cc}
    \includegraphics[width=3in]{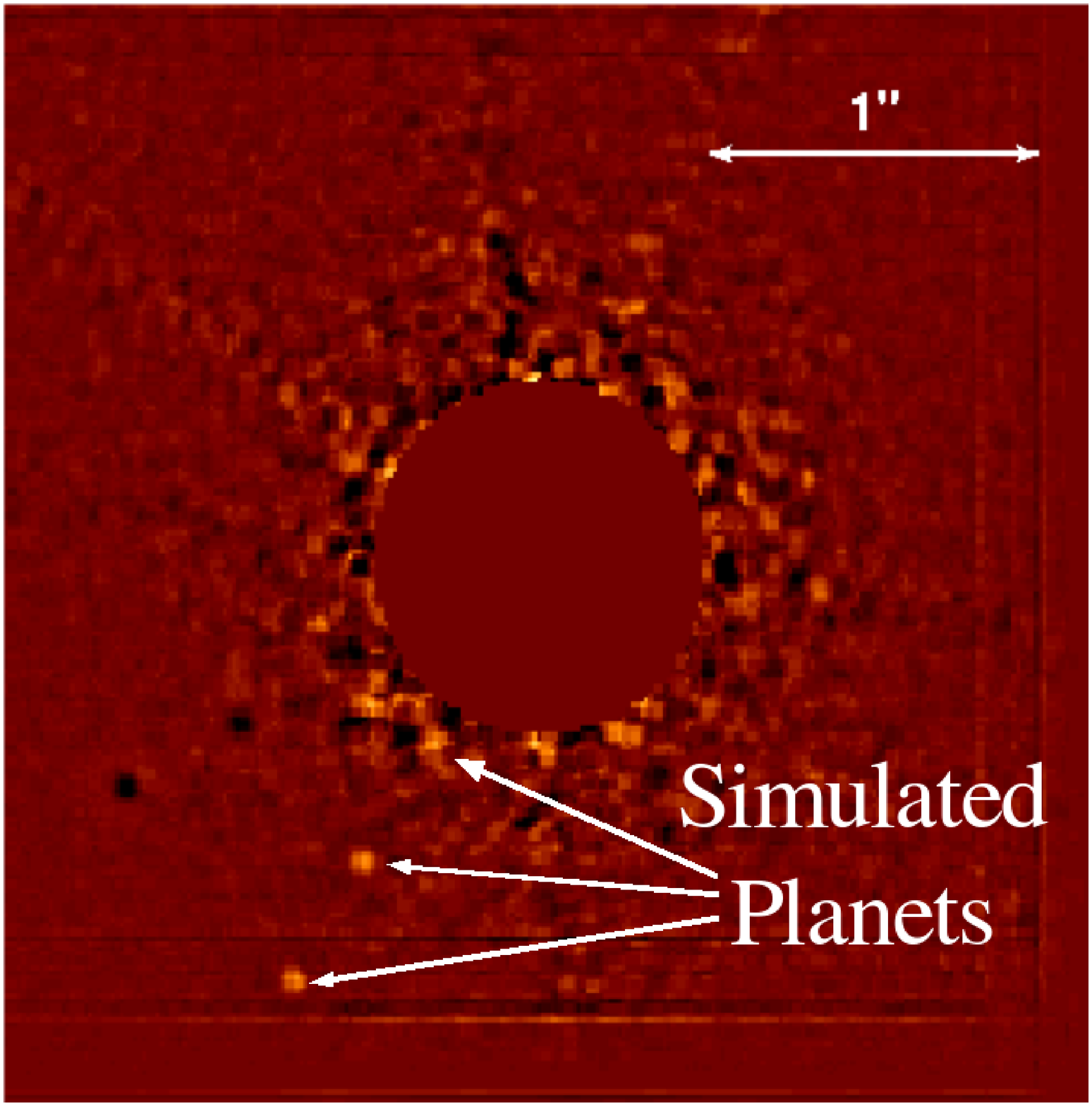} &
    \includegraphics[width=3in]{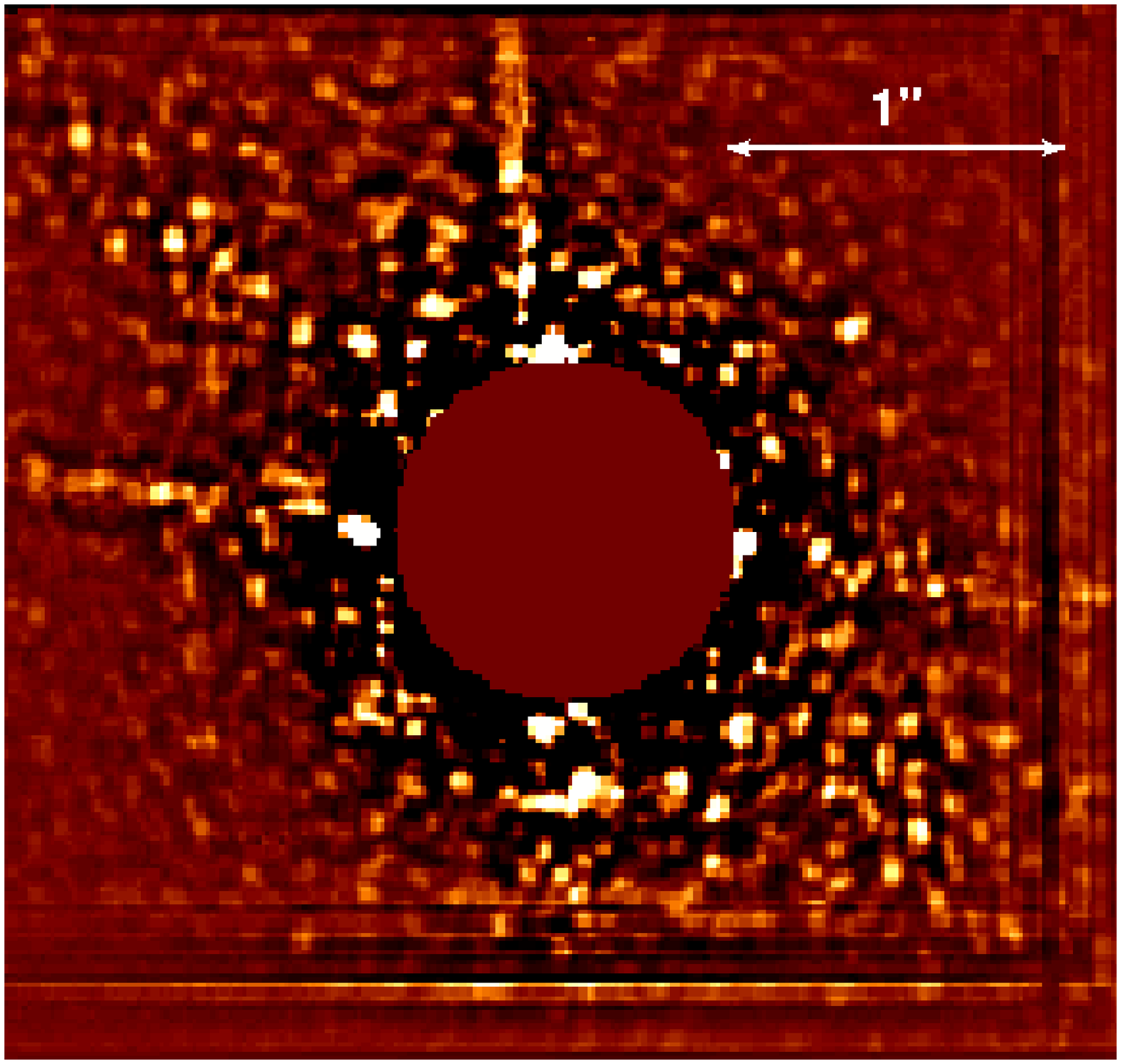} \\
   \end{tabular}
   \end{center}
   \caption[Reduced VLT SDI data]
   { \label{fig:SDIRED} {\bf Left:} A complete reduced dataset 
(28 minutes of data at a series of rotator angles (``roll angles'') -- 
0$^{\circ}$, 33$^{\circ}$, 33$^{\circ}$, 0$^{\circ}$) from the VLT SDI device.
Simulated planets have been added at separations of 
0.55, 0.85, and 1.35'' from the primary, with $\Delta$F1(1.575$\mu$m) = 10 mag 
(star-planet contrast in magnitudes) fainter than the primary.
  These planets are scaled from unsaturated images of the example star 
(AB Dor A) taken right before the example dataset (and have fluxes and 
photon noise in each filter 
appropriate for a T6 effective temperature).  
Past 0.7'', the simulated planets are 
detected in both roll angles with S/N $>$ 10.
Observing at two different roll angles produces two 
independent detections, and hence makes the chance of 
detecting a ``false positive'' almost null.
{\bf Right:} Standard AO data reduction of the same
dataset.  
Filter images have been coadded (rather than subtracted), 
flat-fielded, sky-subtracted, and 
unsharp-masked.  Simulated planets have been added with the 
same properties and at the same separations as before.  None of the simulated
planets are clearly 
detected in the standard AO reduction.  Additionally, many more
bright super speckles remain in the field.
}
\end{figure}

A fully reduced $\sim$30 minute dataset of AB Dor A (70 Myr K1V star 
at a distance of 14.98 pc, V=6.88)  from the VLT SDI device 
is presented in Fig.~\ref{fig:SDIRED}.  Simulated planets have been 
added at separations of 0.55, 0.85, and 1.35'' from the primary, 
with $\Delta$F1(1.575$\mu$m) = 10 mag 
(attenuation in magnitudes in the 1.575 $\mu$m 
F1 filter) fainter than the primary.  For details and further discussion
 of these planet simulations see Section 3.4.

\section{The SDI Survey}

\subsection{Survey Design / Target Selection}

Survey objects were selected primarily on the basis of youth and proximity.  
With a number of exceptions, our 54 survey objects are within 50 pc of the 
Sun and less than 250 Myr in age.  (The 9 exceptions include three 
somewhat older stars with known radial velocity planets, 
2 more distant ($<$150 pc) 
stars with extreme youth indicators, and 4 older nearby young solar 
analogues which were initially misclassified as young objects.)  
Distances were obtained for 48 of our objects from Hipparcos parallax
measurements \citep[parallaxes of $>$0.02'', corresponding to distances 
$<$50 pc,][]{per97}.  Stars were age-selected according to two methods: 
1) if possible, 
according to young cluster membership (and adopting the established age 
for that cluster) for clusters with well established ages such as the Beta
Pic, TW Hya, AB Dor and Tuc-Hor moving groups
or 2) according to other age indicators including 
the strength of spectral age indicators (for instance, the Li 6707, 
the Calcium H and K lines, and H$\alpha$ emission)  
as well as from X-ray emission, variability, and rotational speed.
As moving group ages are generally more robust than measurements for 
individual stars, we expect the ages of stars in these associations, on 
average, to have greater accuracy.  
Our survey covers stars in the Beta Pic, TW Hya, AB Dor, IC 2391, 
and Tucanae/Horologium moving groups. 

\begin{figure}
\begin{center}
\begin{tabular}{c}
\includegraphics[width=\columnwidth]{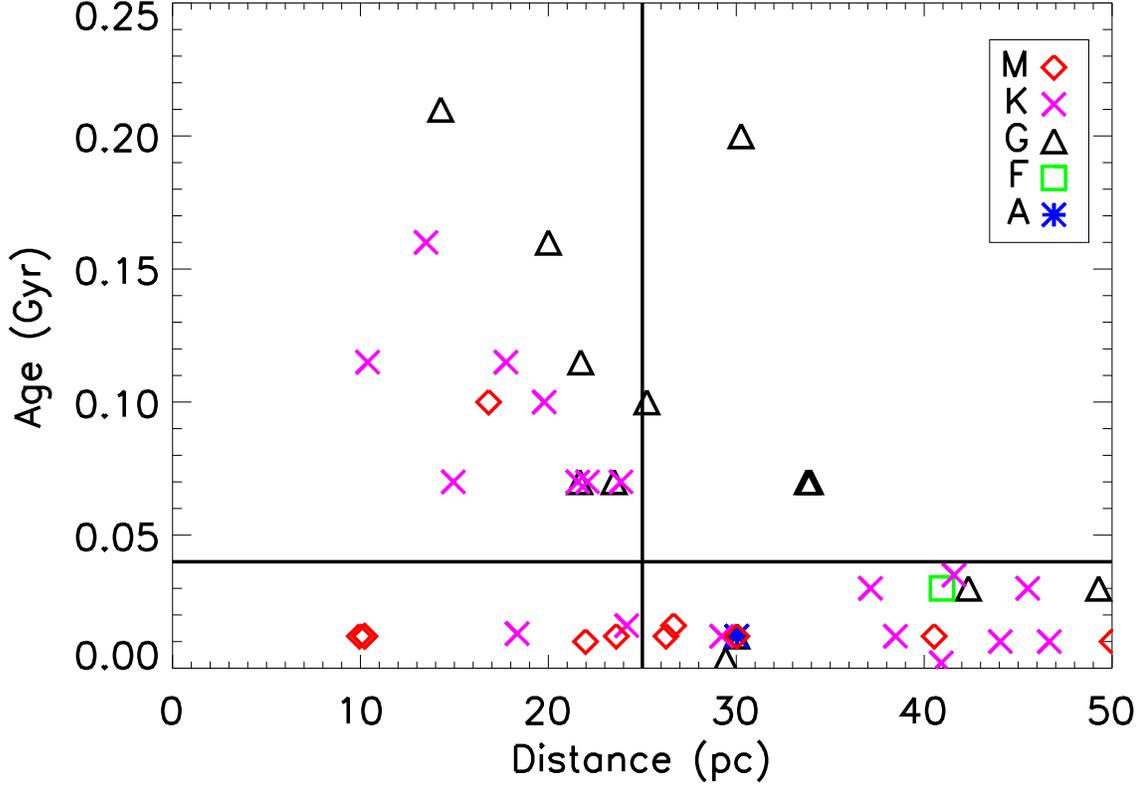}	
\end{tabular}
\end{center}
\caption[Age vs. Distance]
	{\label{fig:agevsdist}Age vs. distance for our survey stars.
Spectral types are delineated by plot symbols.  Objects were selected 
according to youth and proximity to the Sun.  
45 of our survey objects are within 50 pc 
of the Sun and less than 250 Myr in age.  Of the remaining objects, 
2 are very young ($<$10 Myr), somewhat more distant ($<$150 pc) objects,
3 are nearby stars with known RV planets, 
and 4 are nearby solar analogues ($<$20 pc) that were initially misclassified 
as young.  
We selected targets according to two overlapping criteria (shown on plot
 as solid black lines) 1) stars within 25 pc and younger than 250 Myr and
2) stars within 50 pc and younger than 40 Myr.
Stars were age-selected according to association membership, or, 
in the case of unassociated stars, age indicators such as 
the strength of the Li 6707 \Angstrom~line, Calcium H and K lines, H $\alpha$
emission, X-ray emission, etc.  Distances were obtained from Hipparcos parallax
measurements (parallaxes of $>$0.02'').
Our ``median'' survey object is a K star with an age of 30 Myr and 
at a distance of 25 pc.} 
\end{figure}

We select targets stars based on two overlapping criteria: 1) stars within 
25 pc and younger than 250 Myr, and 2) stars within 50 pc and younger than 
40 Myr (see Fig.~\ref{fig:agevsdist}).  Our original list has 
been modified according to the amount of allocated time at the telescope,
the unavailability of GTO targets, as well as severe weather 
constraints for the MMT portion of our survey.
At the VLT, our observing runs spanned the months of August through February 
over 2004 and 2005.  Thus, due to the spacing of observing runs,
in the south, the survey is close to complete from $\sim$17 - $\sim$13 
hours RA.  At the MMT, we had two observing runs, one 
in May 2005 and one in February 2006.  Thus, in the north, the survey is 
complete for the RA range 11 - 21 hours.

Survey objects are presented in Table 1.  A detailed table of observations
is presented in Table 2.   
Survey objects are plotted as a function of distance and 
age in Fig.~\ref{fig:agevsdist}.  Our ``median'' survey object is a K
star with an age of 30 Myr and at a distance of 25 pc.

\subsection{The Performance of the SDI Filters as Spectral Indices}

It is important to carefully consider the expected strength of 
the 1.62 $\mu$m methane absorption break utilized by the SDI device.
The stronger the break strength, the more companion light is preserved after
SDI filter subtraction.  For a candidate object with a weak break strength,
SDI subtraction may effectively attenuate the candidate object
itself, rendering it 
undetectable (although, at separations $>$ 0.15'', a bright object may still be
detectable due to the characteristic dark-light radial pattern 
produced by any real object after pipelining, see Section 2.2.)

To determine the methane break strength expected for a candidate
object (and thus, the expected performance of SDI for that candidate), 
we define an SDI methane spectral index calculated from 
our SDI F1(1.575 $\mu$m) and F3(1.625 $\mu$m) filter images 
\citep[similar~to~the~methane~spectral~index~defined~by~][]{geb02}.



\begin{equation}
index(\frac{F1}{F3}) = 
\frac{\int^{\lambda_2 = 1.5875 \mu m}_{\lambda_1 = 1.5625 \mu m} S_{\lambda} F1({\lambda}) d\lambda}{\int^{\lambda_4 = 1.6125 \mu m}_{\lambda_3 = 1.6375 \mu m} S_{\lambda} F3({\lambda}) d\lambda}
\end{equation}

Each SDI filter was manufactured by 
Barr Associates to have a precise bandwidth of  
0.025 $\mu$m, so the wavelength
intervals ($\lambda_2$ - $\lambda_1$ = $\Delta \lambda$ = 
$\lambda_4$ - $\lambda_3$) in the numerator and denominator have 
the same length for the SDI methane index.

\begin{figure}
\includegraphics[angle=0,width=6in]{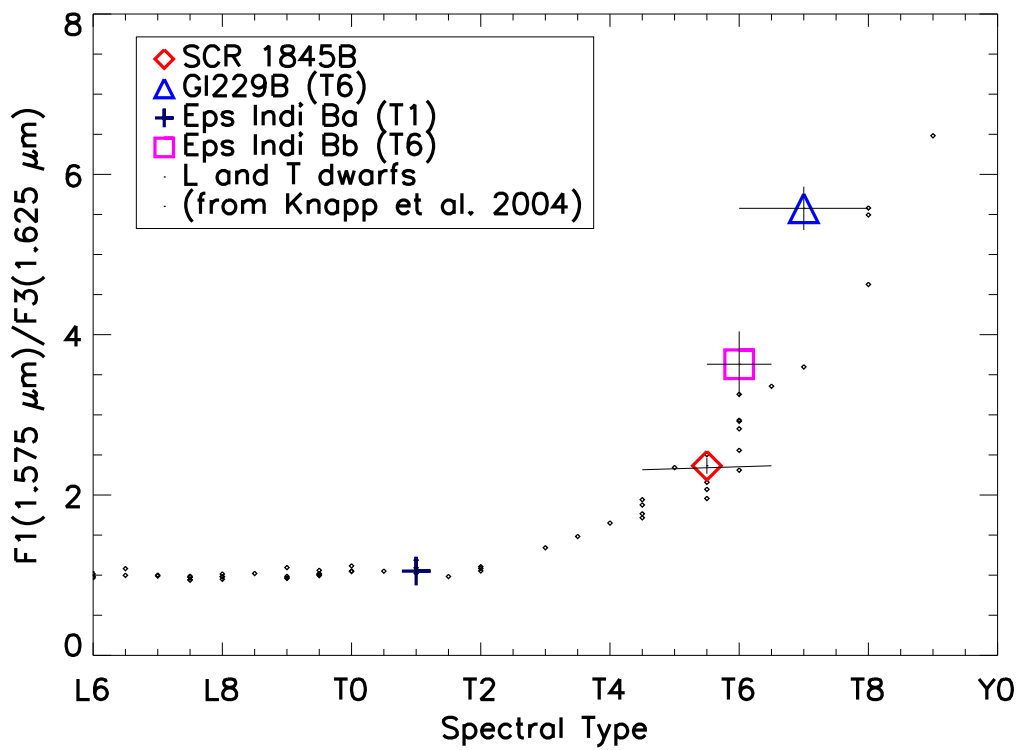}
\caption{ SDI methane spectral indices for the T dwarfs SCR 1845B,
Gl 229B, $\epsilon\,$Ind Ba, and $\epsilon\,$Ind Bb (from Biller et al. 2006b).
As a comparison, SDI methane spectral indices calculated from spectra for 94
L and T dwarfs \citep[spectra~from~][]{kna04} are overplotted. 
SCR 1845B, Gl 229B, and $\epsilon\,$Ind Bb show strong methane indices, 
whereas $\epsilon\,$Ind Bb (T1) is relatively constant in flux 
across the SDI filters and has a much lower methane index.  \citet{geb02}
note that Gl 229B has an anomalously high
methane index for its spectral type.  While \citet{geb02} find an 
overall spectral type of T6$\pm$1 for Gl 229B, they assign Gl 229B a 
spectral type of T7 based on the methane index (which we adopt here).  }
\label{fig:fluxes}
\end{figure}

We calculated SDI spectral indices for the four brown dwarfs 
which have been observed with SDI -- the T6 Gl 229B 
\citep[][]{nak95}, 
the T5.5 SCR 1845B \citep[][]{bil06b} and $\epsilon\,$Ind Ba-Bb 
(T6 + T1) \citep[][]{mcc04}.  
Since we only possess SDI data on a limited number of T dwarfs,  
we calculated the same SDI spectral 
indices from spectra of 56 L dwarfs and 35 T dwarfs \citep{kna04}
in order to evaluate the performance of the SDI for  
a wide range of L and T dwarf objects.
Spectra for these objects were obtained from 
Sandy Leggett's L and T dwarf archive
\footnote[4]{http://www.jach.hawaii.edu/$\sim$skl/LTdata.html}.  In order to make an accurate comparison, 
SDI filter transmission curves were convolved into these
calculations (see Fig.~\ref{fig:SDIFILT}).  
Since we have full spectral data for these objects, 
we also calculated the 1.62 $\mu$m methane spectral
index defined by \citet{geb02}, which were found to be similar to our 
SDI methane spectral indices.
SDI methane spectral indices are plotted for 
both the M9 and T6 components of SCR 1845, the T dwarfs
Gl 229B, $\epsilon\,$Ind Ba, $\epsilon\,$Ind Bb, 
and 94 other L and T dwarfs in Fig.~\ref{fig:fluxes}.
\citet{geb02}
note that Gl 229B has an anomalously high
methane index for its spectral type and assign a large 
uncertainty to Gl 229B's spectral type -- T6$\pm$1 -- which is also 
reflected in its anomalously large SDI spectral index compared to other T6
dwarfs.  From this analysis, we conclude that the SDI device can effectively 
detect objects with spectral type later than T3.  Since T dwarfs with spectral
type earlier than T3 are relatively uncommon compared to later T dwarfs, the
SDI device can effectively detect the full range of 
extrasolar giant planet / brown dwarf spectral types of interest.
According to the models of Burrows et al. 2003 and Marley
et al. 2006, planets $>$10 Myr old should possess T$_{eff}$ $<$ 800 K and
have spectral type of T8 or greater.  

\subsection{Contrast Limits and Minimum Detectable Planet Separation}

\begin{figure}
\begin{center}
\begin{tabular}{cc}
\includegraphics[height=4cm]{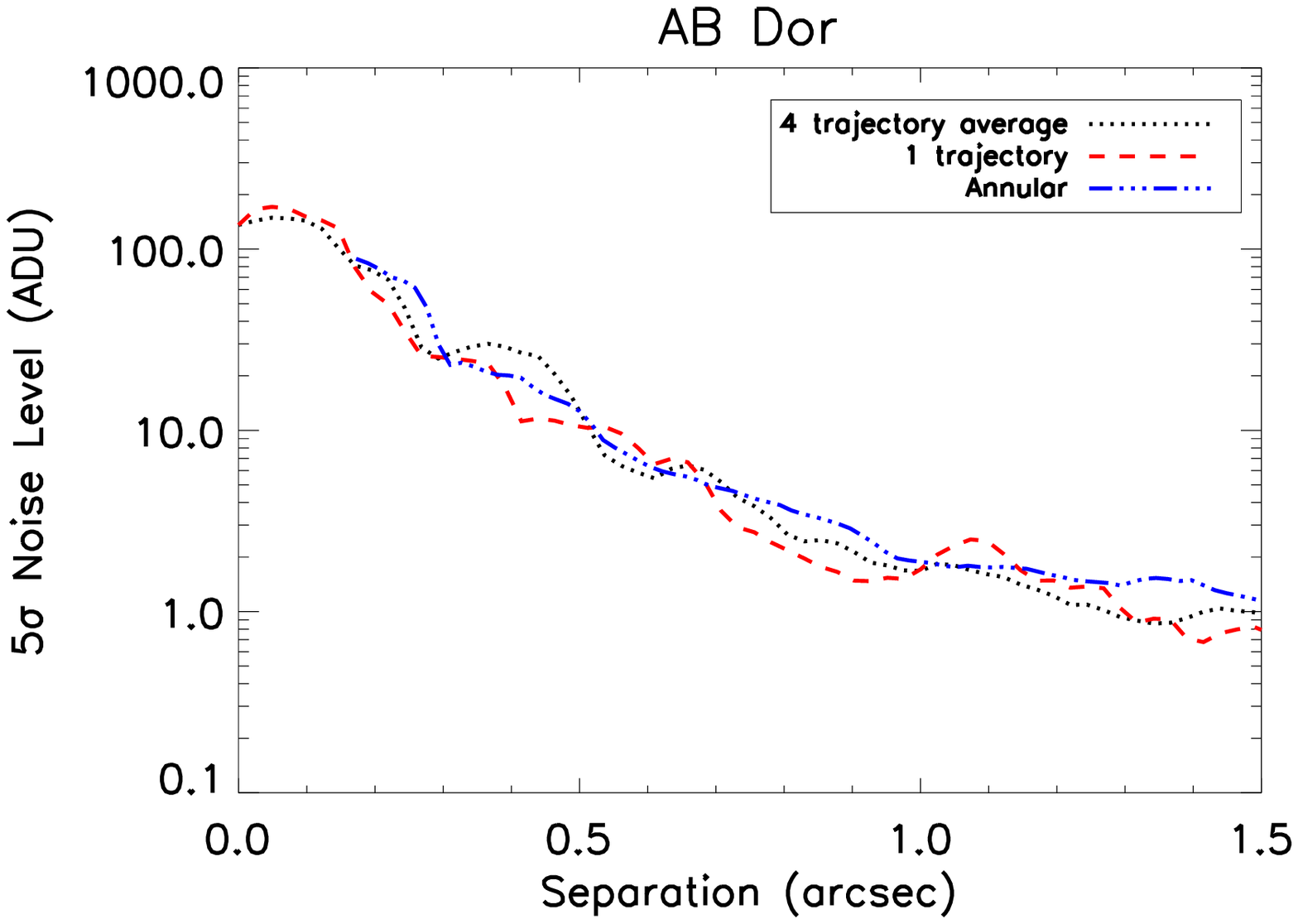} & 
\includegraphics[height=4cm]{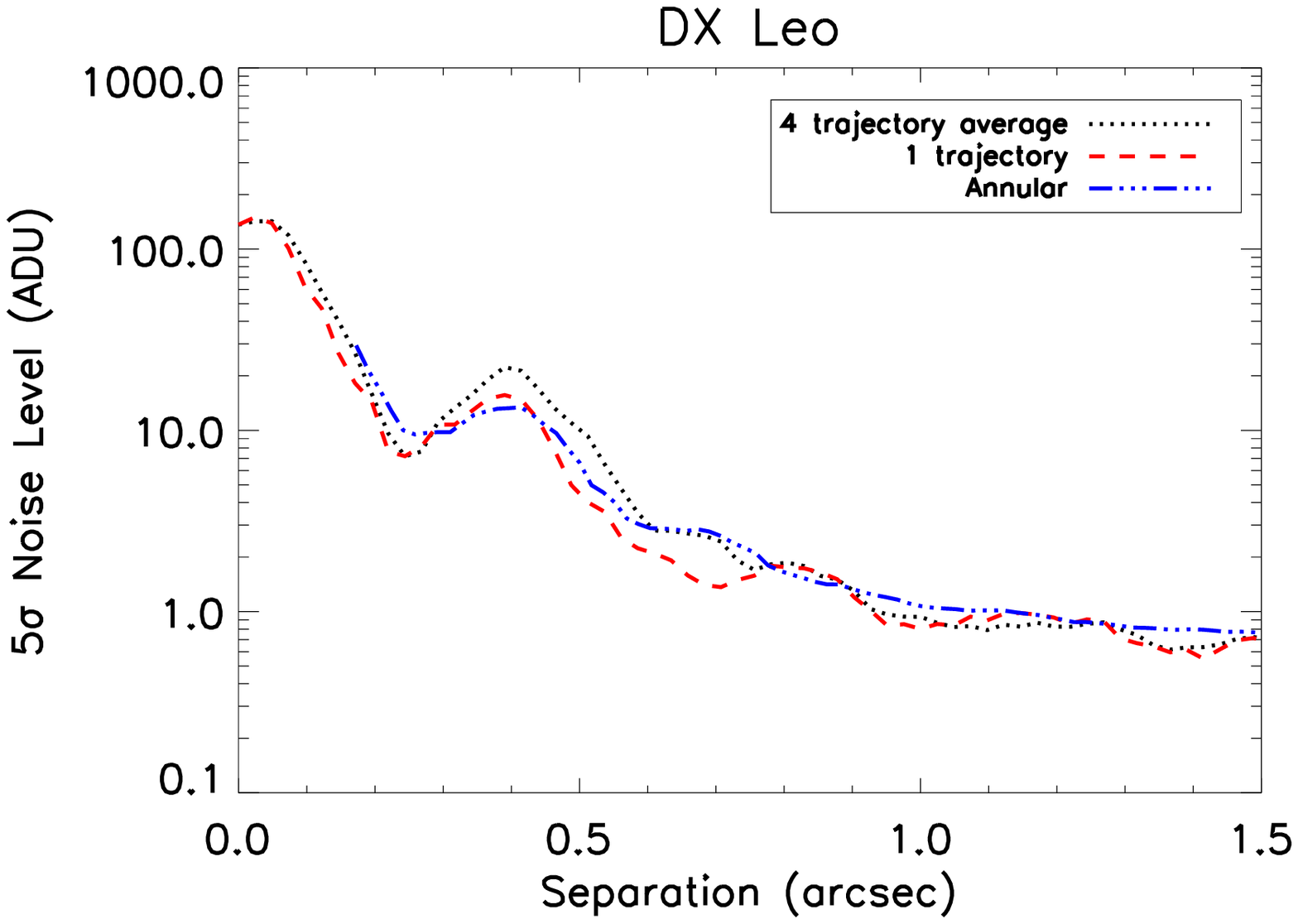} \\ 
\includegraphics[height=4cm]{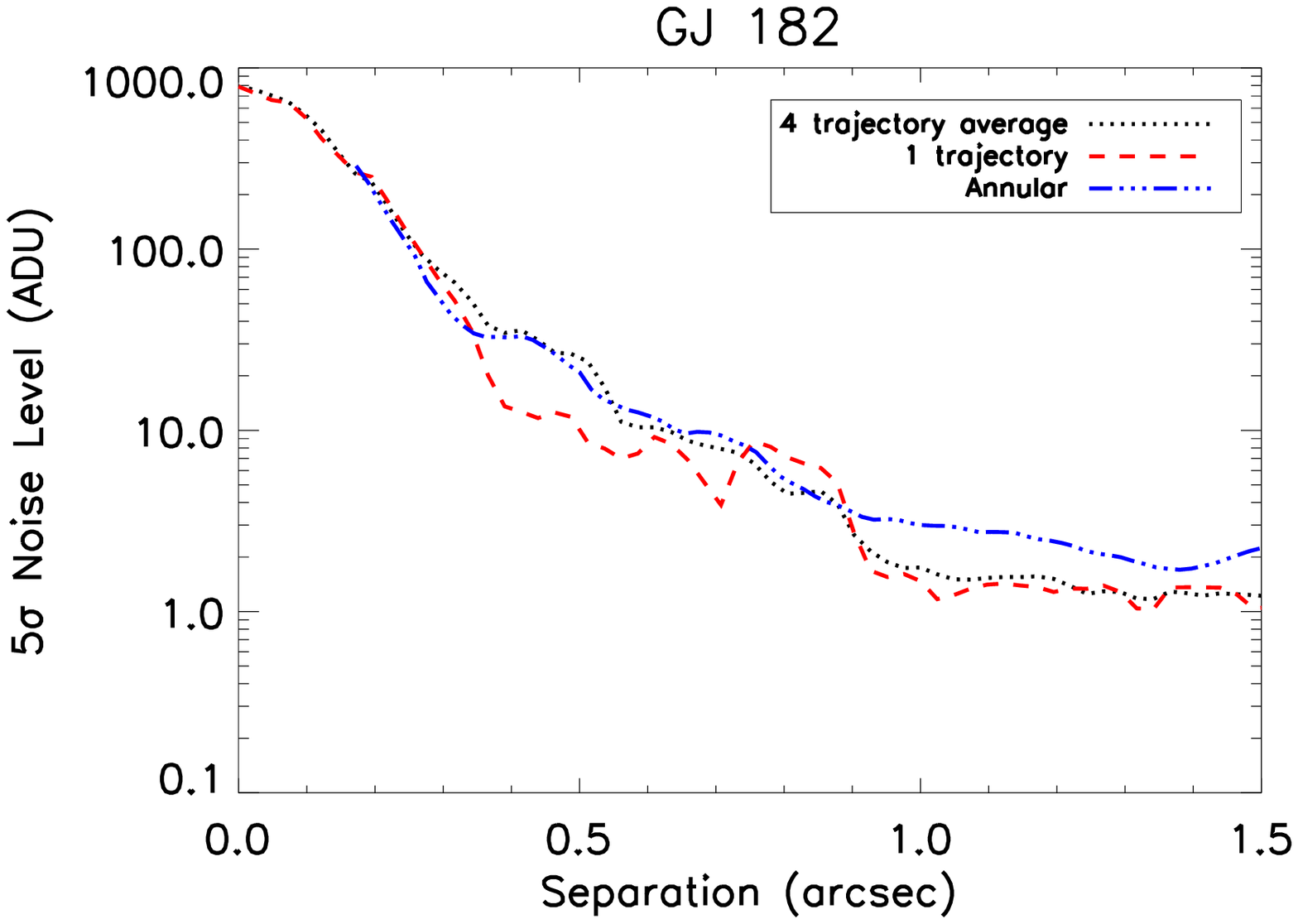} & 
\includegraphics[height=4cm]{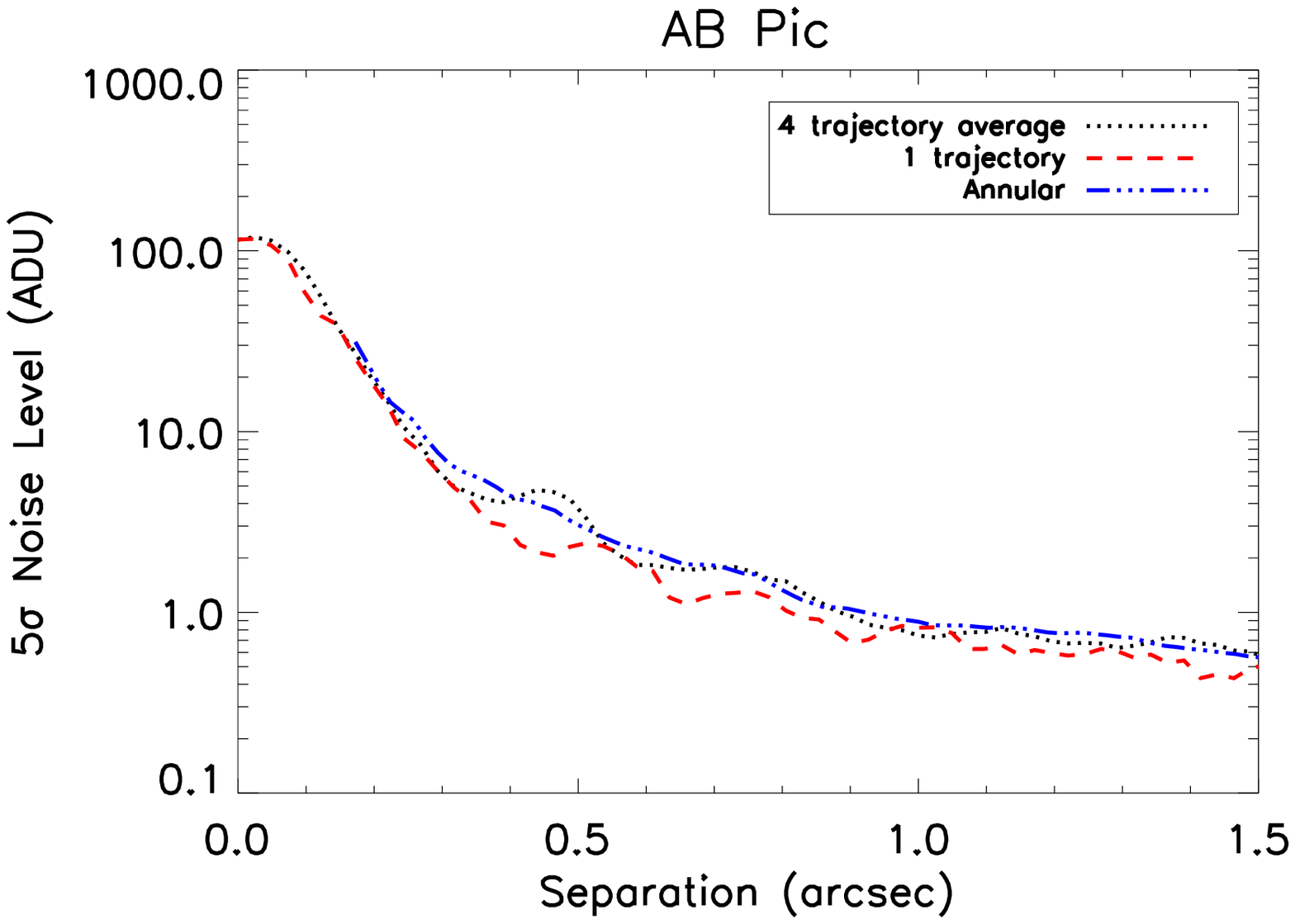} \\
\includegraphics[height=4cm]{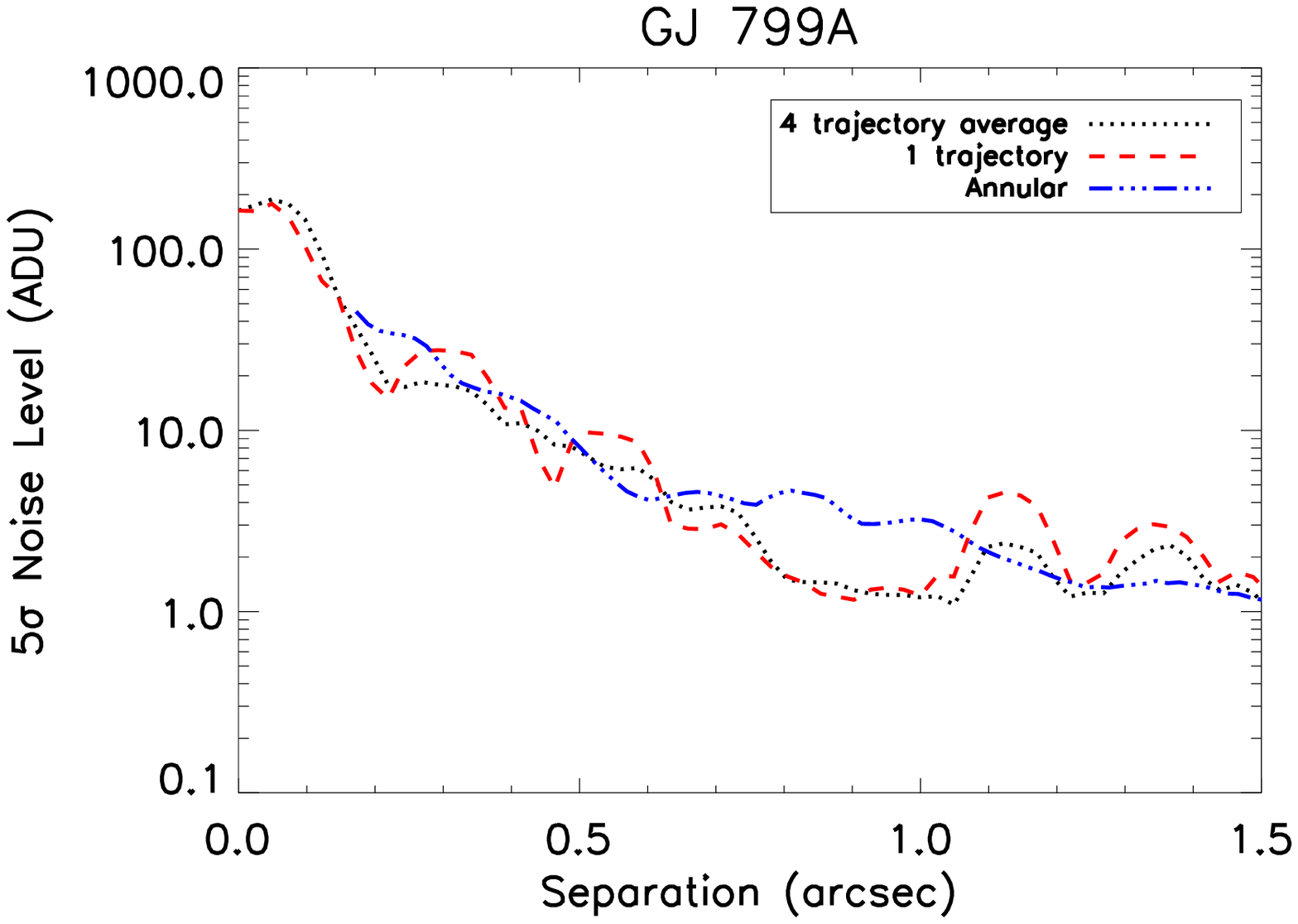} & 
\includegraphics[height=4cm]{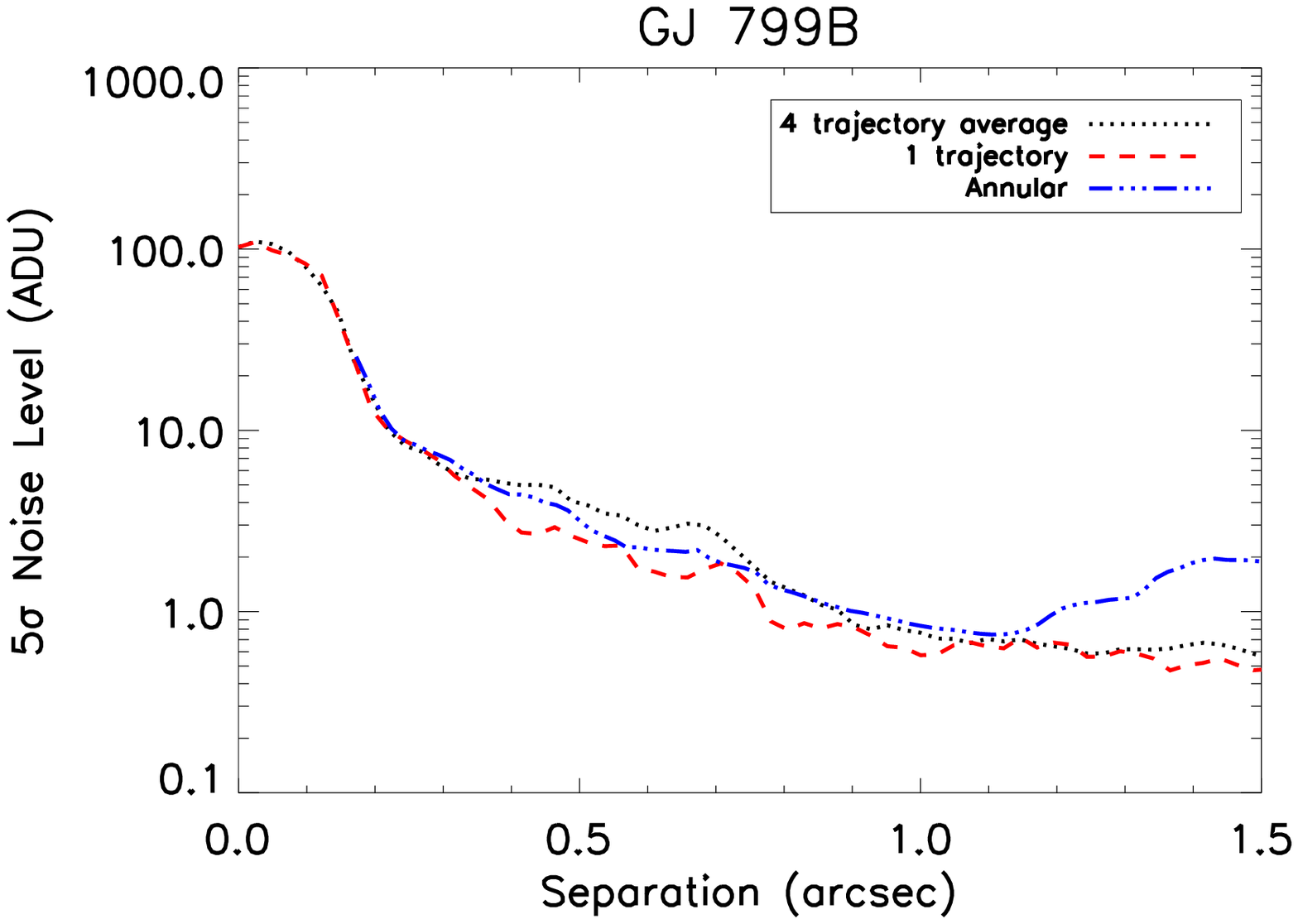} \\ 
\end{tabular}
\end{center}
\caption{Comparison of Noise Curves generated in 3 different manners for 
a set of 6 typical program stars (upper left: AB Dor, upper right: DX Leo, 
middle left: GJ 182, middle right: AB Pic, lower left: GJ 799A, lower right: GJ 799B).  
Noise curves were generated by:
1) translating a 6$\times$6 pixel (0.1''$\times$0.1'') box
along a particular radial trajectory away from the center of the star
image (typical PSF FWHM was 3-5 pixels) then calculating the standard 
deviation in the box at each point along this trajectory, 2)
averaging noise curves generated along four such trajectories, 
and 3) calculating the standard deviation within annular regions 6 pixels 
in width centered on the primary PSF (spider diffraction spikes were not 
masked out in this case because they are already well removed by the spectral
difference).  In general, all three methods produce remarkably
similar noise curves and are equally suitable for characterizing 
the noise properties of an observation.  Since it preserves pixel to pixel 
contrast variations due to speckle noise, the single trajectory method
better simulates the S/N issues encountered in searching for faint 
companions.}
\label{fig:contcomp}
\end{figure}

To determine the range of possible star-planet contrasts achieved in our 
survey, we generated noise curves as a function of radius for every 
survey star.  We tested three different methods of generating noise curves:
1) translating a 6$\times$6 pixel (0.1''$\times$0.1'') box
along a particular radial trajectory away from the center of the star
image (typical PSF FWHM was 3-5 pixels) then calculating the standard 
deviation in the box at each point along this trajectory, 2)
averaging noise curves generated along four such trajectories, 
and 3) calculating the standard deviation within annular regions 6 pixels 
in width centered on the primary PSF (spider diffraction spikes were not 
masked out in this case because they are already well removed by the spectral
difference).  Noise curves generated in these three manners are presented
for a set of 6 typical program stars (AB Dor, DX Leo, GJ 182, AB Pic, 
 GJ 799A, and GJ 799B) 
in Fig.~\ref{fig:contcomp}.  In general, all three methods produce remarkably
similar noise curves and are equally suitable for characterizing 
the noise properties of an observation.  However, 
we choose to utilize the single trajectory 
method because it best simulates the particular signal to noise issues 
encountered when searching for faint companions among super-speckles of 
similar intensity and FWHM (since it preserves pixel to pixel noise variations
due to super-speckles).  The annular method averages out 
speckle noise properties azimuthally.  This produces somewhat unrealistic
results in the case of a faint companion search where one is concerned
only with the speckle structure within the local area of a candidate faint
companion -- speckle structure on the other side of the image is 
unimportant.  In addition, we have tried to choose very ``typical'' 
trajectory per star -- ideally, trajectory to trajectory variations will 
average out across the entire survey.

Noise curves for each program star 
were calculated along a trajectory 45$^{\circ}$ from the image x axis in 
the first quadrant.  The 45$^{\circ}$ was selected as one of many possible 
representative trajectories which was unaffected by instrumental effects
such as spider arms, vibrations along azimuth or altitude mounts, etc.    
At each point along this trajectory, the standard deviation was calculated
(except for the PSF noise curve, for which the mean was calculated).  

\begin{figure}
 \includegraphics[angle=0,width=6in]{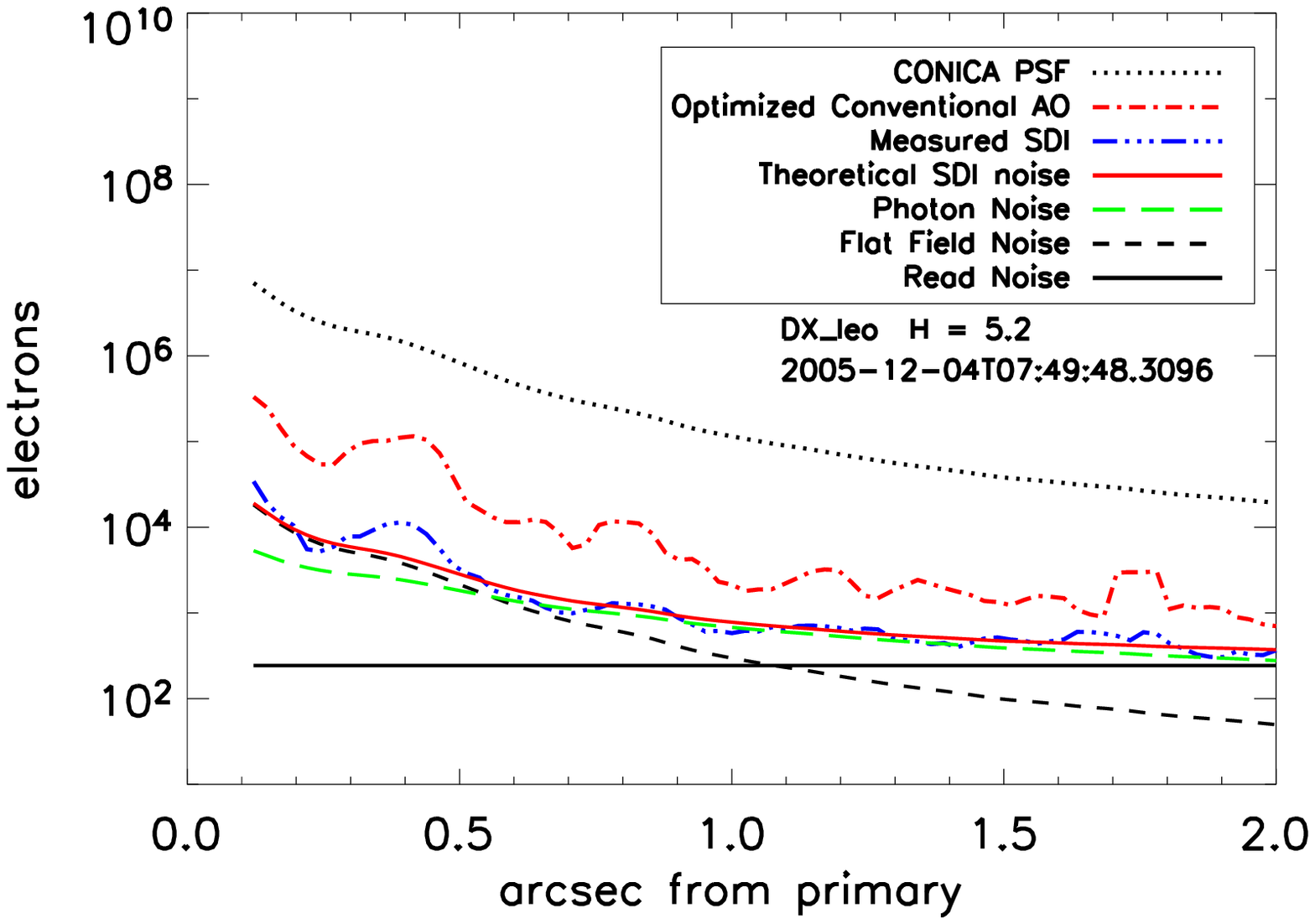}
\caption{Sensitivity curve for DX Leo (18 pc, K0V, 115 Myr, V=7.05, H=5.242).
This is 28 minutes of VLT SDI data.  The CONICA PSF curve is the 
median combination 
of all the F1(1.575 $\mu$m) filter images for this dataset (with a gain
correction applied which accounted for the  
number of exposures, dithers, and roll angles). 
The ``optimized conventional AO'' curve was generated by averaging images from 
all three filters at each roll angle, unsharp masking to remove 
low spatial frequencies, then subtracting the 
combinations at different roll angles from each other.
The ``measured SDI'' data curve is the full 
reduced and differenced SDI data for this object 
(F1(1.575 $\mu$m) - F3a(1.625 $\mu$m) for two roll angles).
The ``theoretical SDI noise'' curve is calculated from photon noise 
(long dashed green curve), flat-field noise (short dashed black curve), 
and read noise (solid black line) added in quadrature.  
Within 0.5'', the SDI data is 
``flat-field'' noise limited.  (In reality,
we are limited by super speckle residuals within this radius.   
Our flat fields are 
accurate to the $\sim$1$\%$ level, but the speckle residuals $<$0.5'' 
vary more than this and thus dominate the SDI noise.)  From
0.5'' onwards, the SDI data is photon-noise limited, asymptotically
approaching the
read-noise limit at separations $>$ 2''.  For a complete set of 
sensitivity curves, 
see: http://exoplanet.as.arizona.edu/$\sim$lclose/SDI.html.}
\label{fig:DXLeo}
\end{figure}

A fully labeled example noise curve for the star DX Leo is presented in
Fig.~\ref{fig:DXLeo}.
Noise curves were generated for a number of cases for each object.
First, a noise curve was generated for the full 
reduced and differenced SDI data (labeled SDI data curve) 
(F1(1.575 $\mu$m) - F3a(1.625 $\mu$m) for two roll angles).
A PSF noise curve 
curve was generated from a  
median combination of all the F1(1.575 $\mu$m) filter images for each dataset
weighted according to the number of exposures, dithers, and roll angles 
in the dataset. 
To recreate the equivalent observation without using the SDI technique 
(and thus characterize the performance of SDI compared to conventional 
AO techniques),  
an ``optimized conventional AO'' curve was generated by combining images from 
all three filters at each roll angle: 

\begin{equation}
Broadband = F1(1.575 \mu m) + F2(1.6 \mu m) + F3(1.625 \mu m)
\end{equation}

then unsharp masking to remove low spatial frequencies, 
and subtracting the ``Broadband'' combinations at different roll angles
from each other.

To characterize the noise level in each observation, we calculated 
an SDI noise curve, which is a combination of photon-noise, flat-field 
noise, and read noise.  Per exposure:

\begin{equation}
\sigma_{SDI} = \sqrt[]{\sigma_{photon}^2 + \sigma_{flatfield}^2 + \sigma_{readnoise}^2}
\end{equation}

Photon-noise was calculated as:

\begin{equation}
\sigma_{photon} = \sqrt{n_{electrons}}
\end{equation}

Readout noise for the CONICA detector at the VLT in Fowler sampling mode is
1.3 ADU (analog-to-digital unit).  
The gain for the latest CONICA detector in the Fowler sampling mode is 
12.1 electrons/ADU so $\sigma_{readnoise}$ = 15.73 electrons.

NACO and ARIES flat fields were found to be accurate to about 1$\%$, so
flat-field noise was estimated as:

\begin{equation}
\sigma_{flatfield} = \epsilon n_{electrons}
\end{equation}

where $\epsilon$=0.01.  The total noise for a full observation (4-5 
dithers, 2-4 roll angles) was then calculated by weighting the SDI noise 
per exposure by the 
number of exposures (NDIT $\times$ number of dithers $\times$ 
number of roll angles):

\begin{equation}
\sigma_{SDI\_fullobs} = \sigma_{SDI} \sqrt{NDIT \times (number~of~dithers) \times (number~of~roll~angles)}
\end{equation}

The PSF curve for a full observation was similarly weighted:

\begin{equation}
PSF = (median PSF) \times NDIT \times (number~of~dithers) \times (number~of~roll~angles)
\end{equation}

For the sample curve shown in Fig.~\ref{fig:DXLeo}, 
the SDI data is ``flat-field'' limited within 0.5'' of the star.  From
0.5'' onwards, the SDI data is photon noise limited, approaching the
read-noise limit at separations $>$ 2''.

We converted our noise in electrons to attainable 
contrasts in magnitudes in the 
F1(1.625 $\mu$m) filter --  
contrast plots in $\Delta$mag
are presented for all non-binary survey objects 
in Figs.~\ref{fig:contrasts1} to ~\ref{fig:contrasts7} 
according to the H magnitude
of the primary for the VLT and according to observing run for the MMT.  
For every observation which possesses an unsaturated 
acquisition image (typically 10$\times$0.1 s images taken over $\sim$30 s), 
the stellar 
peak in the unsaturated acquisition image was used to scale the saturated
stellar peak in the saturated data images and thus attain accurate
contrasts in magnitudes.  For observations lacking an unsaturated
acquisition image, contrast curves for other stars which had similar 
peaks, read noise values, and shape to the contrast curve in question 
were selected from the library of contrast plots in electron units.  The
peaks utilized for these matching contrast curves were then used to scale  
the observation missing an acquisition image.  A peak of 2.2$\times$10$^5$
was adopted for $\epsilon\,$Eri (Kellner et al. 2007, Janson et al. 2007) 
and $\epsilon\,$Ind A (Gei{\ss}ler et al. 2007).  We present contrast 
curves for 48 stars in this paper; the remaining six survey stars were 
either very close binaries, making it difficult to generate a contrast curve, 
or had particularly low quality datasets.  

For the VLT data, attainable contrast depends on primary 
star H magnitude as well as seeing FWHM 
and Strehl ratio during the observation.  
For the brightest stars in 
the survey (H$<$4.5), we attain 5$\sigma$ contrasts of 
$\Delta$F1$\sim$12 mag at 
separations of $>$1'' from the star.  For the faintest survey stars,
we only attain 5$\sigma$ contrasts of 
$\Delta$F1$\sim$10 mag $>$1'' from the star.  However, considerable 
spread in attained contrast is observed in each H magnitude bin -- 
most likely due to variations in observing conditions (seeing, Strehl 
ratio, etc.) across multiple observations.  To quantify the effect of 
seeing on attainable contrast, in Fig.~\ref{fig:seeing} 
we plot the seeing FWHM (averaged over the observation -- the error bars on seeing are the 
seeing variations as measured by the standard deviation of 
the seeing over each observation) vs. attained 
5$\sigma$ contrast at 0.5$\arcsec$ 
for 10 of the stars presented in Fig.~\ref{fig:contrasts2}
with H magnitudes between 4.5 -- 5.5.  For this sample of stars with 
similar H magnitudes, achievable contrast is roughly inversely 
proportional to the seeing FWHM.  A fair amount of scatter is apparent in this
plot and is due in part to seeing variations over the course of 
each observations.  Seeing FWHM can vary  
considerably over the 20-40 minute timescale of a typical SDI 
observation, affecting the AO system performance 
and thus the achievable contrast.

However, higher attained contrast does not necessarily translate across
the board to a lower minimum detectable planet mass. 
Although one might be able to attain a very high contrast 
(5$\sigma$ contrast $>$11 mag at 1'' limited by photon noise) for a bright 
young A star, one 
would have more luck searching for low luminosity planets 
around an intrinsically 
faint young M star (5$\sigma$ contrast $\sim$9 mag at 1'' limited by 
read noise), since the 
inherent contrast difference expected between star and planet is 
considerably smaller.  We obtained
contrasts of $\Delta$H$>$10 mag (5 $\sigma$) at 0.5$\arcsec$ for 45$\%$ of 
target objects at the VLT and contrasts of $\Delta$H$>$9 mag (5 $\sigma$) at 
0.5$\arcsec$ for 80$\%$ of our targets.  This is more a statement on the 
spectral types in our sample than a performance related issue.  

In general, the MMT SDI device performed at a slightly lower level than 
the VLT SDI device -- attaining 5$\sigma$ contrasts 0.5-1 magnitude less 
than those achieved at the VLT for similar separation and primary star H
magnitude.  The lesser performance of the MMT system can be attributed to 
two factors.  First, the diameter of the MMT is 6.5m versus the VLT which has
an 8.2 m diameter -- resulting in a considerable decrease in sensitivity.
Additionally, the seeing sampled by the 
MMTAO system was not as stable as for the NACO AO system -- 
Strehl ratios often changed dramatically over an observation, limiting the 
attainable contrast.  However, the MMT SDI results still probe a higher 
contrast regime at separations $<$1'' 
than is possible with standard AO techniques.

In order to determine what objects realistically can be detected
for our survey stars, 
we must convert between our instrumental F1(1.625 $\mu$m) filter magnitudes and
H band magnitudes and then compare the H magnitudes to those expected
from models of young planets 
(such as Burrows et al. 2003).  To accomplish this, the spectra of both 
the primary and secondary components of each target must be taken into account.
 To convert from our F1 filter magnitudes into calibrated H band magnitudes
we must calculate the H band magnitude offsets for both the primary star
and a potential methane companion (Offset$_A$ and Offset$_B$ respectively):

\begin{equation}
 \Delta H = H_A - H_B = (Offset_B + F1_B) - (Offset_A + F1_A) = (Offset_B - Offset_A) + \Delta F1 
\end{equation}

For primary stars with spectral types F-K, we assume that the star has very
little chromatic variation within the middle of the 
H band, so Offset$_A$ is zero
(see Fig.~\ref{fig:SDIFILT}).  For lower mass M stars, which are very red, 
the magnitude offset is not 
negligible.  To take an extreme example, a very low mass M8 primary will
have a magnitude offset of Offset$_A$=-0.12$\pm$0.08 mag (calculated 
using the spectrum of the M8 star VB10, an H transmission
curve, and our F1 filter transmission curve).  The latest stars in our 
survey have spectral type M0- M5, so Offset$_A$ will be $<$0.1 mag for 
these cases.  

Any T3 or later companion to one of our survey stars 
will be blue compared to the primary and will appear ``brighter'' 
in the F1 filter than in the H band 
(in other words, it will have a higher 
``flux'' in the F1 filter ($\#$ photons per unit bandwidth) -- 
see Fig.~\ref{fig:SDIFILT})
-- so Offset$_B$ will definitely 
be non-negligible.  We calculated Offset$_B$ for 
18 objects with spectral types of T4.5-T8 
\citep[spectra~from][]{kna04}, then averaged together by spectral
type to derive an average offset for each spectral type. 
For a T5 companion, Offset$_{T5}$ = 
0.5$\pm$0.05 mag, for a T6 companion, Offset$_{T6}$ = 0.6$\pm$0.07 mag, and 
for a T8 companion, Offset$_{T8}$ = 0.87$\pm$0.04 mag.
While we do not convert our full $\Delta$F1 contrast plots to $\Delta$H
contrast plots, for every survey star 
we calculate limiting $\Delta$H contrasts (5$\sigma$ values), 
 at 0.5'' and 1.0'', equivalent separation in AU, apparent
 H magnitude, and absolute H magnitude for a T8 spectral type companion 
(since extrasolar planets are expected to have spectral type $\ga$ T8,
Burrows et al. 2003).  These results are presented in Tables 4 and 5.  
However, it is difficult to translate our absolute
H magnitudes into model planet masses since we have assumed a T8 spectral
type in our conversion between $\Delta$F1 and $\Delta$H contrasts -- but 
a companion which actually has the
 limiting absolute H magnitude we find (combined with 
the known age and distance of the system) may have a very different spectral
type.   

\begin{figure}
\includegraphics[width=\columnwidth]{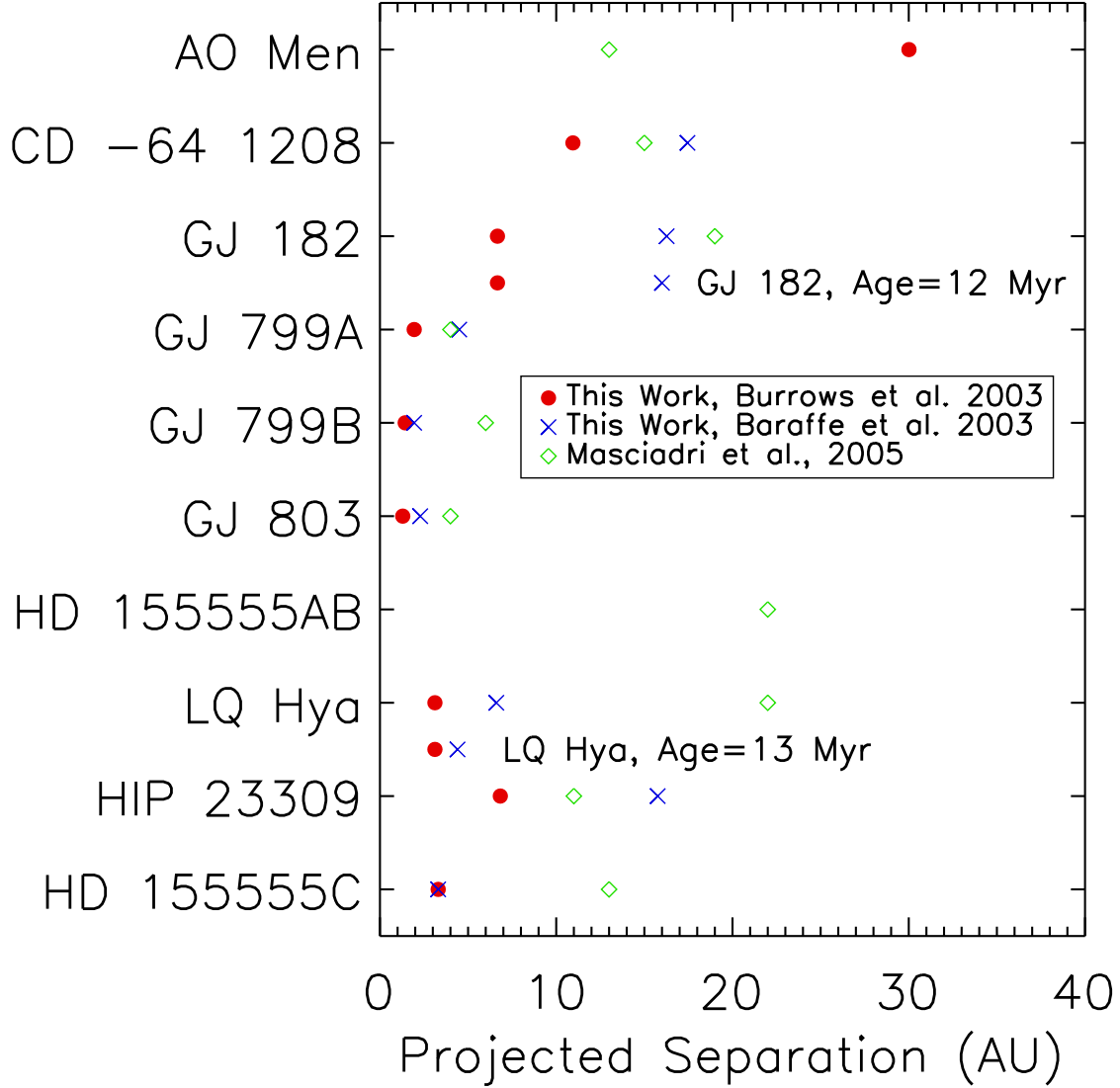}
\caption[Minimum Separations]
{Minimum detectable planet separations for a 5 M$_{Jup}$ planet for the  
10 objects in common between this survey and Masciadri et al. (2005) 
who used VLT NACO without SDI. 
For the purpose of comparison, we have adopted ages from Masciadri et al. 
(2005); we note our preferred age on the figure where our adopted ages
differ from Masciadri et al. (2005). 
We translated theoretical 5 M$_{Jup}$ 
planet models (Burrows et al. 2003, Baraffe et al. 2003) 
into H magnitudes for these 10 cases
then determined the minimum separation at which such a companion 
could be detected (at the 5$\sigma$ level) in our survey. 
For the 10 objects in common between the surveys, our SDI survey attains lower 
minimum separations for 8 out of 10 objects (we note also 
that the two objects for which we did not attain lower separations were 
particularly low quality AO/SDI datasets).  
}
\label{fig:elenacomp}
\end{figure}

Since we cannot translate our H magnitudes directly into planetary mass 
companions, we followed the analysis of Masciadri et al. (2005) and 
translated theoretical planet models (Burrows et al. 2003, 
Baraffe et al. 2003) 
into H magnitudes 
then determined the minimum separation at which such a companion 
could be detected (at the 5$\sigma$ level) in our survey. 
The minimum separation at which a 5 M$_{Jup}$ or a 10 M$_{Jup}$ companion 
could be detected for each of our survey stars is shown in 
Table 6.  Using the Burrows et al. (2003) models, 
for 50$\%$ of the cases in our survey
we detect no 5 M$_{Jup}$ planets at separations larger than 18.6 AU
and no 10 M$_{Jup}$ planets are separations larger than 7.5 AU.  While 
these numbers are comparable to those found in Masciadri et al. (2005), 
our current survey actually attains higher contrasts on a case by case basis 
than Masciadri et al. (2005).    
Our median survey object has an age of 50 Myr whereas the median survey object of 
Masciadri et al. (2005) has a considerably younger age of 12 Myr -- 
the star-planet contrast is less at younger ages, 
thus one would expect a younger object
to have a lower minimum separation at a given attained contrast 
 than a similar but older object.  
For the 10 objects in common between the surveys, our survey attains lower 
minimum separations for 8 out of 10 objects (we note also 
that the two objects for which we did not attain lower separations were 
particularly low quality SDI datasets).  Minimum detectable separations 
for a 5 M$_{Jup}$ object for the 
10 objects in common are plotted in Fig.~\ref{fig:elenacomp} (using the
ages adopted by Masciadri et al. 2005).  Our
survey is generally more sensitive than Masciadri et al. (2005) on 
shared stars because the SDI technique allows us to achieve higher 
contrasts closer to the star (separations of 0.3'' - 1.0'') compared to 
the deep broad-band imaging technique of Masciadri et al. (2005), thus allowing 
us to potentially detect companions at tighter separations.
We also 
shared 4 survey objects in common with Lowrance et al. (2005) and 1 object 
($\epsilon\,$Eri) in common with Luhman and Jayawardhana (2002).  In all 
of these cases, our limiting contrasts at 0.5'' 
($\Delta$H$\sim$10-11 mag) are considerably higher than those attained in 
these previous surveys ($\Delta$H$\sim$6.5-7.6 mag), thus we are
sensitive to planets at much smaller separations with SDI.
 
\subsection{Survey Completeness}

One would not expect a planet to be detectable at all phases of its orbit -- 
to really understand the types of planets to which we are sensitive, we must take
orbital motion into account and translate separations on the sky into 
orbital semi-major axes ({\it a}).  To this end, we generated contour 
plots of fractional completeness 
as a function of mass and semi-major axis.   
For every survey star, we simulate 10000 planets for 
each combination of mass and semi-major axis.  Eccentricities are 
drawn from a distribution of eccentricities consistent with known radial 
velocity planets.   Standard distributions were used to randomly compute 
viewing angle and orbital phase, giving an instantaneous separation between 
star and planet.  We use the distance, age, spectral type, and H-band 
magnitude of the star, and luminosity as a function of mass, calculated from 
the Burrows et al. (2003) 
models, to provide each simulated planet a separation 
on the sky in arcseconds, and an H-band flux ratio compared to its parent 
star.  Combining this with the SDI contrast curve for each star in the survey,
we can then determine the percentage of simulated planets detected as a
function of mass and semi-major axis for each survey star.  
Contour plots for a set of 4 typical 
program stars (AB Dor, DX Leo, GJ 182, and GJ 799B) are presented 
in Fig.~\ref{fig:contour}.  Note that we conservatively assume only T-type objects 
can be detected, hence masses $>$ 10 M$_{Jup}$ are not considered for many
young targets.  The value attached to each contour level defines 
the completeness of our observation to detecting (at the 5$\sigma$ confidence 
level) a planet with the given semi-major axis and mass.  It is worth noting 
that the only assumptions necessary for the generation of these plots is the 
eccentricity distribution of planets and the Burrows et al. 2003 models.  

We use this method to summarize our survey completeness in 
Fig.~\ref{fig:50percent}.  Having computed the completeness for each star to 
planets at various masses and semi-major axes, we take slices at representative
values of the semi-major axis, and present the number of stars in our 54 star 
survey which are at least 50\% complete to such a planet.  Our survey 
places the strongest constraints on planets between 6-10 M$_{Jup}$ with 
semi-major axes between 20-40 AU.  With 20 such stars (with 50$\%$ or 
greater completeness in this mass/semi-major axis range) surveyed without 
a detection of a planet, a simple way of interpreting our results (though 
without statistical rigor) is that we would expect the frequency of such 
planets to be of order 10\% or less.

The evolutionary models of \citet{bur03} utilize a ``hot start'' initial
condition which, while appropriate for brown dwarfs, is possibly significantly
different from the actual initial origins of planets.  The \citet[][]{bur03}
models begin with a high-temperature, high-entropy hydrogen-helium sphere 
which is allowed to radiate and cool over time.  In 
contrast, a planet forms when gas
accretes onto a rocky core, according to the core-accretion models of Ida
 and Lin (2005) and the disk instability models of Boss (2003). 
Recently, \citet[][]{marl06} simulated model planets with more realistic 
(lower entropy) initial conditions.  These model planets have significantly
lower luminosities at young ages ($<$1 Gyr).  Model planets also converge to 
the ``hot start'' evolutionary tracks at different times according to mass --
a 1 M$_{Jup}$ model converges to traditional tracks by 20 Myr, while a 10 
M$_{Jup}$ requires up to 1 Gyr to match traditional tracks.  Currently, H band
magnitudes for these models are not yet available, but will be available in
Spring 2007 (private communication, J. Fortney).  When H band magnitudes are 
available, we will repeat this analysis using these new models.

\subsection{Sensitivity Case Study: AB Dor with Simulated Planets}

Since our survey data are highly saturated in the core of the image,  
it is difficult to place simulated objects in our data with a 
high degree of positional accuracy, as there is no external reference
for position between data taken at different dithers and roll angles.  
However, as part of the 
SDI survey observations, our team discovered a close-in 
(0.156$\arcsec$) companion (hereafter AB Dor C) to the young star 
AB Dor \citep[][]{clo05b}.  While this companion is a very low mass M star 
\citep[0.090$\pm$0.005~M$_{Sun}$,~M5.5$\pm$1,][]{clo05b,clo07b} and 
hence, does not possess methane absorption features, it it still clearly 
detected in our SDI data.  In our second AB Dor dataset where AB Dor C is 
separated from its primary by 0.2'' (Nielsen et al. 2005), the AB Dor C source can be 
used to our advantage as a reference position from which to offset -- 
allowing us to 
add simulated planets into this dataset with highly accurate positions
and relative fluxes independent of our ``pipeline'' calculated centroids.  

Simulated planets were produced by scaling $\sim$10$\times$0.1 s 
unsaturated images of AB Dor A 
taken right before the example dataset.  Planets were simulated with 
$\Delta$F1(1.575$\mu$m) = 9, 10, 11, and 12 mag and with methane break
strengths appropriate for T5, T6, and T8 spectral types.  Methane break 
strengths were calculated using the methane spectral index defined in 
Section 3.2.  Photon noise and zero points 
appropriate for each object was added using 
the IRAF artdata/mkobject tool.  The photometric zero point was 
calculated from AB Dor C.  

\begin{figure}
   \includegraphics[width=\columnwidth]{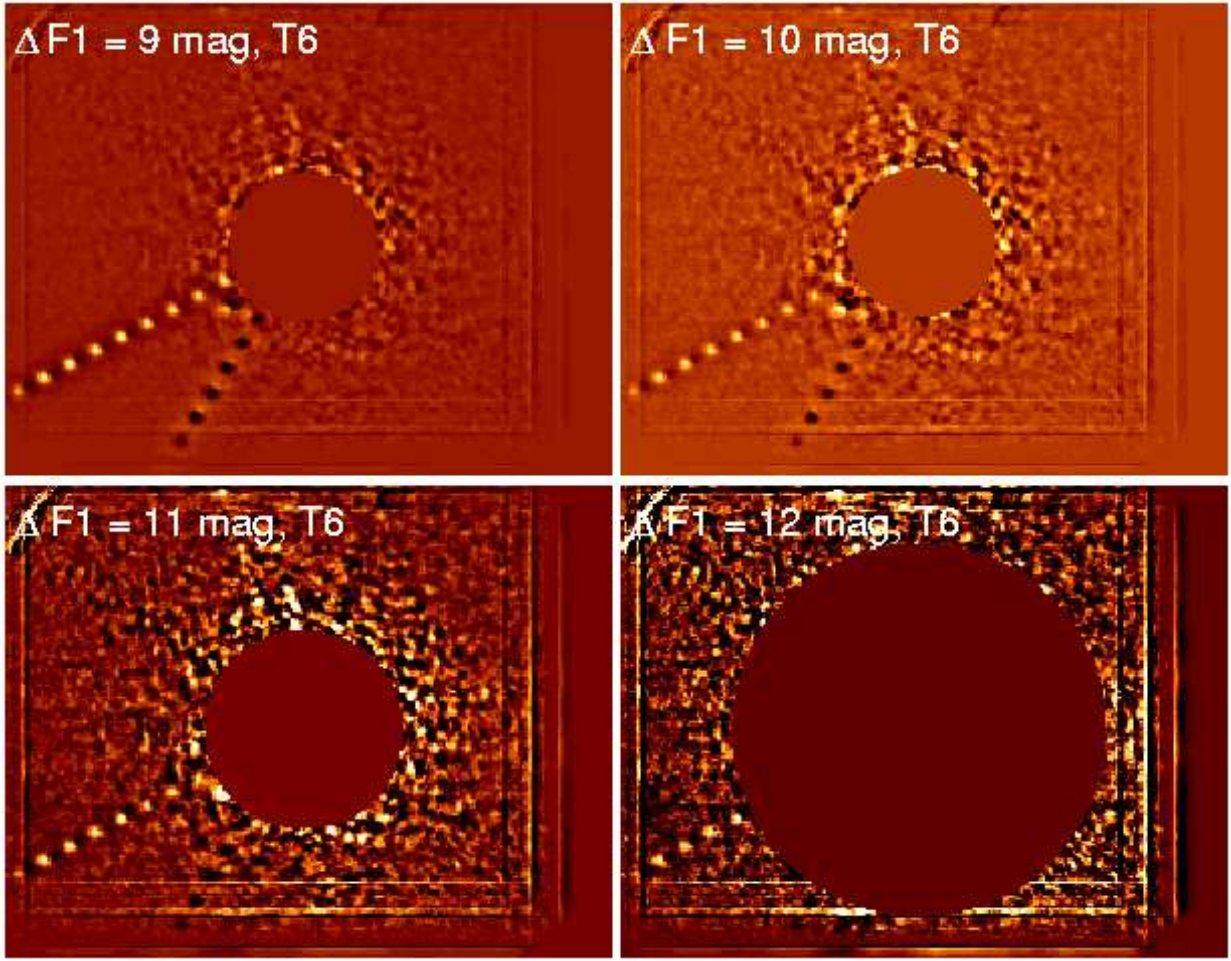} 
   \caption[Reduced VLT SDI data]
   { \label{fig:SDIREDplan} A complete reduced dataset of AB Dor A
(28 minutes of data at a series of rotator angles -- 
0$^{\circ}$, 33$^{\circ}$, 33$^{\circ}$, 0$^{\circ}$) from the VLT SDI device.
Simulated planets have been added every 0.2'' from the star (0.4'', 0.6'', 
0.8'', 1.0'', 1.2'', 1.4'', 1.6'', 1.8'', 2.0'', and 2.2'')   
with $\Delta$F1(1.575$\mu$m) = 9 mag (upper left,
(attenuation in magnitudes in the 1.575 $\mu$m 
F1 filter), 10 mag (upper right), 11 mag (lower left) and 12 mag 
(lower right) fainter than the star.  The 0.4'' object 
falls within the inner dark circle (dark circle 
radius of 0.5'', 0.5'', 0.7'', and 1.3'' respectively
for the 9, 10, 11, and 12 mag objects); the 2.2'' object falls outside the 
frame aperture in a number of dither images and thus is detected with lower
S/N than the other objects.
These simulated 
planets are scaled from unsaturated images of AB Dor A 
taken right before the example dataset (and have fluxes and 
photon-noise in each filter 
appropriate for a T6 object).  
}
\end{figure}

\begin{figure}
   \begin{center}
   \begin{tabular}{c}
   \includegraphics[width=\columnwidth]{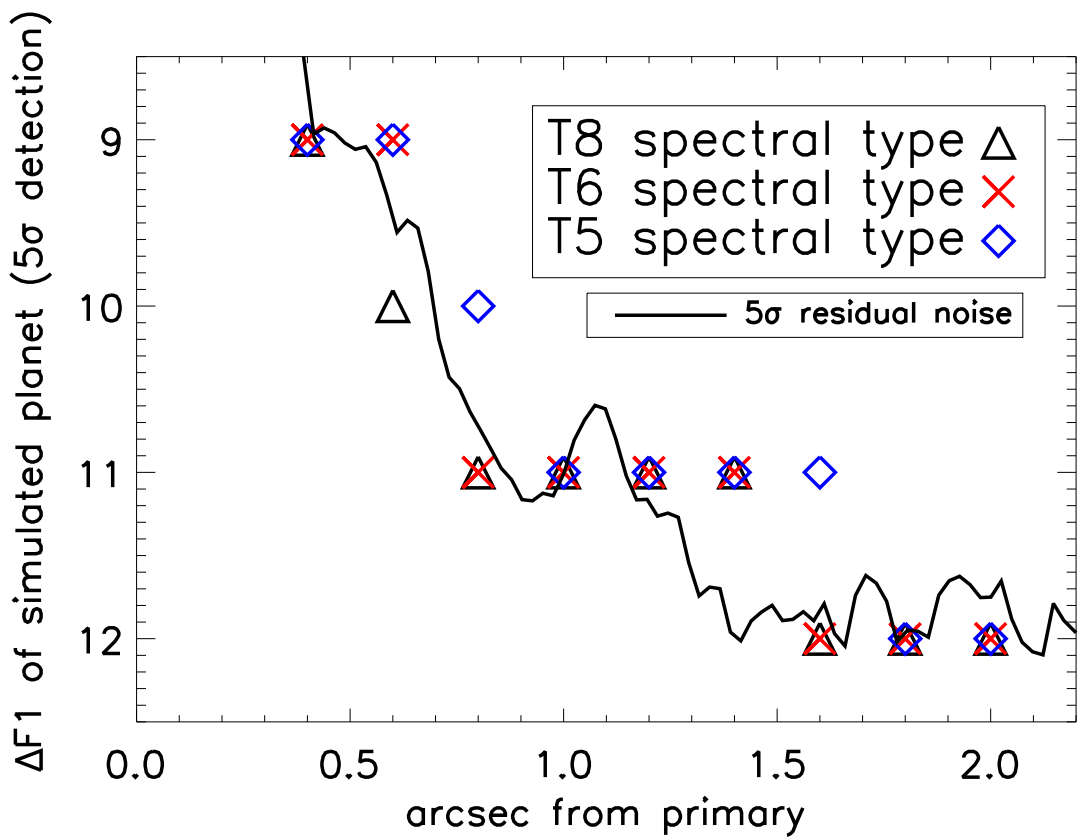}
   \end{tabular}
   \end{center}
   \caption[Maximum Achievable Planet Contrast vs. Separation]
   { \label{fig:simcontrast1} Maximum achievable planet contrast (5$\sigma$
detection) vs. separation for 28 minutes of VLT SDI data for AB Dor A.  To
determine the maximum achievable planet contrast as a function of separation,
we inserted and then attempted to retrieve simulated planets with a variety
of separations and $\Delta$F1 contrasts appropriate for T5, T6, and T8 
spectral types.  The residual SDI noise  
curve for AB Dor A is also overplotted; the two curves agree well, giving
us confidence in our measured contrast limits.  
}
   \end{figure}

\begin{figure}
   \begin{center}
   \begin{tabular}{c}
   \includegraphics[width=\columnwidth]{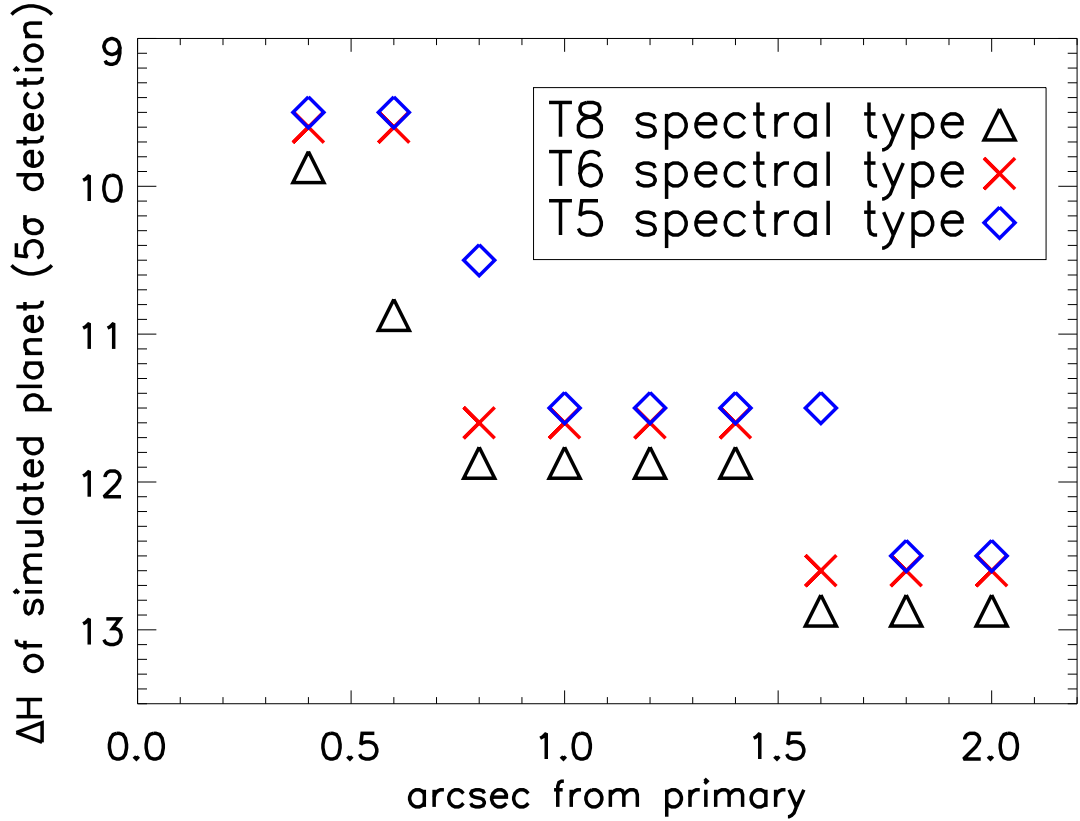}
   \end{tabular}
   \end{center}
   \caption[Maximum achievable H band planet contrast vs. separation]
   { \label{fig:simcontrastH1} Maximum achievable H band 
planet contrast (5$\sigma$
detection) vs. separation for 28 minutes of VLT SDI data for AB Dor A.  To
determine the maximum achievable planet contrast as a function of separation,
we inserted and then attempted to retrieve simulated planets with a variety
of separations and $\Delta$F1 contrasts appropriate for T5, T6, and T8 
spectral types.  $\Delta$F1 contrasts were converted
to $\Delta$H magnitudes using the magnitude offsets calculated in section 3.3.
}
   \end{figure}

A fully reduced 28 minute dataset of AB Dor A (70 Myr K1V at a distance 
of 14.98 pc, V=6.88) from the VLT SDI device 
is presented in Fig.~\ref{fig:SDIREDplan} with simulated planets 
added at separations of 0.4'', 0.6'', 0.8'', 1.0'', 1.2'', 1.4'', 
1.6'', 1.8'', 2.0, and 2.2'' from the primary 
($\Delta$F1(1.575$\mu$m) = 9, 10, 11, and 12 mag and spectral type T8).
Past 0.7'', the $\Delta$F1(1.575$\mu$m) = 10 simulated planets are detected 
with S/N $>$ 10.  The 2.2'' object falls off the edge of the aperture in 
several dithers and thus appears somewhat attenuated compared to the other
simulated objects.
Maximum achievable companion contrast at the 5$\sigma$ level as a
function of distance from the star is plotted in Fig.~\ref{fig:simcontrast1}.
The residual noise curve for this star (see section 3.3) is also 
overplotted.  Contrast curves (5$\sigma$) calculated 
with both techniques agree well with each other.  Using
the magnitude offsets developed in section 3.4, we convert our 
$\Delta$F1(1.575$\mu$m) contrasts into $\Delta$H for each spectral type.
We adopt Offset$_A$ = 0 mag, Offset$_B$ = 0.5 mag for a T5 object, 
Offset$_B$ = 0.6 mag for a T6 object, and Offset$_B$ = 0.87 mag for a 
T8 object.  $\Delta$H vs. separation 
in arcsec is presented in Fig.~\ref{fig:simcontrastH1}.

\begin{figure}
   \begin{center}
   \begin{tabular}{c}
   \includegraphics[width=\columnwidth]{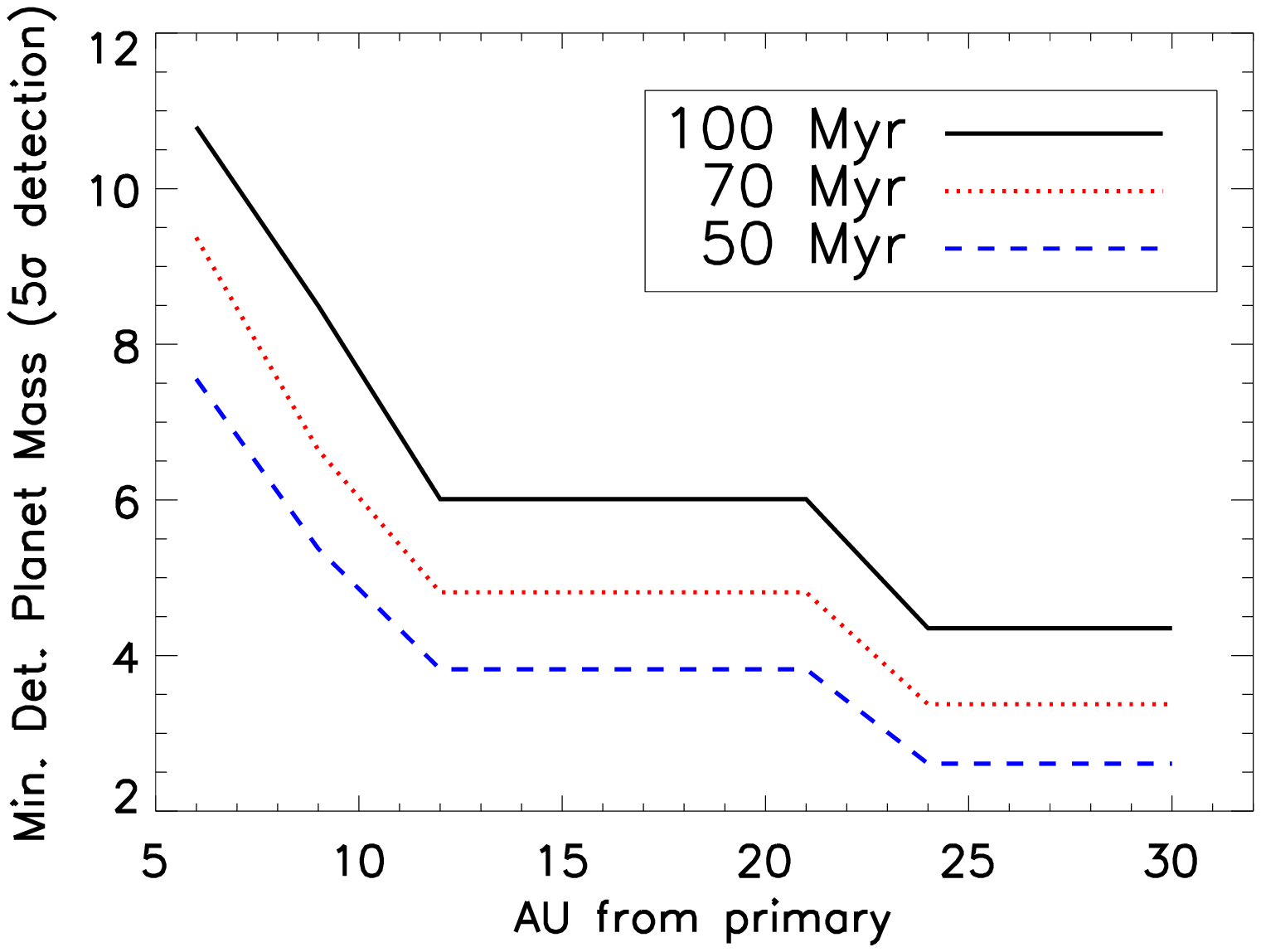}
   \end{tabular}
   \end{center}
   \caption[Minimum Detectable Planet Mass vs. Separation]
   { \label{fig:simcontrastmass} Minimum detectable planet mass
(5$\sigma$ detection) vs. separation (AU) for 28 minutes of VLT SDI 
data for AB Dor A.  To
determine the minimum detectable planet mass as a function of separation,
we inserted and then attempted to retrieve simulated planets with a variety
of separations and $\Delta$F1 contrasts appropriate for T5, T6, and T8 
spectral types.  $\Delta$F1 contrasts were converted
to $\Delta$H magnitudes using the magnitude offsets calculated in section 3.3
and were then converted to absolute H magnitudes using the 2MASS apparent
H magnitude and the Hipparcos distance for each star.  
Absolute H magnitudes were converted into planet
masses using the models of \citet[][]{bur03} and adopting a range of 
system ages from 50 - 100 Myr.  For AB Dor, we should be able to 
image (5$\sigma$ detection) a 5 M$_{Jup}$ planet 12 AU from the star.  
For a complete set of minimum detectable planet mass vs. separation curves, 
see: http://exoplanet.as.arizona.edu/$\sim$lclose/SDI.html.}
   \end{figure}

$\Delta$F1 contrasts were
translated into planet masses using the 100 Myr models of \citet[][]{bur03}.  
According to the 100 Myr old model, objects with mass $\leq$ 10 M$_{Jup}$ will 
have T$_{eff}$ $<$ 900 K -- these objects are reliably of spectral
types later than T7 (temperature scale from Burgasser et al. 2006).  
Thus, we adopt the T8 spectral type curve for this
analysis.  AB Dor has a likely age of 50-70 Myr (Nielsen et al. 2005, 
Close et al. 2007b) -- we interpolate the models 
of \citet[][]{bur03} to derive masses at these ages as well.  
The minimum detectable planet mass as a function of distance from the star 
is plotted in Fig.~\ref{fig:simcontrastmass}.  Adopting an age of 70 Myr 
for AB Dor A, we can detect a 5 M$_{Jup}$ planet 12 AU from the star.
However, as noted above, the Burrows et al. 2003 models utilize a hot 
start initial condition which may be inappropriate for a young planet. 
The Marley et al. (2006) 
models utilize more appropriate initial conditions and 
when H band magnitudes become available for these models, we will repeat 
this analysis. 

\subsection{Comparison with Other Direct Detection Methods}

\begin{figure}
\includegraphics[width=\columnwidth]{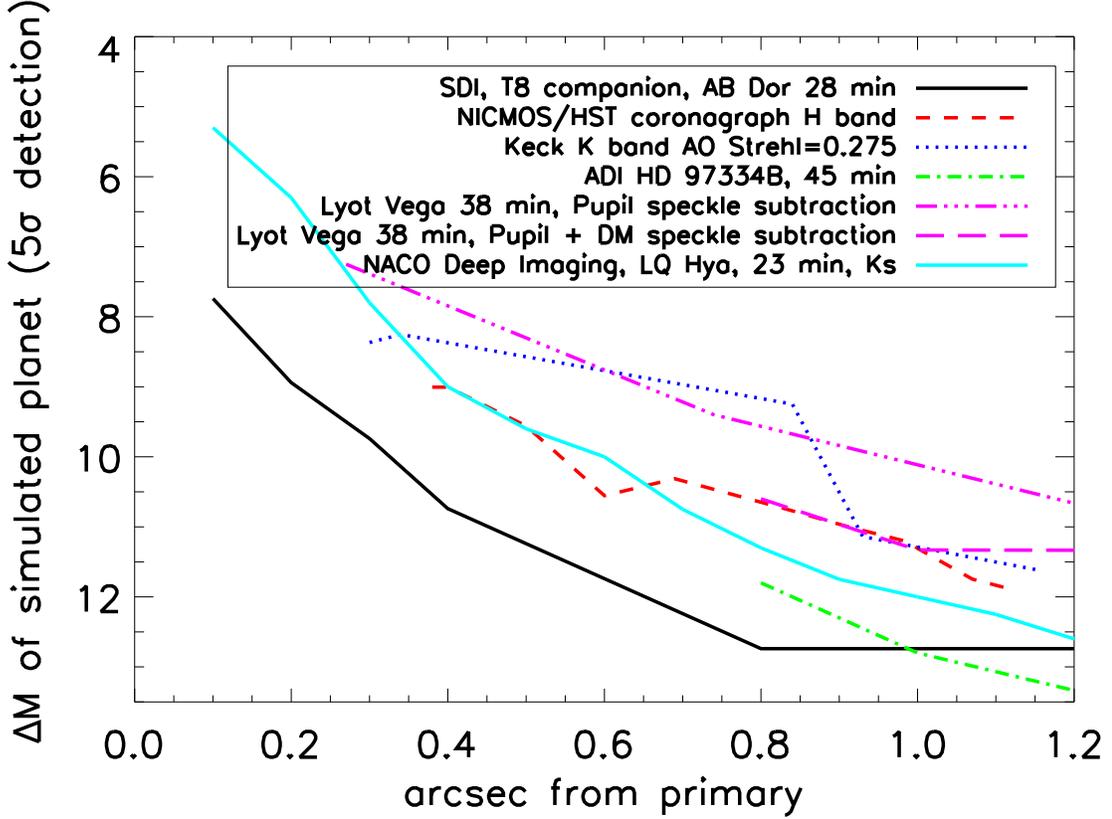}
\caption[Comparison with other direct detection methods]
{\label{fig:contcomp2} Comparison of SDI contrast curve with other
methods.  The Lyot curve is for the 3.6m AEOS telescope 
(Hinkley et al. 2007) and the NICMOS
curve coronograph curve is from HST (Schneider et al. 2003); 
otherwise curves are all from $\geq$8m class
telescopes.  We use the LQ Hya contrast curve from Masciadri et
al. (2005) because this star (K2, 18 pc vs. K1, 15 pc) is the closest
match from that work to AB Dor A (our SDI comparison star.)  The SDI
contrast curve has been converted from $\Delta$F1=1.575$\mu$m to
$\Delta$H contrasts appropriate for a T8 spectral type object.  Inside
0.4'', SDI contrasts are derived from the 1-trajectory SDI contrast
plot of AB Dor A; outside of 0.4'', SDI contrasts are derived from our
in-depth planet simulation case study of AB Dor A.  For methanated
companions, SDI provides improved contrast by 1-4 mag within
1$\arcsec$ as compared to other methods.  Past 1$\arcsec$, narrowband
imaging becomes less efficient and broad-band techniques (such as ADI;
Marois et al. 2006) reach higher contrasts.}
\end{figure}

We believe that our SDI images are the highest contrast astronomical
images ever made from ground or space for methane rich companions
$\leq$1'' from their star.  To substantiate this claim, we compare our
SDI contrast curves with those produced using a variety of other
competing methods (Azimuthal Differential Imaging (ADI), Marois et
al. 2006, Lyot Coronagraph, Hinkley et al. 2007, HST NICMOS, Schneider
et al. 2003, K-band Keck AO, Schneider et al. 2003, and NACO deep
imaging in the Ks band, Masciadri et al. 2005).  Comparison contrast
curves are presented in Fig.~\ref{fig:contcomp2}.  Apart from the Lyot
and NICMOS curves, all curves are from $\geq$8m class telescopes.  For
ease of comparison, we convert our $\Delta$F1=1.575 $\mu$m SDI
contrast curve into the equivalent $\Delta$H contrast appropriate for
a T8 spectral type companion.  For methanated companions, SDI provides
improved contrast by 1-4 mag within 1$\arcsec$ as compared to other
methods.

\subsection{New and Confirmed Close Binary Stars}

A number of close binary stars were discovered or confirmed during our survey.
In Table 7, we present separations and position angles 
measured from unsaturated SDI images of these 
stars acquired before each full SDI dataset was taken.  
These values are meant as estimates, hence, no error estimate is provided.
We discovered close stellar companions to HIP 9141 (0.15'' measured SDI separation), 
AB Dor A (0.16'' measured SDI separation, see Close et al. 2005a), 
HD 48189A (0.14'' measured SDI separation), HD 135363 (0.26'' measured SDI separation) 
and CD-64 1208 (0.18'' measured SDI separation).  The $<$0.5'' separation between the primary 
stars and these object makes it highly improbable that they are 
background objects.
Additionally, we confirmed the close binary RXJ 1243.6-7834 (0.068'' 
measured SDI separation) discovered by Brandner et al. (2000), the visual double LH 98 062 
(2.4'' measured SDI separation) discovered by Mochnacki et al. (2002), the spectroscopic 
binary TWA 4 (0.78'' measured SDI separation) discovered by Torres et al. (1995) and the close binary
EK Dra (0.67'' measured SDI separation) discovered by Metchev and Hillenbrand (2004).

\subsection{Candidate Identification / Elimination}

Survey data were examined for planet candidates by eye and also using
automated detection algorithms; generally, the human eye proved
more effective for detecting candidates.  
We identified 8 very tentative planet candidates at the VLT which passed 
the following tests:

1) Candidate must appear at the appropriate positions in the full 
reduced data.
(i.e. candidate image position must jump by the appropriate roll angle.)  

2) Candidate must appear (at least marginally) 
at the appropriate position in each of the separate roll angle images

3) Candidates detected in the F1(1.575 $\mu$m) - F3a(1.625 $\mu$m) difference
should also be detected in the F2(1.6 $\mu$m) - F3a(1.625 $\mu$m) difference
as well. 

These extremely tentative ($<$2$\sigma$) 
candidates are noted in the comments column of Table 1, with the 
predicted mass (from the models of Burrows et al. 2003) 
and separation had it been real. 
No candidates were detected with $>$ 3$\sigma$.  None of the 8 
tentative candidates were detected at a 
second epoch, thus the survey reached a null result for extrasolar planets
at the $\sim$3$\sigma$ level and certainly at the 5$\sigma$ level analyzed 
here.

\subsection{Planet Detectability}

\begin{figure}
   \includegraphics[width=\columnwidth]{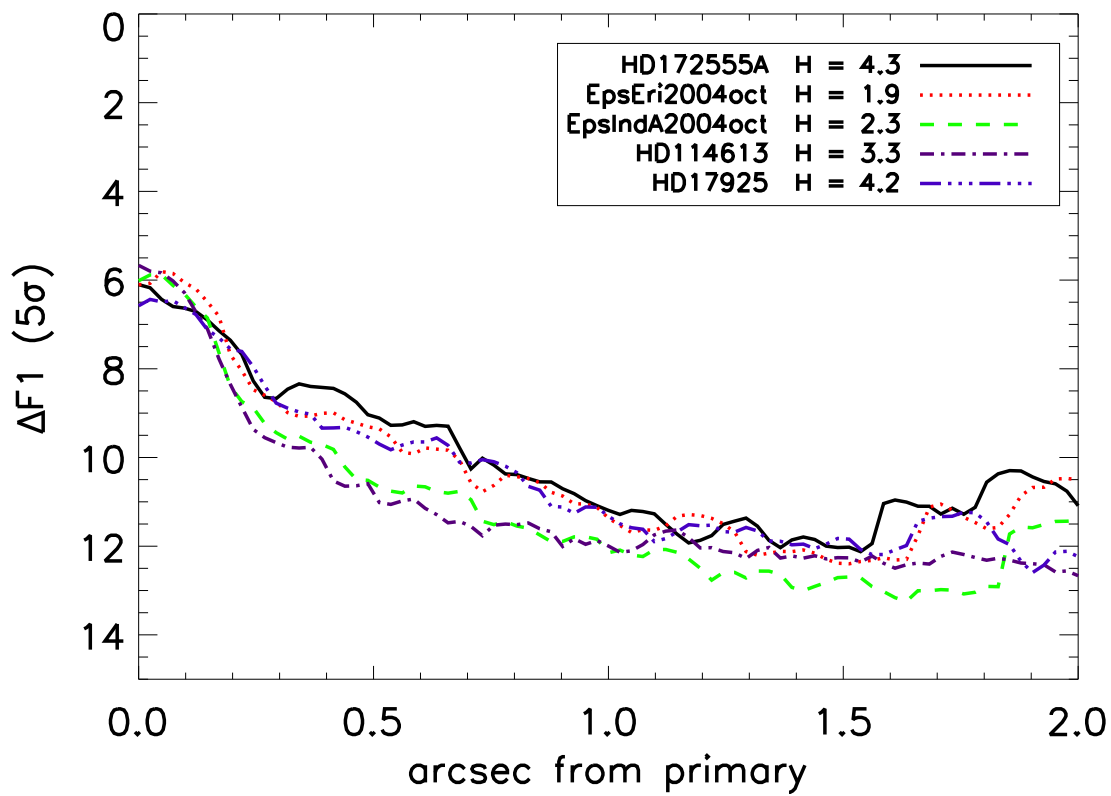} 
   \caption[Contrasts for VLT survey objects with H $<$ 4.5]
   { \label{fig:contrasts1}  5$\sigma$ 
Contrasts for VLT SDI survey objects with H $<$ 4.5 in the 
F1(1.575 $\mu$m) filter.  These contrast curves were 
generated by translating a 6$\times$6
pixel (0.1''$\times$0.1'') 
box along a particular radial trajectory away from the center of the 
star and then calculating the standard deviation within that box as a
function of radius.  Curves were generated from the full 
reduced and differenced SDI data for each object 
(F1(1.575 $\mu$m) - F3a(1.625 $\mu$m) for two roll angles).  
}
\end{figure}

\begin{figure}
   \includegraphics[width=\columnwidth]{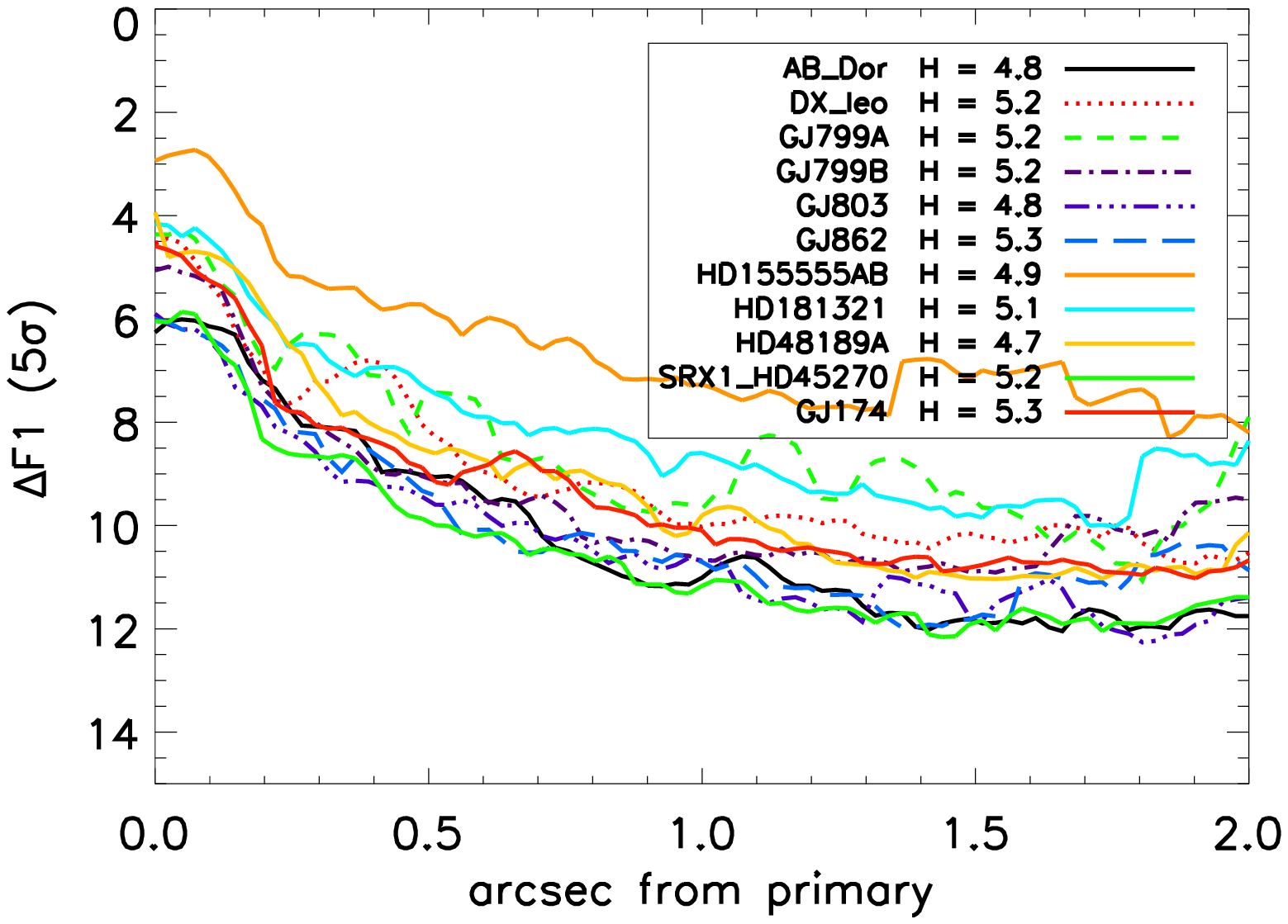} 
   \caption[Contrasts for VLT survey objects with 5.5 $>$ H $>$ 4.5]
   { \label{fig:contrasts2}  Same as Fig.~\ref{fig:contrasts1} but
for VLT SDI survey objects with 5.5 $>$ H $>$ 4.5.  
}
\end{figure}

\begin{figure}
   \includegraphics[width=\columnwidth]{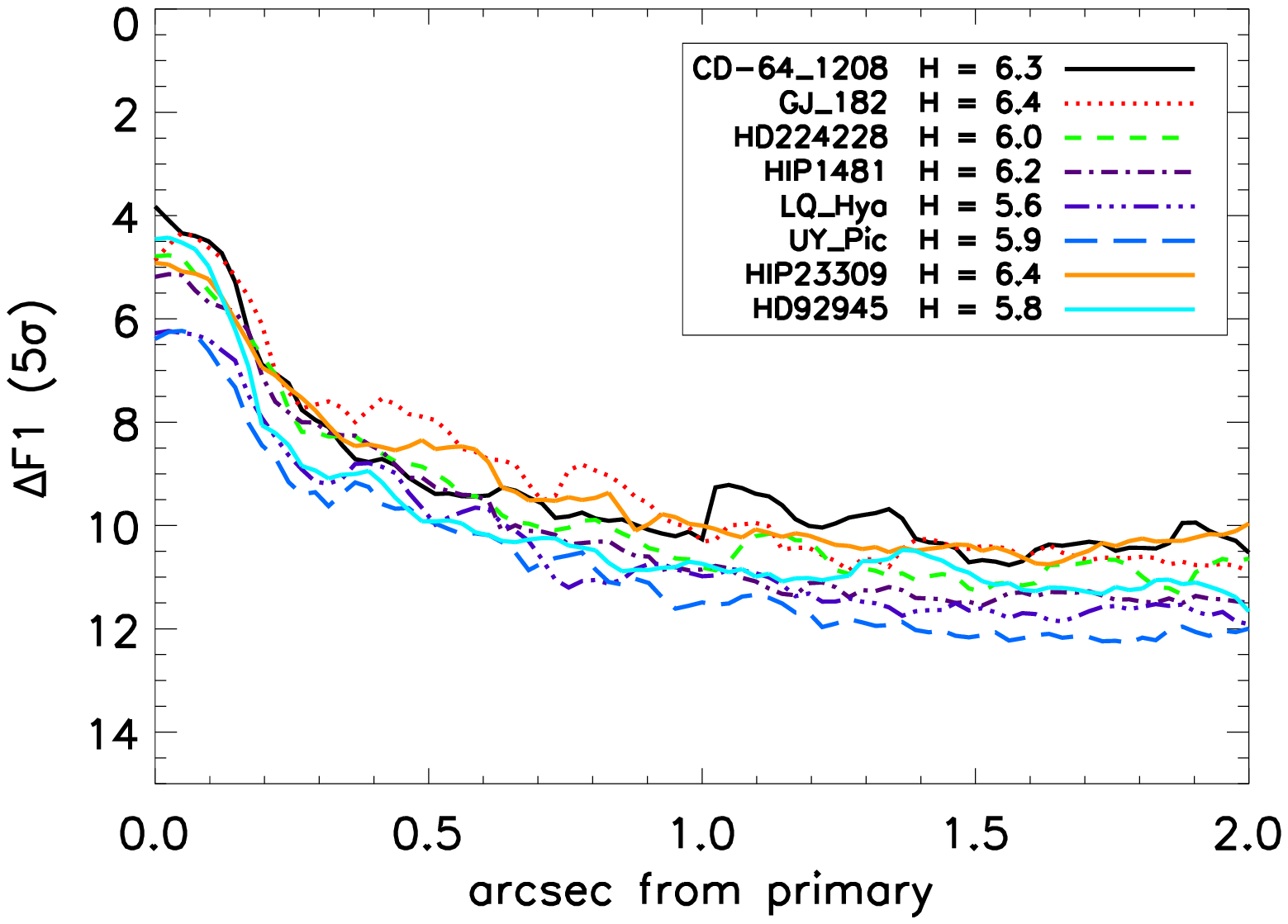} 
   \caption[Contrasts for VLT survey objects with 6.5 $>$ H $>$ 5.5]
   { \label{fig:contrasts3}  Same as Fig.~\ref{fig:contrasts1} but
for VLT SDI survey objects with 6.5 $>$ H $>$ 5.5.  
}
\end{figure}

\begin{figure}
   \includegraphics[width=\columnwidth]{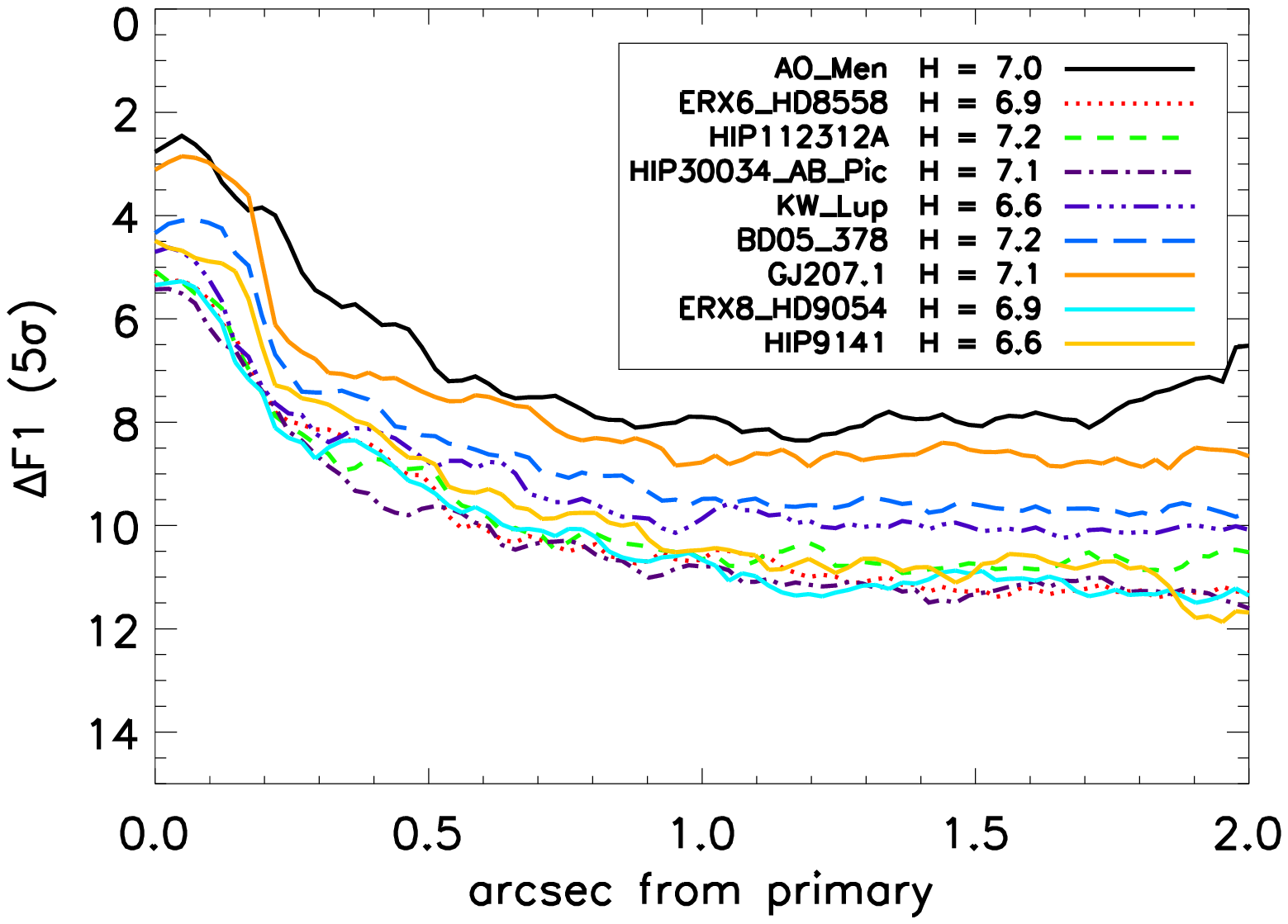} 
   \caption[Contrasts for VLT survey objects with 7.5 $>$ H $>$6.5]
   { \label{fig:contrasts4}  Same as Fig.~\ref{fig:contrasts1} but
for VLT SDI survey objects with 7.5 $>$ H $>$ 6.5.   
}
\end{figure}

\begin{figure}
   \includegraphics[width=\columnwidth]{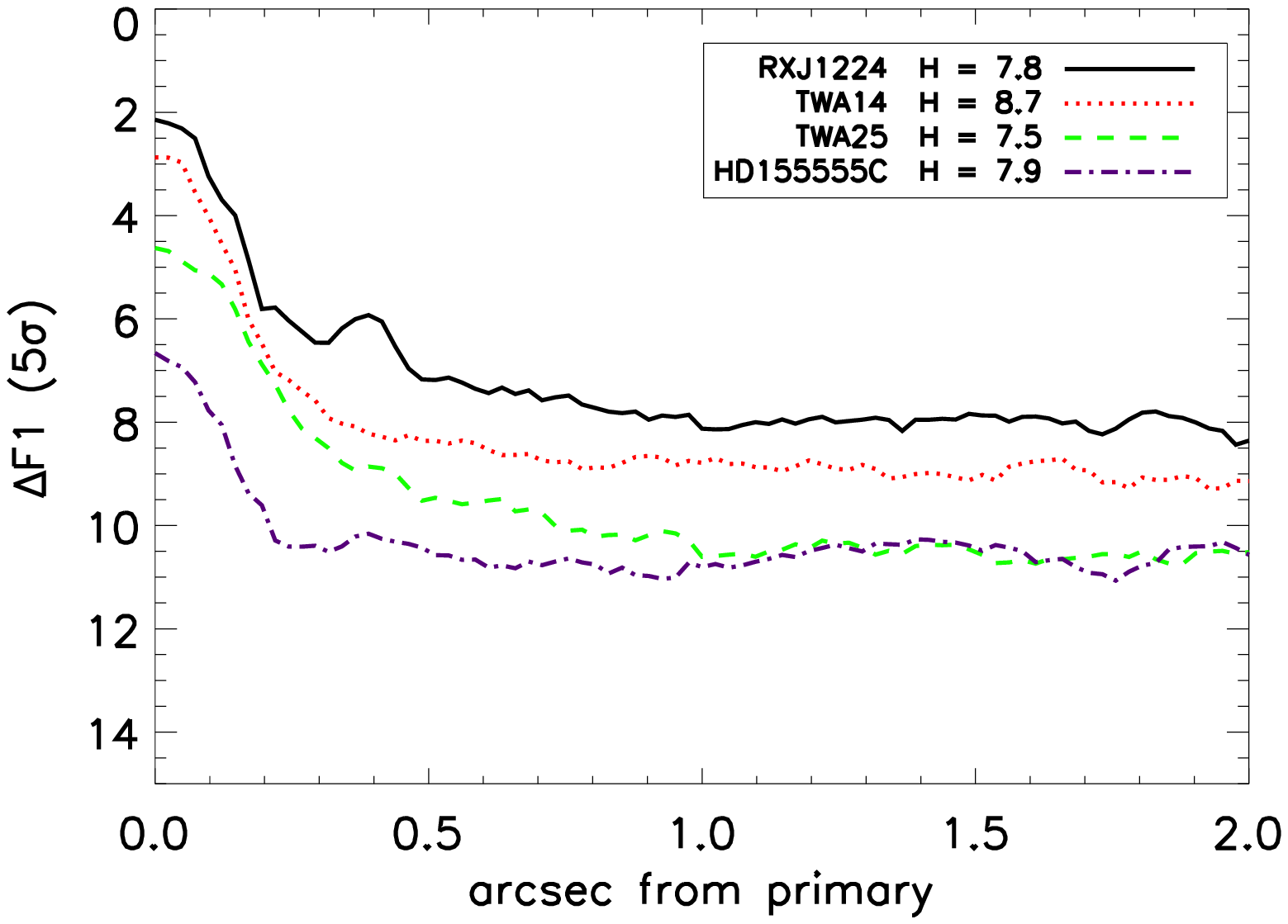} 
   \caption[Contrasts for VLT survey objects with H $>$7.5]
   { \label{fig:contrasts5}  Same as Fig.~\ref{fig:contrasts1} but
for VLT SDI survey objects with H $>$ 7.5. 
}
\end{figure}

\begin{figure}
   \includegraphics[width=\columnwidth]{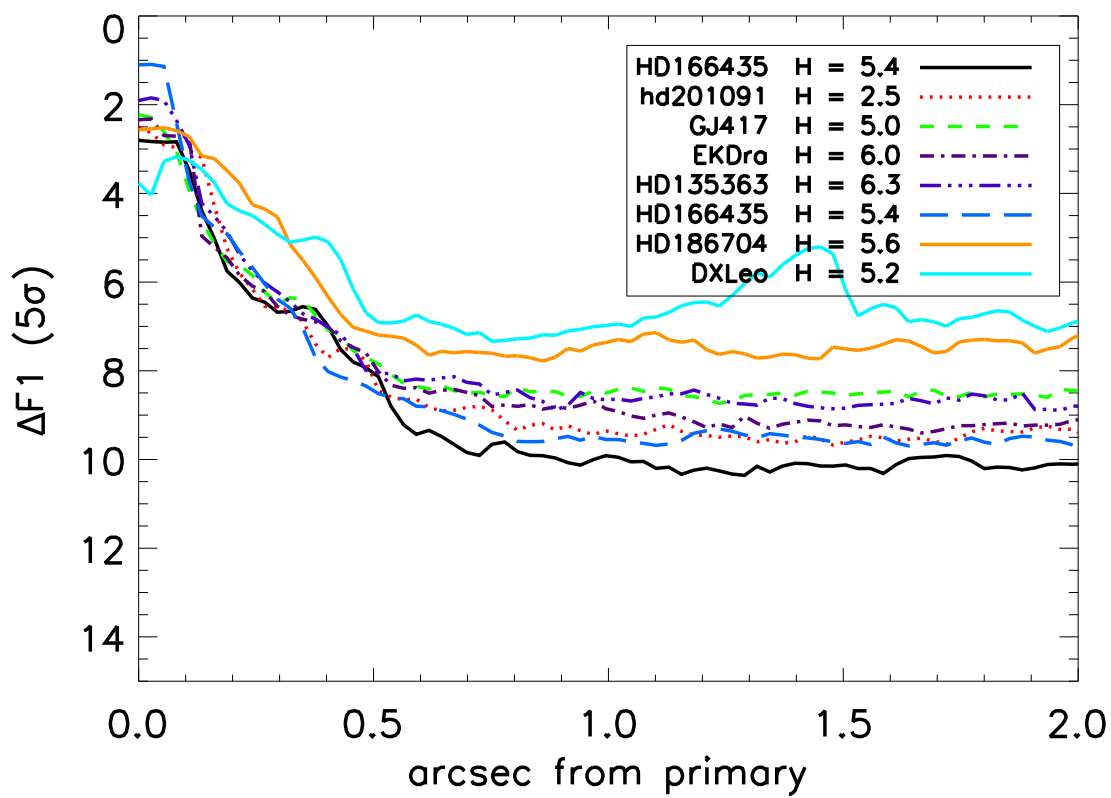} 
   \caption[Contrasts for MMT survey objects observed in May 2005]
   { \label{fig:contrasts6}  5$\sigma$ 
contrasts for MMT SDI survey objects observed in May 2005.  
These contrast curves were generated by translating a 6$\times$6 
(0.1''$\times$0.1'') 
pixel box along a particular radial trajectory away from the center of the 
star and then calculating the standard deviation within that box as a
function of radius.  Curves were generated from the full 
reduced and differenced SDI data for each object 
(F1(1.575 $\mu$m) - F3a(1.625 $\mu$m) for two roll angles).
}
\end{figure}

\begin{figure}
   \includegraphics[width=\columnwidth]{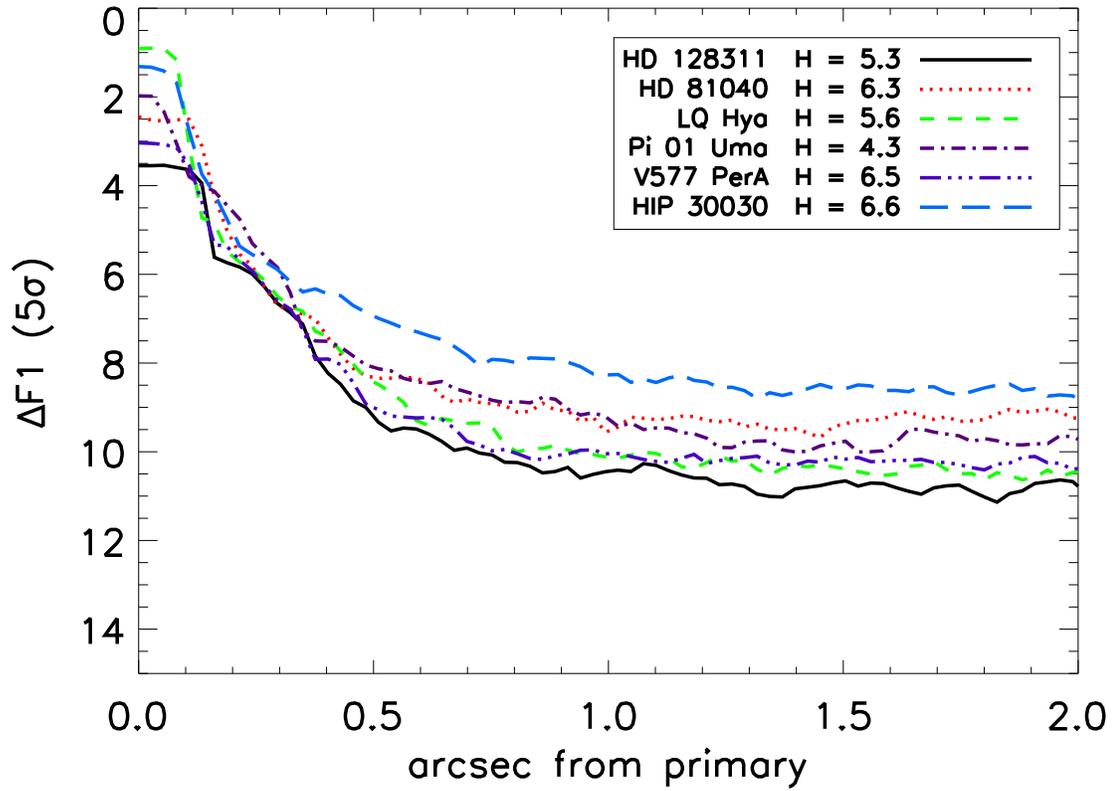} 
   \caption[Contrasts for MMT survey objects observed in February 2006]
   { \label{fig:contrasts7}  Same as Fig.~\ref{fig:contrasts6} but for 
MMT SDI survey objects observed in February 2006.}
\end{figure}

\begin{figure}
   \includegraphics[width=\columnwidth]{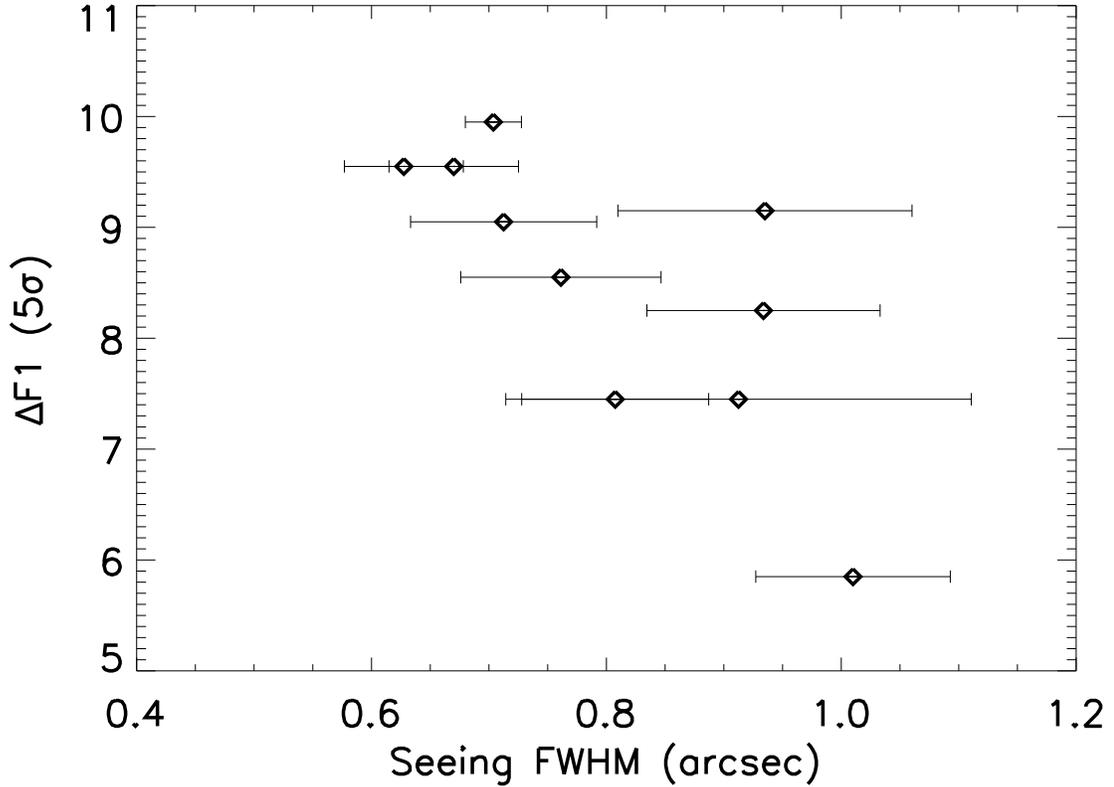} 
   \caption[FWHM vs. Contrast at 0.5'']
   { \label{fig:seeing}  Seeing FWHM (averaged over 
each observation) vs. attained 
5$\sigma$ contrast at 0.5$\arcsec$ separation from the primary star  
for 10 of the stars presented in Fig.~\ref{fig:contrasts2}
with H magnitudes between 4.5 -- 5.5.  The error bars on seeing are the 
seeing variation (as measured by the standard deviation of 
the seeing) over each observation.
For this sample of stars with 
roughly the same H magnitude, achievable contrast varies 
roughly inversely with the average seeing FWHM.  
Scatter in this plot is in part due to the fact that
seeing FWHM can change considerably over a 20-40 minute long observation.    
}
\end{figure}

\begin{figure}
\begin{center}
\begin{tabular}{cc}
\includegraphics[height=6cm]{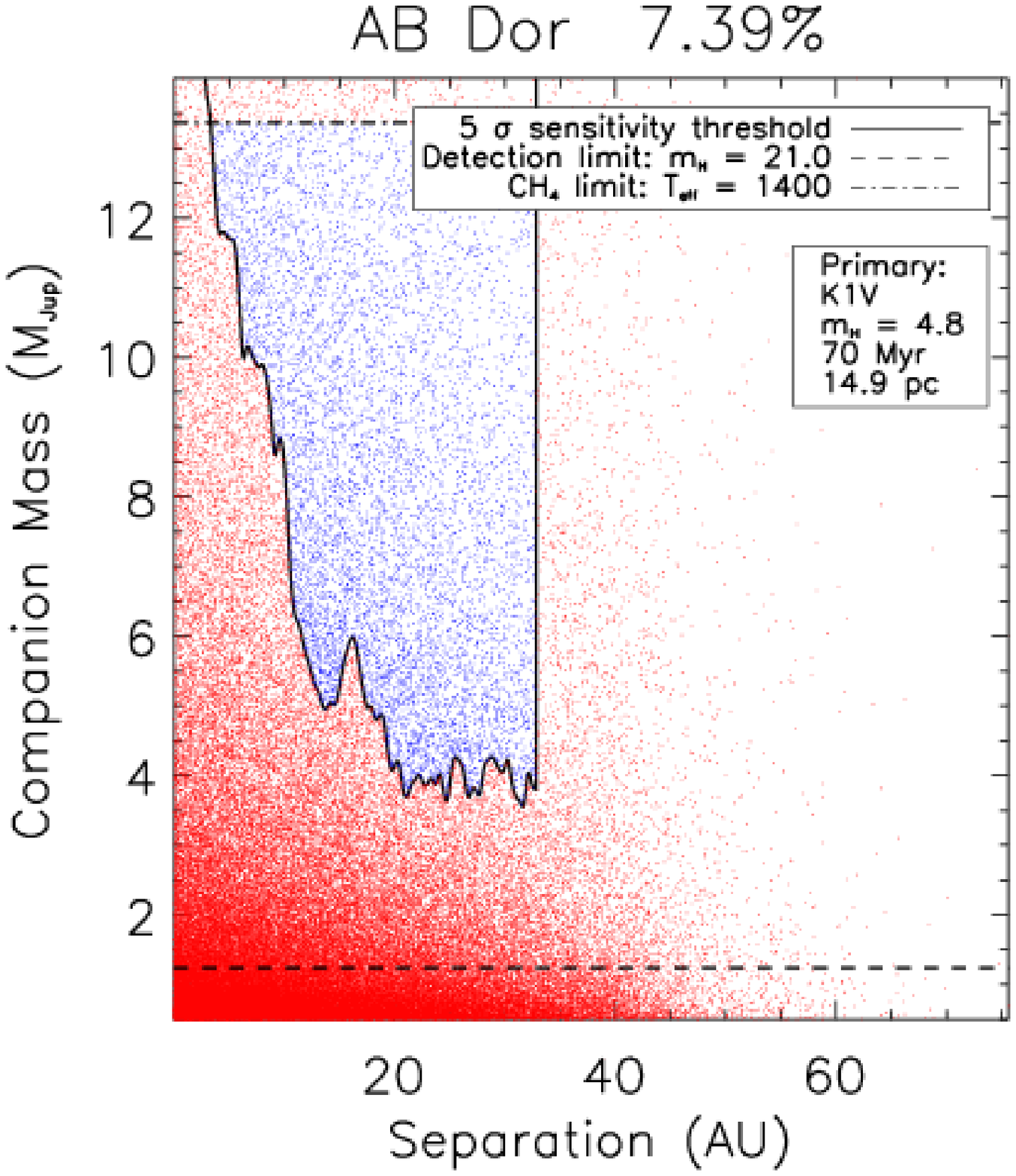} & 
\includegraphics[height=6cm]{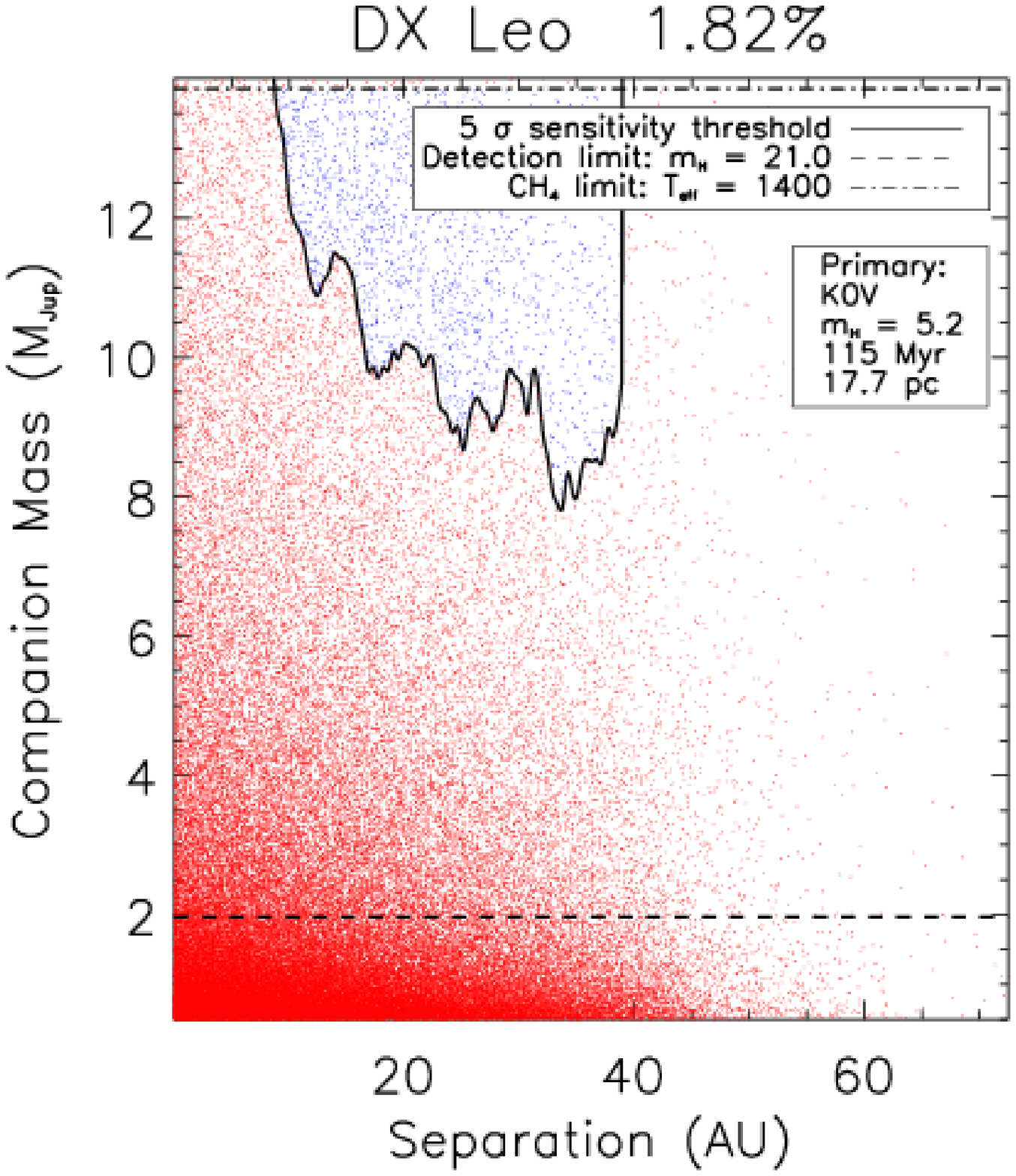} \\ 
\includegraphics[height=6cm]{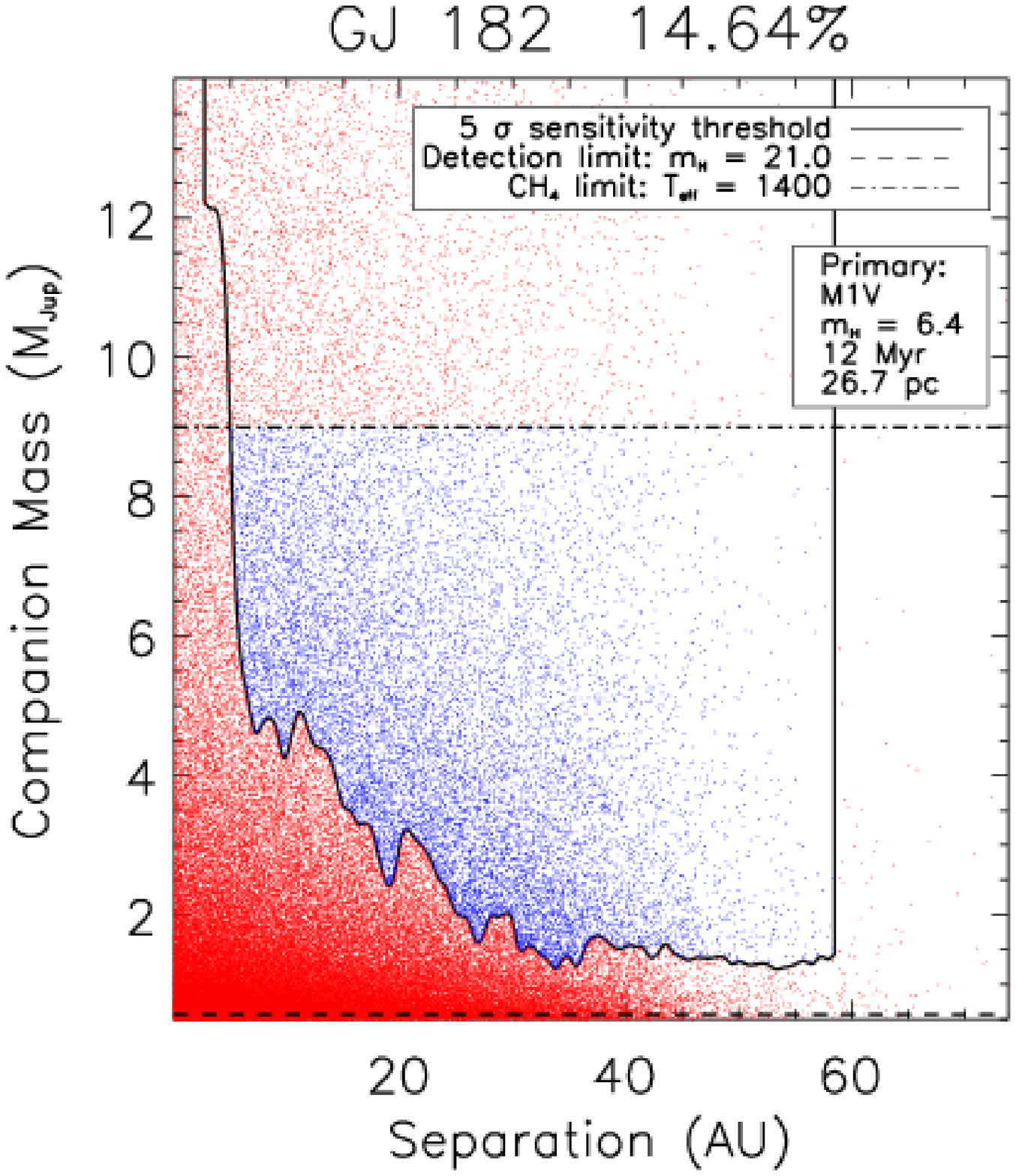} & 
\includegraphics[height=6cm]{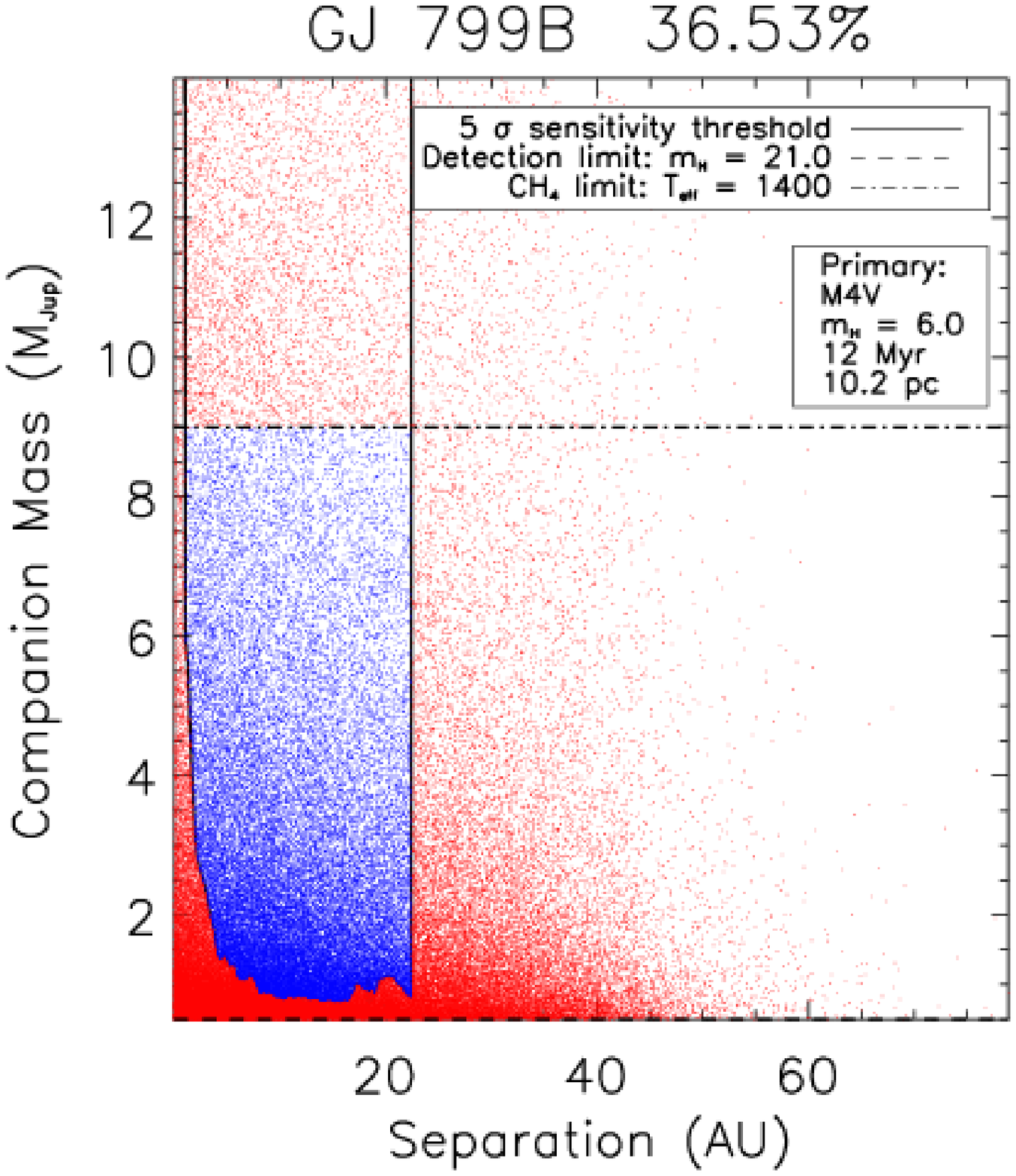} \\
\end{tabular}
\end{center}
\caption[Minimum Detectable Mass vs. Separation]
{ 
Minimum detectable mass vs. separation for 
a set of 4 typical program stars (upper left: AB Dor, upper right: DX Leo, 
lower left: GJ 182, lower right: GJ 799B).  We convert our contrast curves
in $\Delta$mag units (from Figs.~\ref{fig:contrasts1} 
to ~\ref{fig:contrasts5})
into minimum detectable mass vs. separation (in AU)
using the models of \citet[][]{bur03} and the distance to the star.  
To characterize the possible planets we expect to detect
around each star, we simulated an 
ensemble of 10$^6$ possible planets per star, assuming distributions for 
mass, eccentricity, and semi-major axis based on known radial velocity 
planets, as well as distributions for orbital phase and viewing angle.  When 
combined with the properties of the individual target star and its measured 
contrast curve, we can determine what fraction of these simulated planets we 
expect to detect at the 5$\sigma$ level (shown above each plot with the 
name of the target star).  
The ensemble of simulated planets is shown as small dots for each star in; 
simulated planets which are detected with the contrast attained by 
SDI are plotted in blue and those that remain undetected are plotted in red.
Assuming each star possesses exactly one planet, we can assign a detection 
probability for that star from the percentage of simulated planets detected.
For our 48 program stars which possess contrast curves, 
the average detection probability is 8.0$\%$, the median detection probability
is 4.1$\%$, and the maximum detection probability is 47$\%$.
For GJ 799B (12 Myr M star at 10 pc), 
we can detect (at 5$\sigma$) 5 M$_{Jup}$ planet at 2 AU.
}
\label{fig:minmasses}
\end{figure}

To determine what sort of planets we can detect in this survey, 
we converted our contrast 
curves in $\Delta$mag units into minimum detectable mass vs. 
separation (assuming
a late T to early Y spectral type for all possible objects
and using the models of \citet[][]{bur03}).
We calculated minimum detectable mass vs. separation for all stars with 
contrast curves in Figs.~\ref{fig:contrasts1} to Figs.~\ref{fig:contrasts7};
minimum detectable mass vs. separation is presented for a set of four
typical survey stars (AB Dor, DX Leo, GJ 182, and GJ 799B) 
in Fig.~\ref{fig:minmasses}.  
However, to detect an object of any given mass
requires that such an object exists around its parent object!  
The likelihood of detecting any object at a given radius 
is a combination of the minimum detectable mass for the parent star at 
that radius and the likelihood of such
an object existing.  Therefore it is very important to fully characterize
and understand the expected distribution of objects around each survey
star.  The results of the survey then also constrain the possible 
distribution of extrasolar planets as a function of radius. 
  
\begin{figure}[]
\begin{center}
\begin{tabular}{cc}
\includegraphics[height=6cm]{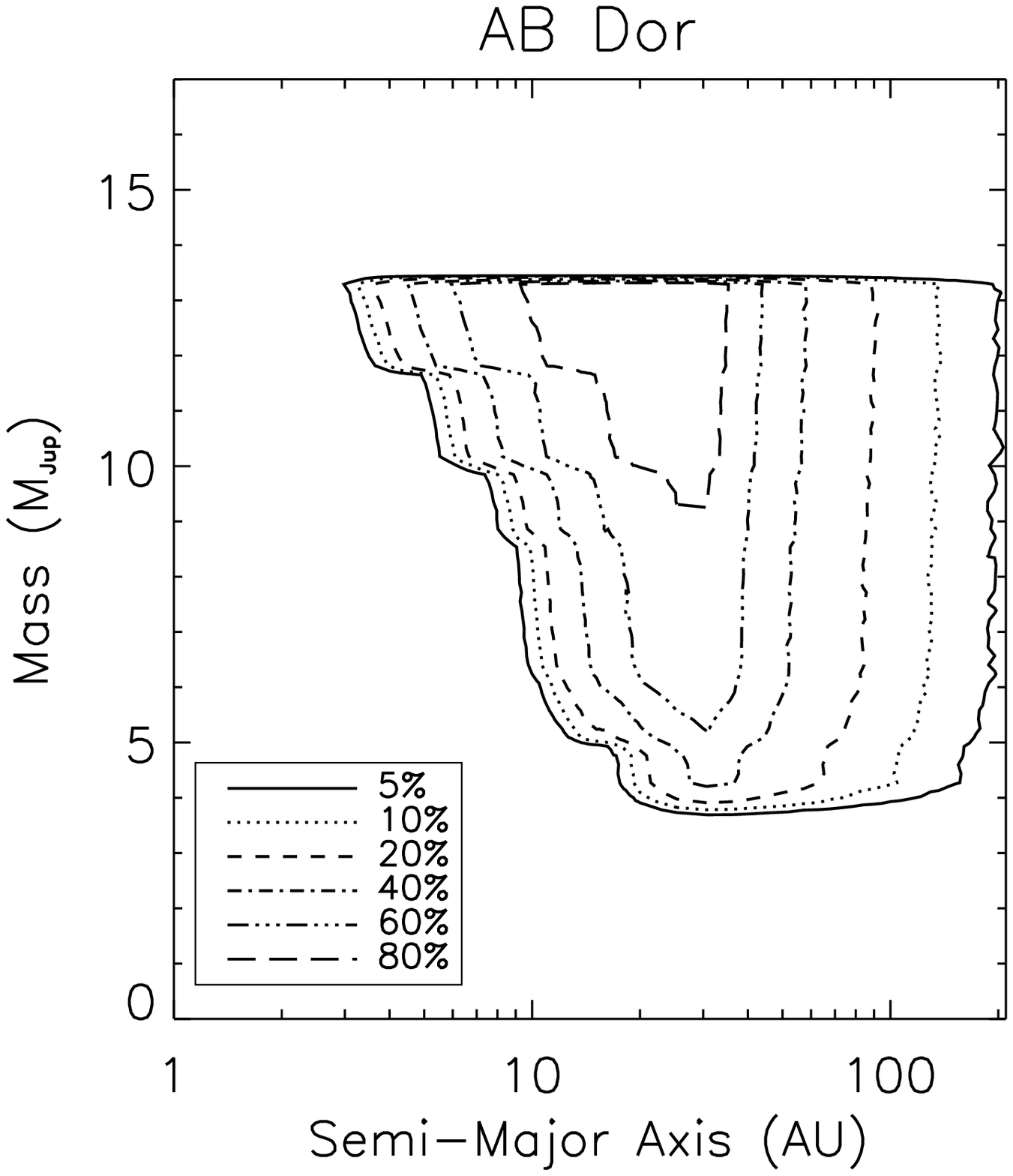} & 
\includegraphics[height=6cm]{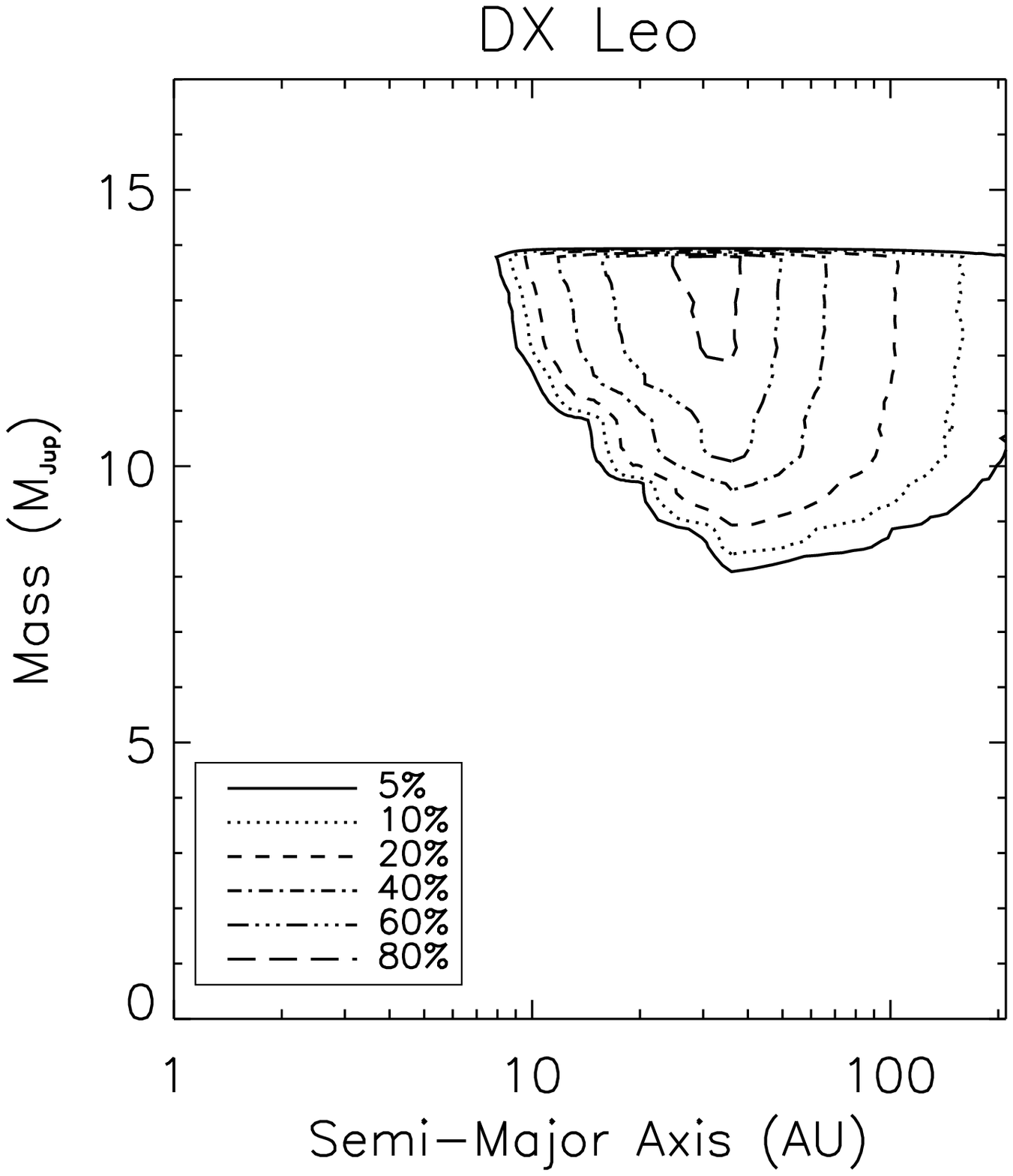} \\ 
\includegraphics[height=6cm]{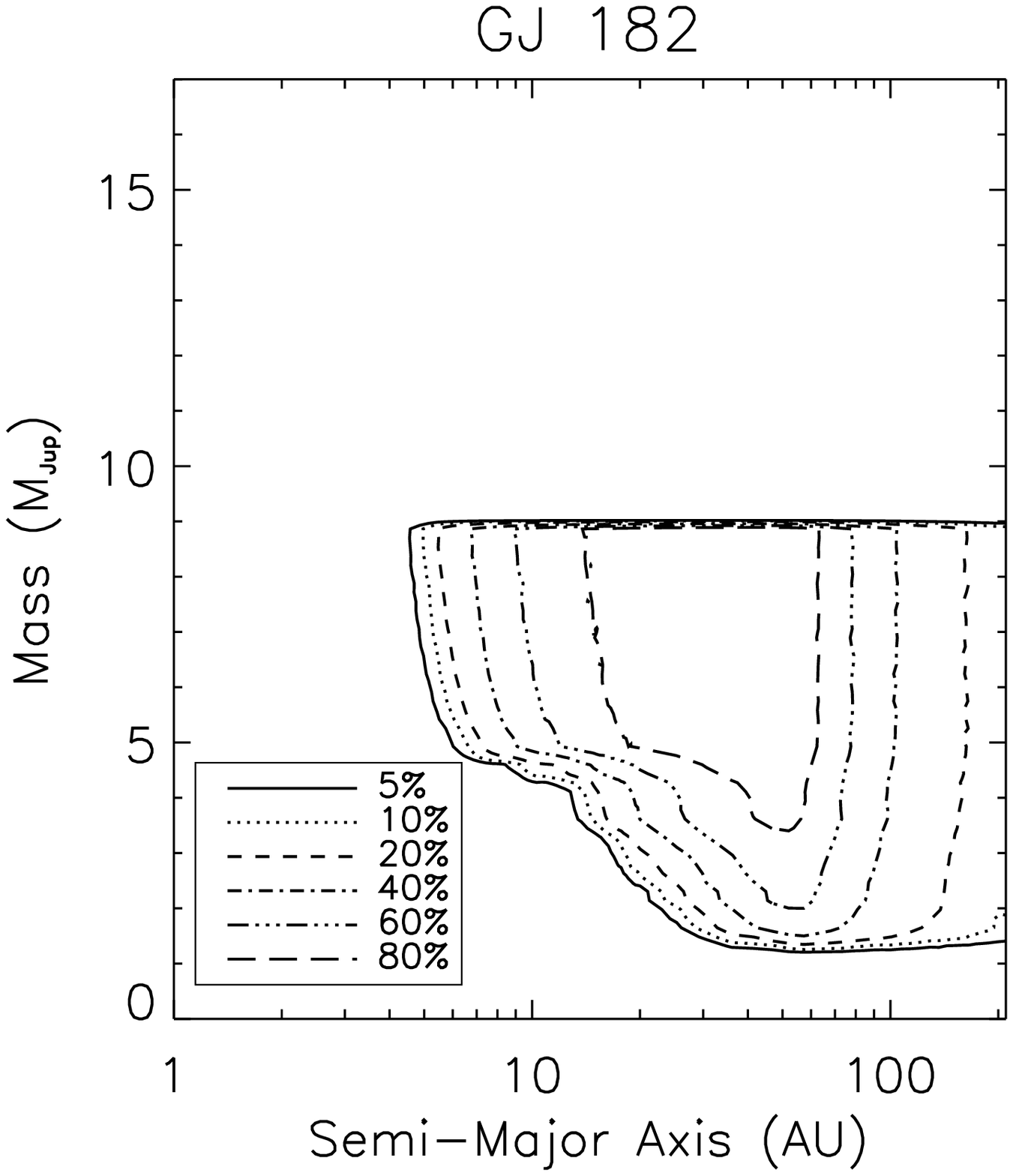} & 
\includegraphics[height=6cm]{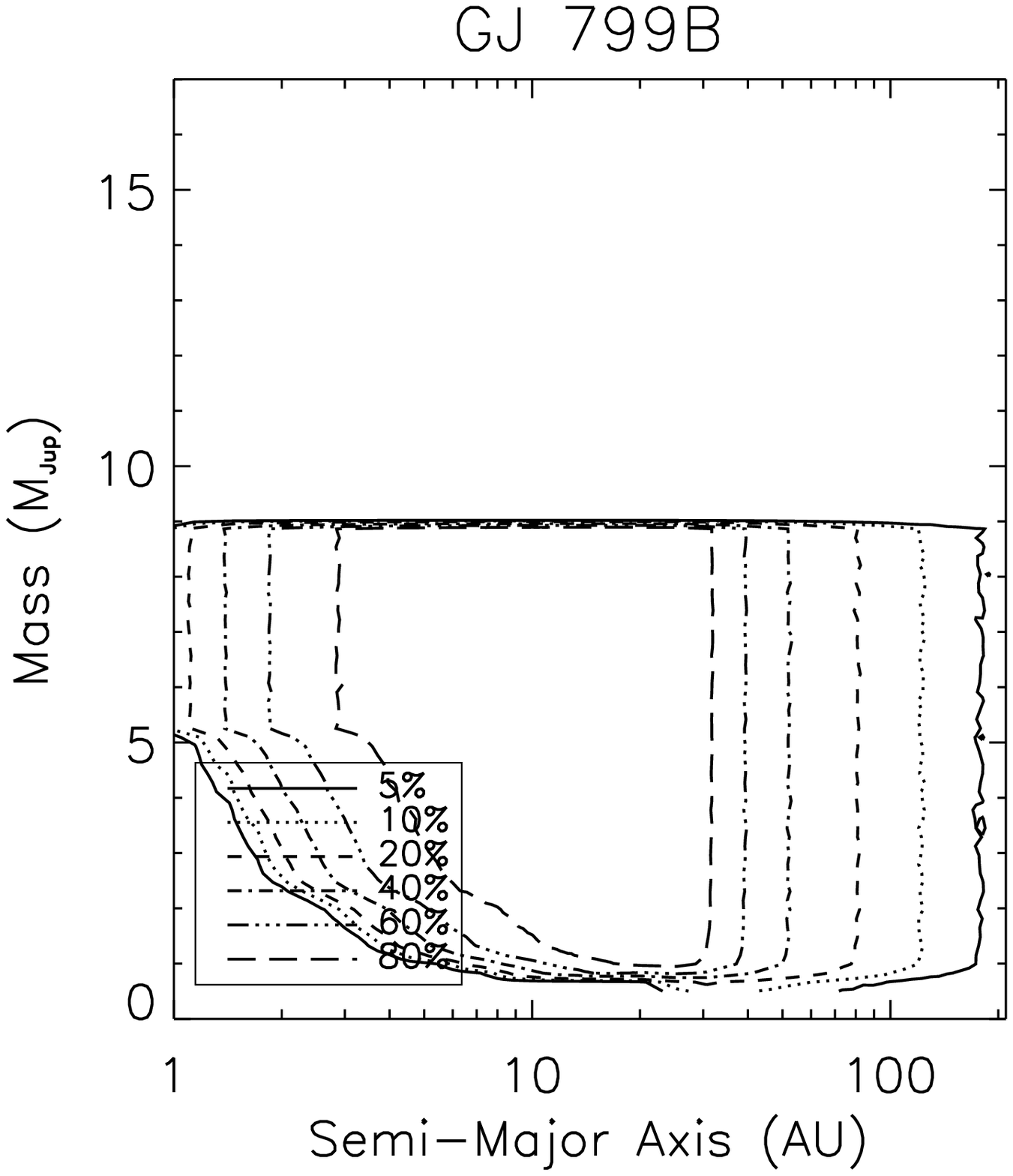} \\ 
\end{tabular}
\end{center}
\caption[Contour Plots]
{ Planet detection completeness contour plots for 
a set of 4 typical program stars (upper left: AB Dor, upper right: DX Leo, 
lower left: GJ 182, lower right: GJ 799B).  For a given mass and 
semi-major axis, 10,000 planets are simulated by our Monte Carlo method, over 
the expected distributions of eccentricity, orbital phase, and viewing 
angle.  Given the parameters of the target star and the models of Burrows 
et al. (2003), we determine what fraction of the simulated planets are 
detectable at the 5$\sigma$ level given the contrast plot for that star.  The 
contours show this detection probability across the 100,000 different 
combinations of mass and semi-major axis considered in this plot.  The strong 
upper limit in mass is set by our conservative $<$1400 K limit for the 
methane break 
required for a robust SDI detection.  In these models, we simply
do not allow an object with T$_{eff}>$1400 K to be detected, when 
in reality SDI can detect such non-methane objects (e.g. AB Dor C, 
Close et al. 2005b, Nielsen et al. 2005).  For a complete set of 
planet detection completeness contour plots, 
see: http://exoplanet.as.arizona.edu/$\sim$lclose/SDI.html.}
\label{fig:contour}
\end{figure}

\begin{figure}
   \includegraphics[width=\columnwidth]{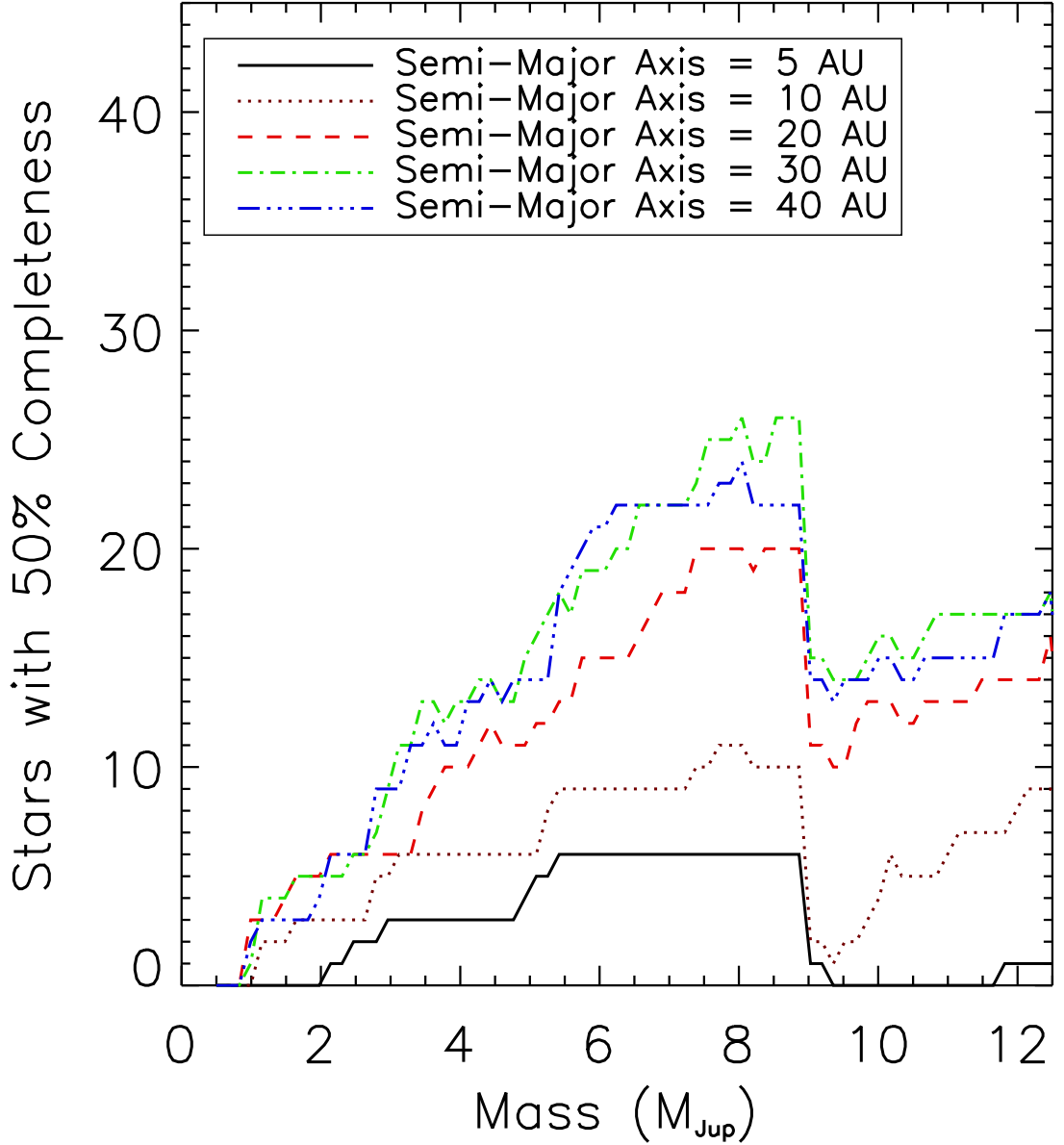} 
   \caption[50$\%$ completeness plot]
   { \label{fig:50percent} Our 
50$\%$ completeness levels.  Combining the results 
of Fig.~\ref{fig:contour}, we consider individual values of the semi-major 
axis across the planetary mass range, and at each combination calculate the 
total number of stars in our survey (out of a total of 54) where the fraction 
of such planets, given by the Monte Carlo Simulation, that can be detected at 
the 5$\sigma$ level is 50$\%$ or greater.  
{\it Clearly, our survey is best able to 
place constraints on planets between 6 and 10 M$_{Jup}$, and with semi-major 
axis between 20 and 40 AU}.  The decrease in sensitivity for masses $>$7 
M$_{Jup}$ is due to the fact that such high mass planets are too hot
to possess significant methane absorption if they are very young 
and, thus, are not ideal SDI targets.  The higher completeness for 7-8 
M$_{Jup}$ planets for semi-major axis of 30 AU vs. semi-major axis of 40 AU
is due to the small field of view of the SDI device; planets with semi-major
axes $>$ 30 AU can fall outside the SDI field in some of these cases. 
}
\end{figure}

To this end, we ran detailed Monte Carlo simulations
to characterize the ensemble of planets expected to exist around 
each star.  We conduct a similar simulation to that used to produce the 
contour plots of Fig.~\ref{fig:contour}, as described in 
Section 3.4 (these simulations are described in much more 
depth in Nielsen et al. 2006).  In contrast to the production of the 
contour plots, we simulate 10$^6$ planets instead of 10$^4$, and mass 
and semi-major axis are now assigned distributions of 
their own.  The mass and semi-major axis distributions, 
like the distribution for eccentricity, are produced by 
considering the population of published radial velocity planets (e.g. 
Butler et al. 2006), with mass and eccentricity both chosen to fit the 
histograms from observed planets.  Semi-major axis has been observed to 
follow a distribution of N({\it a}) $\propto$ {\it a}$^{-1}$ 
for radial velocity planets (Wright 
et al. 2005).  Since the radial velocity method has an inherent bias toward 
close-in planets (which have shorter orbital periods and larger radial 
velocity amplitudes), we attempt to correct for this by assuming a power-law 
distribution that is constant in semi-major axis -- i.e. 
N({\it a}) $\propto$ constant.  We consider the results 
of Fischer and Valenti's (2005) volume-limited sample, and choose an outer 
limit for the semi-major axis distribution such that, for stars in the 
metallicity range in our sample, each star is expected to host one planet.  
This is done by integrating the semi-major axis distribution from 0.02 AU 
(corresponding to HD 41004Bb, the closest-in exoplanet known thus far) to 
2.5 AU, the detection limit for the sample of Fischer and Valenti (2005), then 
noting the fraction of stars with planets in the metallicity range 
(-0.5 $<$ [Fe/H] $<$ 0.25) of our target stars (4.1\%) and choosing an 
upper cut-off to the distribution when the integral reaches 100\%.  This 
gives us a constant probability distribution for semi-major axis between 
0.02 and 45 AU that contains the same number of planets found in the 
$<$2.5 AU radial velocity survey.

\begin{figure}
\includegraphics[width=\columnwidth]{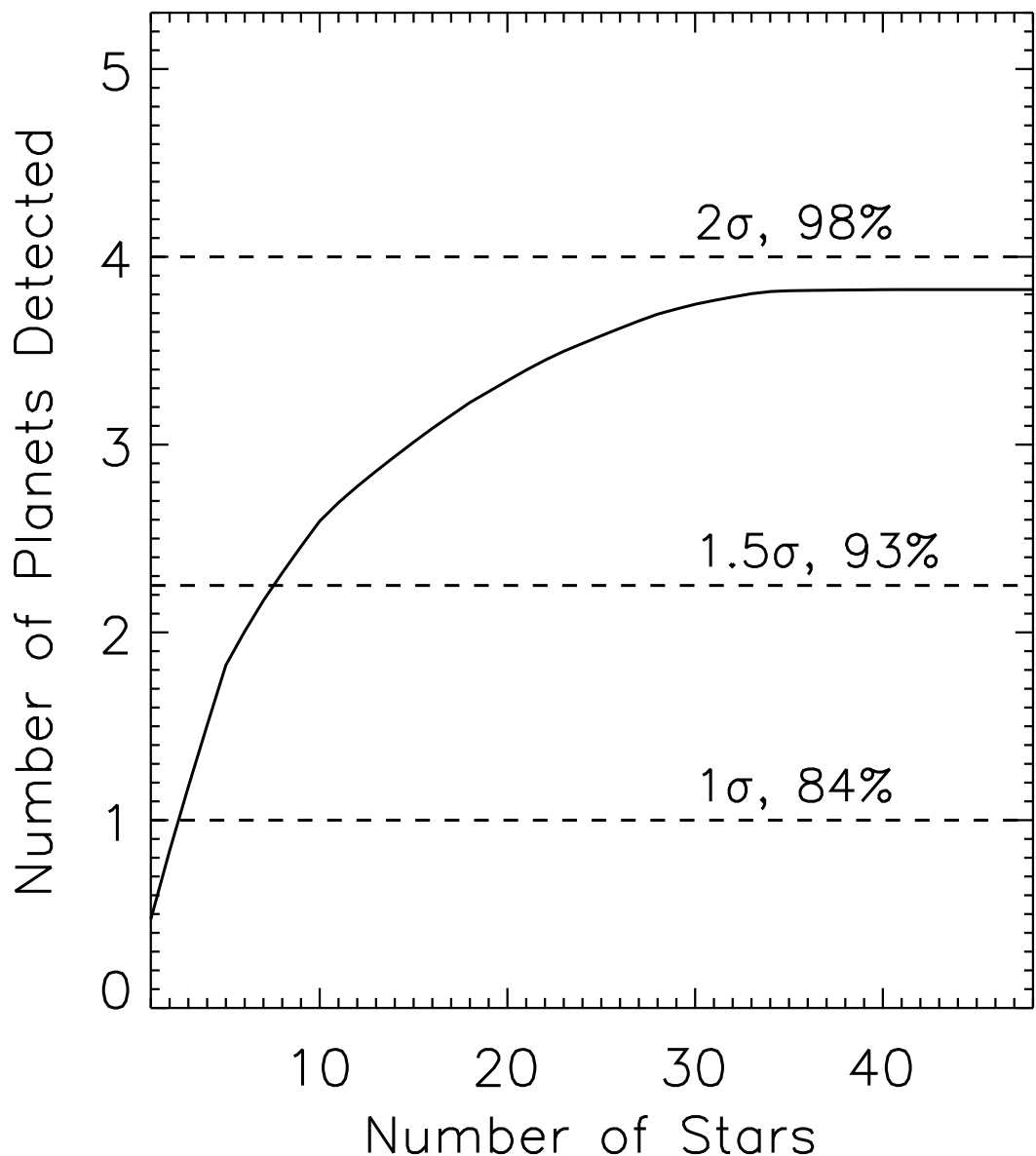}
\caption{Expected number of planets detected.  
By taking the results of our Monte Carlo simulations, and 
assuming that each program star
possesses exactly one planet, we can assign a detection probability for that
star from the percentage of simulated planets detected.  By adding these 
detection fractions for each star, we can compute the expected number of 
planets detected from our survey.  We order the target stars by decreasing 
detection probability, and plot the total number of planets expected to be 
detected as a function of the number of stars.  Over the entire survey, we 
expect to detect 3.8 planets, a 2$\sigma$ null result.  Thus, our assumed 
distribution for the frequency (1 planet per star, hence 100$\%$), 
semimajor axis distribution (N(a) $\propto$ constant), 
and luminosities (Burrows et al. 2003) of extrasolar planets is 
excluded at the 98$\%$ level by our extrasolar planet survey null result.}
\label{fig:ep}
\end{figure}

The ensemble of simulated planets is shown for our set of 
four typical stars in 
Fig.~\ref{fig:minmasses}.  Simulated planets which are detected are plotted
as blue dots and those that remain undetected are plotted as red dots.  
In addition to 
the contrast plot, we also consider a planet ``undetectable'' 
when its apparent 
H magnitude drops below 21 mag (a limit set by our total integration time), or when the 
planet's temperature rises above 1400 K (given as a function of age and 
planet mass by Burrows et al. 2003).  Above this temperature, the strength of 
the 1.62 $\mu$m methane break weakens to the point that the SDI method loses 
effectiveness.  Since we assume that each program star possesses exactly 
one planet that 
follows the distributions given above, we can assign a detection probability 
for that star from the percentage of the simulated planets that are 
detectable at the 5$\sigma$ level.   
For our 48 program stars (consisting of 40 stars with ages $>$250 Myr 
and closer than 50 pc, 1 10 Myr old star at a distance of 67 pc, 
3 stars with known RV planets and 4 nearby solar analogues)
which possess contrast curves, 
the average detection probability is 8.0$\%$, the median detection probability
is 4.1$\%$, and the maximum detection probability is 47$\%$.  We have 
chosen to leave the older stars in this sample in our statistics 
even though their detection probabilities are essentially zero.
Integrating over the probability distribution of our program stars, 
in Fig.~\ref{fig:ep} we 
plot the number of planets we expect to detect as a function 
of total stars observed, ordering the results so that the best stars (highest 
detection probabilities) are considered first.  For the 48 stars in our 
surveys for which we acquired contrast curves, 
we expect to detect a total of 3-4 planets
(3.8 to be exact) based on the above assumptions.  
Thus, our survey null detection 
rules out this exoplanet distribution at the 98\% (2.0$\sigma$) level.  
It is important to note that this null result 
shows that this particular combination of assumptions (mass distribution, 
eccentricity distribution, constant semi-major axis distribution, upper limit 
to semi-major axis at 45 AU, assumption that each star has a planet, and the 
mass-luminosity conversion from the models of Burrows et al. 2003) is ruled 
out to this confidence level; determining which individual assumptions are 
incorrect will required data beyond that of the current survey.  These 
simulations (including a variety of other possible exoplanet distributions) 
are discussed in more detail in Nielsen et al. (in prep.)
Nevertheless, our null detection in this survey sets strong upper limits
on the distribution of 
young massive extrasolar planets $>$5 AU from their primaries and  
provides valuable constraints for theories of planet formation and migration.

\section{Conclusions} 

We obtained datasets for 54 stars 
(45 stars were observed in the southern sky at the VLT,  
11 stars were observed in the northern sky at the MMT, and 2 stars 
were observed at both telescopes).
In our VLT data, we achieved H band contrasts
$>$ 10 mag (5$\sigma$) at a separation of 1.0" from
the primary star on 45$\%$ of our targets
and H band contrasts of $>$ 9 mag at a separation of 0.5'' 
on 80$\%$ of our targets.  
With this degree of attenuation, we should be able to 
image (5$\sigma$ detection) a 5 M$_{Jup}$ planet 15 AU from the star
around a 70 Myr K1 star at 15 pc or a 5 M$_{Jup}$ planet at 2 AU from 
a 12 Myr M star at 10 pc.  We believe that our SDI images are the highest 
contrast astronomical 
images ever made from ground or space for methane rich companions within
1'' of their primary star.  

Eight tentative candidates were identified (none with S/N $>$ 2 $\sigma$). 
Had these candidates been real, they would have possessed 
separations of 3 - 15.5 AU and 
masses of 2-10 M$_{Jup}$.  However, none of the candidates were 
detected in second epoch observations.
Thus, we find a null result from our survey.  Nonetheless, 
our result still has serious implications for the distribution
of extrasolar planets.  In the course of our survey, we also discovered 
5 new close stellar binary systems with measured separations of 0.14'' to 0.26''.

For 20 of our survey stars, 
we attained 50$\%$ completeness for 6-10 M$_{Jup}$ planets at semi-major
axes of 20-40 AU.  Thus, our completeness levels are sufficient to 
significantly test theoretical planet distributions.  From our survey null
result, we can rule out (at the 98$\%$/2.0$\sigma$ level) 
a model planet population using a constant 
distribution (N(a) $\propto$ constant) of planet semi-major 
axis out to a distance of 45 AU (a number
of further exoplanet distribution models are considered in Nielsen et al. 
in prep).  Our null detection in this survey sets strong upper limits
on the distribution of 
young massive extrasolar planets $>$5 AU from their primaries and  
provides valuable contraints for theories of planet formation and migration.

\acknowledgements{This publication is based on observations made with 
the MMT and the ESO VLT at Paranal Observatory under programme ID's 
074.C-0548, 074.C-0549, and 076.C-0094.
This publication makes use of data products from the Two Micron All Sky Survey,
which is a joint project of the University of Massachusetts and the Infrared
Processing and Analysis Center/California Institute of Technology, funded 
by the National Aeronautics and Space Administration and the National 
Science Foundation.
We thank Ren\'{e} Racine for refereeing this paper and for useful suggestions and  
Remi Soummer for suggesting the method of countour plots to present our detection 
limits.  BAB is supported by the NASA GSRP grant NNG04GN95H and NASA Origins 
grant NNG05GL71G.  LMC is supported by an NSF CAREER award and the NASA Origins of the Solar
System program.  ELN is supported by a Michelson Fellowship.
}

\clearpage

\clearpage

\clearpage

\clearpage

\clearpage

\clearpage

\begin{deluxetable}{lcccccccccc}
\rotate
\tablecolumns{11}
\tablewidth{0pc}
\tabletypesize{\tiny}
\tablecaption{Properties of SDI Survey Stars}
\tablehead{
\colhead{Target} & \colhead{RA\tablenotemark{a}} & \colhead{DEC} & \colhead{Distance(pc)\tablenotemark{a}} 
& \colhead{SpT\tablenotemark{*}}  & \colhead{Age(Gyr)\tablenotemark{b}} & 
\colhead{Age Ref\tablenotemark{b}} &
\colhead{V\tablenotemark{c}} & \colhead{H\tablenotemark{d}} & 
\colhead{Detectability} & \colhead{Comments}}
\startdata
Nearby Young Stars \\\hline
HIP 1481 & 00 18 26.1 & -63 28 39.0 & 41 & F8/GOV\tablenotemark{e} & 0.03 & Tuc & 7.5 & 6.2 & 4.2$\%$ & \\ 
ERX6 & 01 23 21.2 & -57 28 50.7 & 49.3 & G6V\tablenotemark{e} & 0.03 & Tuc/Hor & 8.5 & 6.9 & 4.1$\%$ &  \\ 
ERX8 & 01 28 08.7 & -52 38 19.2 & 37.1 & K1V\tablenotemark{e} & 0.03 & Tuc/Hor & 9.3 & 6.9 & 8.6$\%$ &  \\ 
HIP 9141 & 01 57 48.9 & -21 54 05.0 & 42.4 & G3Ve/G5V\tablenotemark{f} & 0.03 & Tuc/Hor & 8.1 & 6.6 & 3.6$\%$ & poss. 0.15'' binary \\ 
BD +05 378 & 02 41 25.9 & +05 59 18.4 & 40.5 & M0\tablenotemark{g} & 0.012 & Beta Pic & 10 & 7.2 & 7.6$\%$ & \\ 
HD 17925 & 02 52 32.1 & -12 46 11.0 & 10.4 & K1V\tablenotemark{f} & 0.115 & Possible Her/Lyr & 6 & 4.2 & 5.7$\%$ & \\ 
LH98 062 & 03 24 06.5 & +23 47 06.1 & 19.8 & K4V\tablenotemark{h} & 0.1 & Li from LH98 & 10 & 6.5 & & 2.4'' binary \\ 
V577 PerA & 03 33 13.5 & +46 15 26.5 & 33.8 & G5IV/V\tablenotemark{j} & 0.07 & AB Dor mg & 8.3 & 6.5 & 2.0$\%$ & 7'' binary \\ 
V834 Tau & 04 41 18.9 & +20 54 05.4 & 13.5 & K3V\tablenotemark{k} & 0.16 & Li from WSH03 & 8.1 & 5.3 & 2.7$\%$ &  \\ 
GJ 182 & 04 59 34.8 & +01 47 00.7 & 26.7 & M1Ve\tablenotemark{l} & 0.016 & Li from FMS97 & 10 & 6.5 & 15$\%$ & very tentative planet candidate \\
  & & & & & & & & & & (4.8 AU, $\sim$4 M$_{Jup}$) \\ 
 & & & & & & & & & & not detected at second epoch \\
HIP 23309 & 05 00 47.1 & -57 15 25.5 & 26.3 & M0/1\tablenotemark{m} & 0.012 & Beta Pic & 10 & 6.4 & 16$\%$ & \\ 
AB Dor & 05 28 44.8 & -65 26 54.9 & 14.9 & K1III\tablenotemark{e} & 0.07 & AB Dor mg & 6.9 & 4.8 & 7.4$\%$ & 0.16'' binary AB Dor C, \\
 & & & & & & & & & & Close et al. 2005 \\ 
GJ 207.1 & 05 33 44.8 & +01 56 43.4 & 16.8 & M2.5e\tablenotemark{n} & 0.1 & Lowrance et al. 2005 & 9.5 & 7.1 & 4.0$\%$ &  \\ 
UY Pic & 05 36 56.8 & -47 57 52.9 & 23.9 & K0V\tablenotemark{o} & 0.07 & AB Dor mg & 8 & 5.9 & 8.1$\%$ & \\ 
AO Men & 06 18 28.2 & -72 02 41.4 & 38.5 & K6/7\tablenotemark{m} & 0.012 & Beta Pic & 9.9 & 7 & 2.5$\%$ & very tentative planet candidate \\
 & & & & & & & & & & (14 AU, $\sim$4 M$_{Jup}$) \\ 
 & & & & & & & & & & not detected at second epoch \\
HIP 30030 & 06 19 08.1 & -03 26 20.0 & 52.4 & G0V\tablenotemark{p} & 0.03 & Tuc/Hor & 8 & 6.6 & 0.2$\%$ & \\ 
AB Pic & 06 19 12.9 & -58 03 16.0 & 45.5 & K2V\tablenotemark{e} & 0.03 & Tuc & 9.1 & 7.1 & 6.8$\%$ & planetary mass companion, \\
& & & & & & & & & & Chauvin et al. 2005 \\ 
 & & & & & & & & & & very tentative planet candidate \\
 & & & & & & & & & &(15.5 AU, $\sim$5 M$_{Jup}$) \\
 & & & & & & & & & & not detected at second epoch \\
SRX1 & 06 22 30.9 & -60 13 07.1 & 23.5 & G1V\tablenotemark{e} & 0.07 & AB Dor & 6.5 & 5.2 & 4.8$\%$ & \\
HD 48189A & 06 38 00.4 & -61 32 00.2 & 21.7 & G1/G2V\tablenotemark{e} & 0.07 & AB Dor & 6.2 & 4.7 & 1.8$\%$ & 0.14'' binary \\ 
BD +23 1978 & 08 36 55.8 & +23 14 48.0 & 41.6 & K5V\tablenotemark{q} & 0.035 & Montes et al. 2001 & 8.7 & 6.5 &  \\ 
$\pi_1\,$UMa & 08 39 11.7 & +65 01 15.3 & 14.3 & G1.5V\tablenotemark{q} & 0.21 & Li from WSH03 & 5.6 & 4.3 & 0.1$\%$ & \\LQ Hya & 09 32 25.6 & -11 11 04.7 & 18.3 & K0V\tablenotemark{q} & 0.013 & Li from WSH03 & 7.8 & 5.6 & 32$\%$ &  \\ 
DX Leo & 09 32 43.7 & +26 59 18.7 & 17.7 & K0V\tablenotemark{q} & 0.115 & Her/Lyra & 7 & 5.2 & 1.8$\%$ & very tentative planet candidate \\
 & & & & & & & & & & (2.6 AU, $\sim$10 M$_{Jup}$)\\ 
 & & & & & & & & & & not detected at second epoch \\
TWA 22 & 10 17 26.9 & -53 54 28.0 & 22 & M5\tablenotemark{g} & 0.01 &  & 14 & 8.1 & & \\ 
HD 92945 & 10 43 28.3 & -29 03 51.4 & 21.6 & K1V\tablenotemark{q} & 0.07 & AB Dor & 7.8 & 5.8 & 6.8$\%$ & very tentative planet candidate \\
 & & & & & & & & & & (10.4 AU, $\sim$6 M$_{Jup}$) \\ 
 & & & & & & & & & & not detected at second epoch \\
GJ 417 & 11 12 32.4 & +35 48 50.7 & 21.7 & G0V\tablenotemark{r} & 0.115 & Her/Lyra & 6.4 & 5 & 0.0$\%$ & \\ 
TWA 4 & 11 22 05.3 & -24 46 39.6 & 46.7 & K4V\tablenotemark{t} & 0.01 &  & 9.1 & 5.8 & & 0.78'' binary \\ 
TWA 25 & 12 15 30.8 & -39 48 42.0 & 44.1 & M0\tablenotemark{g} & 0.01 & TW Hydra & 11 & 7.5 & 14$\%$ & \\ 
RX J1224.8-7503 & 12 24 47.3 & -75 03 09.4 & 24.2 & K2\tablenotemark{u} & 0.016 & Li from AKCWM95 & 11 & 7.8 & 13.5$\%$ & \\ 
RX J1231.9-7848 & 12 31 56.0 & -78 48 36.0 & 50 & M1\tablenotemark{u} & 0.01 & Li from AKCWM95 & 14 & 9.6 &  & \\ 
EK Dra & 14 39 00.2 & +64 17 30.0 & 33.9 & G0\tablenotemark{w} & 0.07 & AB Dor & 7.6 & 6 & 0.67$\%$ & binary, \\
  & & & & & & & & & & Metchev $\&$ Hillenbrand 2004 \\ 
HD 135363 & 15 07 56.3 & +76 12 02.7 & 29.4 & G5V\tablenotemark{q} & 0.0032 & Li from WSH03 & 8.7 & 6.3 & 5.2$\%$ &  0.26'' binary \\ 
KW Lup & 15 45 47.6 & -30 20 55.7 & 40.9 & K2V\tablenotemark{v} & 0.002 & Li from NB98 & 9.4 & 6.6 & 3.9$\%$ &  \\ 
HD 155555AB & 17 17 25.5 & -66 57 04.0 & 30 & G5IV+KOIV/V\tablenotemark{m} & 0.012 & Beta Pic & 7.2 & 4.9 & 0.06$\%$ & \\ 
HD 155555C & 17 17 27.7 & -66 57 00.0 & 30 & M4.5\tablenotemark{m} & 0.012 & Beta Pic & 13 & 7.9 & 47$\%$ & \\ 
HD 166435 & 18 09 21.4 & +29 57 06.2 & 25.2 & G0\tablenotemark{x} & 0.1 & RHK from Wright+ 04 & 6.8 & 5.4 & 1.1$\%$ &   \\ 
HD 172555A & 18 45 26.9 & -64 52 16.5 & 30 & A5IV/V\tablenotemark{e} & 0.012 & Beta Pic & 4.8 & 4.3 & 5.8$\%$ & \\ 
CD-64 1208 & 18 45 37.0 & -64 51 44.6 & 29.2 & K7\tablenotemark{m} & 0.012 & Beta Pic & 10 & 6.3 & 9.9$\%$ & 0.18'' binary \\ 
HD 181321 & 19 21 29.8 & -34 59 00.5 & 20 & G1/G2V\tablenotemark{v} & 0.16 & Li from WSH03, & 6.5 & 5 & 0.09$\%$ & very tentative planet candidate \\
 & & & & & & RHK from Gray+ 06 & & & & (7 AU, $\sim$5 M$_{Jup}$) \\ 
  & & & & & & & & & & not detected at second epoch \\
HD 186704 & 19 45 57.3 & +04 14 54.6 & 30.3 & G0\tablenotemark{x} & 0.2 & RHK from Wright+ 04 & 7 & 5.6 & 0.0$\%$ &  \\ 
GJ 799B & 20 41 51.1 & -32 26 09.0 & 10.2 & M4.5e\tablenotemark{n} & 0.012 & Beta Pic & 13 & --- & 36$\%$ & \\ 
GJ 799A & 20 41 51.2 & -32 26 06.6 & 10.2 & M4.5e\tablenotemark{n} & 0.012 & Beta Pic & 11 & 5.2 & 18$\%$ & \\ 
GJ 803 & 20 45 09.5 & -31 20 27.1 & 9.94 & M0Ve\tablenotemark{n} & 0.012 & Beta Pic & 8.8 & 4.8 & 33$\%$ & very tentative planet candidate \\
 & & & & & & & & & & (3 AU, $\sim$2 M$_{Jup}$) \\ 
 & & & & & & & & & & not detected at second epoch \\
HIP 112312A & 22 44 57.8 & -33 15 01.0 & 23.6 & M4e\tablenotemark{g} & 0.012 & Beta Pic & 12 & 7.2 & 34$\%$ & very tentative planet candidate \\
  & & & & & & & & & & (6.2 AU, $\sim$8 M$_{Jup}$) \\ 
 & & & & & & & & & & not detected at second epoch \\
HD 224228 & 23 56 10.7 & -39 03 08.4 & 22.1 & K3V\tablenotemark{v} & 0.07 & AB Dor & 8.2 & 6 & 5.8$\%$ &  \\ \hline
More Distant Young Stars \\ \hline
TWA 14 & 11 13 26.5 & -45 23 43.0 & 66.7 & M0\tablenotemark{s} & 0.01 & TW Hydra & 13 & 8.7 & 7.8$\%$ &  \\ 
RXJ 1243.6-7834 & 12 43 36.7 & -78 34 07.8 & 150 & M0\tablenotemark{u} & 0.008 & Li from AKCWM95 & 13 & 8.7 & & 0.068'' binary \\ \hline
Stars with known RV planets \\\hline
$\epsilon\,$Eri & 03 32 55.8 & -09 27 29.7 & 3.22 & K2V\tablenotemark{i} & 0.8 & Benedict et al. 2006 & & 1.9 & 0.1$\%$ & Kellner et al. 2007, \\
 & & & & & & & & & & Janson et al. 2007 \\ 
HD 81040 & 09 23 47.1 & +20 21 52.0 & 32.6 & G0V\tablenotemark{q} & 2.5 & Li from SUZT06 & 7.7 & 6.3 & 0.0$\%$ &  \\ 
HD 128311 & 14 36 00.6 & +09 44 47.5 & 16.6 & K0\tablenotemark{q} & 0.63 & RHK from Gray+ 03 & 7.5 & 5.3 & 0.0$\%$ &  \\ \hline
Nearby Solar Analogues \\\hline
HD 114613 & 13 12 03.2 & -37 48 10.9 & 20.5 & G3V\tablenotemark{v} & 4.2 & Li from RGP93, & 4.8 & 3.3 & 0.0$\%$ &  \\ 
 & & & & & & RHK from Gray+06 & & & & \\ 
HD 201091 & 21 06 53.9 & +38 44 57.9 & 3.48 & K5Ve\tablenotemark{n} & 2.0\tablenotemark{1} & RHK from Gray+ 06 & 5.2 & 2.5 & 0.0$\%$ &  \\ 
$\epsilon\,$Ind A & 22 03 21.7 & -56 47 09.5 & 3.63 & K4.5V\tablenotemark{n} & 1.3 & Lachaume et al. 1999 & 4.7 & 2.3 & 0.09$\%$ & Gei{\ss}ler et al. 2007 \\ 
GJ 862 & 22 29 15.2 & -30 01 06.4 & 15.4 & K5Ve\tablenotemark{n} & 6.3\tablenotemark{1} & RHK from Gray+ 06 & 7.7 & 5.3 & 0.0$\%$ &  \\ 
\enddata
\tablenotetext{1}{In general, we have only determined Ca R'HK ages for stars with spectral types K1 
or earlier, but in the case of these two K5 stars, \\ 
we have only the R'HK measurement on which to rely 
for age determination.  The calibration of Mt. Wilson \\ S-index to R'HK for K5 stars (B-V $\sim$ 1.1 mag) 
has not been well-defined (Noyes et al. 1984; specifically the photospheric subtraction), and hence \\
applying a R'HK vs. age relation for K5 stars is unlikely to yield useful ages.  Although we adopt specific 
values for the ages of these stars, it would be more accurate to state simply that these stars have ages \\
$>$1 Gyr.  As a result, almost all simulated planets are too faint to detect around these stars, so the 
precise error in the age does not significantly affect our final results.}
\tablenotetext{a}{derived from the Hipparcos survey, Perryman et al.(1997)}
\tablenotetext{b}{ages for stars with cluster memberships from Zuckerman
and Song (2004), otherwise, ages are either lithium ages, calcium RHK ages, or an average of both. \\
Acronyms for lithium and calcium age references: AKCWM95: Alcala, Krautter, Schmitt, Covino, Wichmann, and Mundt 1995, \\
FMS97: Favata, Micela, and Sciortino 1997, LH98: Li and Hu 1998, NB98: Neuhauser and Brandner 1998, \\
RGP93: Randich, Gratton, and Pallavicini 1993, SUZT06: Sozetti, Udry, Zucker, Torres, Beuzit, et al. 2006, \\
WSH03: Wichmann, Schmitt, and Hubrig 2003}
\tablenotetext{c}{from the CDS Simbad service}
\tablenotetext{d}{from the 2MASS Survey, Cutri et al. (2003)}
\tablenotetext{*}{Spectral reference key: e: Houk and Cowley 1975, f: Houk and Cowley 1988, \\
g: Zuckerman and Song 2004, h: Li and Hu 1998, i: Cowley, Hiltner, and Witt 1967, \\
j: Christian and Mathioudakis 2002, k: Leaton and Pagel 1960, l: Favata, Barbera, Micela, and Sciortino 1995, \\
m: Zuckerman, Song, Bessell, and Webb 2001, n: Gliesse and Jahrei{\ss} 1991, o: Houk 1978, \\
p: Cutispoto, Pallavicini, Kuerster, and Rodono 1995, \\
q: Montes, Lopez-Santiago, Galvez, Fernandez-Figueroa, De Castro, and Cornide 2001, \\
r: Bidelman 1951, s: Zuckerman, Webb, Schwartz, and Becklin 2001, t: Houk and Smith-Moore 1988, \\
u: Alcala, Krautter, Schmitt, Covino, Wichmann, and Mundt 1995, v: Houk 1982, \\
w: Gliesse and Jahrei{\ss} 1979, x: Henry Draper Catalog}
\end{deluxetable}

\clearpage

\begin{deluxetable}{lcccc}
\tablecolumns{5}
\tablewidth{0pc}
\tablecaption{VLT SDI Observation Log}
\tablehead{
\colhead{Object} & \colhead{Date} & \colhead{DIT} & \colhead{NDIT} & \colhead{Total Exp (minutes)} }
\startdata
HIP 1481 & 2004-11-15 & 14 & 6 & 56 \\ 
 & 2005-11-24 & 16 & 5 & 26.7 \\ 
 & 2005-11-25 & 16 & 5 & 26.7 \\ 
 & 2005-11-27 & 16 & 5 & 26.7 \\ \hline
ERX6 & 2004-11-14 & 22 & 4 & 29.3 \\ 
  & 2004-11-16 & 22 & 4 & 58.7 \\ \hline
ERX8 & 2004-11-17 & 22 & 4 & 58.7 \\ \hline 
HIP9141 & 2004-09-27 & 14 & 6 & 56 \\ \hline
BD +05 378 & 2005-02-01 & 32 & 3 & 25.6 \\ \hline 
HD 17925 & 2003-08-14 & 7.5 & 16 & 40 \\ 
  & 2003-08-16 & 4 & 30 & 40 \\ 
 & 2003-08-17 & 4 & 30 & 20 \\ 
 & 2004-02-02 & 1 & 120 & 20 \\ 
 & 2004-11-16 & 4.1 & 17 & 46.5 \\ 
 & 2004-11-17 & 4.1 & 17 & 46.5 \\ \hline
LH98 062 & 2004-02-03 & 14 & 9 & 21 \\ \hline
$\epsilon\,$Eri & 2004-09-19 & 0.6 & 160 & 64 \\ \hline
V834 Tau & 2005-01-25 & 10 & 9 & 24 \\ 
 & 2005-02-01 & 10 & 9 & 24 \\ \hline
GJ 182 & 2004-02-02 & 7 & 17 & 39.7 \\ 
  & 2005-11-22 & 20 & 4 & 26.7 \\ 
  & 2005-11-24 & 20 & 4 & 26.7 \\ 
  & 2005-11-27 & 20 & 4 & 26.7 \\ \hline
HIP23309 & 2005-01-30 & 24 & 4 & 25.6 \\ 
  & 2005-01-31 & 24 & 4 & 51.2 \\ \hline
AB Dor & 2004-02-02 & 5 & 24 & 20 \\ 
  & 2004-09-28 & 12 & 7 & 28 \\ 
  & 2004-11-16 & 10.4 & 8 & 27.7 \\ \hline
GJ 207.1 & 2005-01-27 & 32 & 3 & 25.6 \\ \hline
UY Pic & 2004-11-16 & 14 & 6 & 28 \\ 
 & 2004-11-17 & 14 & 6 & 56 \\ \hline
AO Men & 2004-02-03 & 14 & 9 & 21 \\ 
 & 2005-11-15 & 30 & 1 & 17.5 \\ 
 & 2005-11-24 & 30 & 1 & 10 \\ \hline
AB Pic & 2004-11-14 & 20 & 4 & 26.7 \\ 
 & 2004-11-15 & 20 & 4 & 26.7 \\ 
 & 2005-11-22 & 20 & 4 & 13.3 \\ 
 & 2005-11-25 & 20 & 4 & 53.3 \\ \hline
SRX1 & 2004-11-18 & 12 & 7 & 28 \\ 
 & 2004-11-19 & 12 & 7 & 28 \\ \hline
HD 48189A & 2004-11-17 & 6.5 & 11 & 23.8 \\ 
 & 2004-11-18 & 6.5 & 11 & 23.8 \\ \hline
BD +23 1978 & 2005-01-27 & 24 & 4 & 25.6\\
 & 2005-01-28 & 24 & 4 & 25.6\\
LQ Hya & 2004-02-02 & 5 & 24 & 40 \\ 
 & 2004-12-08 & 14 & 6 & 28 \\ 
 & 2004-12-14 & 14 & 6 & 28 \\ \hline
DX Leo & 2004-02-05 & 3 & 38 & 19 \\ 
 & 2005-12-04 & 14 & 6 & 28 \\ 
 & 2005-12-19 & 14 & 6 & 28 \\ \hline
TWA 22 & 2005-01-25 & 32 & 1 & 48.5 \\ \hline
HD92945 & 2004-02-05 & 5 & 24 & 60 \\ \hline
TWA 14 & 2005-01-28 & 32 & 3 & 25.6 \\ 
 & 2005-01-29 & 32 & 3 & 25.6 \\ \hline
TWA 4 & 2004-02-02 & 7 & 17 & 9.92 \\ \hline
TWA25 & 2005-01-28 & 32 & 3 & 25.6 \\ \hline
RX J1224.8-7503 & 2004-02-02 & 40 & 3 & 20 \\ 
 & 2005-01-16 & 30 & 3 & 60 \\ 
 & 2005-01-27 & 30 & 3 & 120 \\ \hline
RX J1231.9-7848 & 2004-02-05 & 20 & 6 & 20 \\ \hline
RXJ 1243.6-7834 & 2004-02-02 & 5 & 24 & 40 \\ \hline
HD 114613 & 2004-02-02 & 1 & 120 & 40 \\ \hline
KW Lup & 2004-09-15 & 22 & 4 & 14.7 \\ 
 & 2004-09-16 & 24 & 4 & 22.7 \\ 
 & 2004-09-17 & 24 & 4 & 24 \\ \hline
HD155555AB & 2003-08-14 & 7.5 & 16 & 10 \\ 
 & 2003-08-15 & 7.5 & 16 & 20 \\ 
 & 2003-08-16 & 7.5 & 16 & 10 \\ 
 & 2003-08-17 & 7.5 & 16 & 10 \\ 
 & 2004-09-16 & 10 & 9 & 30 \\ 
 & 2004-09-18 & 14 & 6 & 28 \\ \hline
HD155555C & 2003-08-14 & 30 & 4 & 40 \\ 
 & 2003-08-16 & 30 & 4 & 40 \\ \hline
HD172555A & 2003-08-17 & 5 & 24 & 20 \\ 
 & 2004-09-17 & 5 & 15 & 25 \\ 
 & 2004-09-18 & 5 & 15 & 6.25 \\ 
 & 2004-09-19 & 5 & 15 & 18.8 \\ \hline
CD-64 1208 & 2003-08-17 & 20 & 6 & 40 \\ 
 & 2004-09-16 & 15 & 6 & 30 \\ \hline
HD181321 & 2003-08-15 & 7.5 & 16 & 40 \\ 
 & 2004-09-18 & 11 & 8 & 29.3 \\ \hline
GJ799B & 2003-08-16 & 20 & 6 & 40 \\ 
 & 2003-08-17 & 20 & 6 & 30 \\ 
 & 2004-09-19 & 15 & 6 & 30 \\ \hline
GJ799A & 2003-08-16 & 20 & 6 & 40 \\ 
 & 2004-09-16 & 10 & 9 & 30 \\ 
 & 2004-09-19 & 15 & 6 & 30 \\ \hline
GJ803 & 2003-08-14 & 7.5 & 18 & 56.2 \\ 
 & 2003-08-15 & 10 & 12 & 40 \\ 
 & 2003-08-17 & 7.5 & 16 & 40 \\ 
 & 2004-09-17 & 6 & 15 & 30 \\ 
 & 2004-09-18 & 10 & 9 & 30 \\ \hline
$\epsilon\,$Ind A & 2004-09-18 & 0.5 & 192 & 48 \\ \hline
GJ 862 & 2003-08-15 & 10 & 12 & 40 \\ 
 & 2003-08-16 & 10 & 12 & 40 \\ 
 & 2004-09-19 & 13 & 7 & 48.2 \\ \hline
HIP112312A & 2004-09-19 & 25 & 4 & 66.7 \\ \hline
HD224228 & 2003-08-16 & 10 & 12 & 40 \\ 
 & 2003-08-17 & 20 & 6 & 40 \\ 
 & 2004-10-08 & 14 & 6 & 28 \\ 
 & 2004-10-20 & 21 & 4 & 28 \\ \hline
\enddata
\end{deluxetable}

\clearpage

\begin{deluxetable}{lcccc}
\tablecolumns{5}
\tablewidth{0pc}
\tablecaption{MMT SDI Observation Log}
\tablehead{
\colhead{Object} & \colhead{Date} & \colhead{DIT} & \colhead{NDIT} & \colhead{Total Exp (minutes)}}
\startdata
V577 PerA & 2006-02-12 & 20 & 7 & 37.3 \\ 
 & 2006-02-13 & 21.5 & 7 & 40.1 \\ \hline
HIP30030 & 2006-02-12 & 30 & 5 & 30 \\ \hline 
$\pi_1\,$UMa & 2006-02-13 & 5.8 & 13 & 40.2 \\ \hline 
HD 81040 & 2006-02-12 & 11.7 & 13 & 40.3 \\ \hline 
LQ Hya & 2006-02-12 & 8 & 19 & 40.5 \\ \hline 
DX Leo & 2005-05-01 & 10 & 13 & 34.7 \\ \hline 
GJ 417 & 2005-04-30 & 7 & 17 & 31.7 \\  \hline
HD 128311 & 2006-02-12 & 4 & 19 & 60.8 \\  \hline
EK Dra & 2005-05-01 & 20 & 7 & 37.3 \\  \hline
HD 135363 & 2005-05-01 & 30 & 5 & 40 \\  \hline
HD 166435 & 2005-04-30 & 7 & 17 & 31.7 \\ 
 & 2005-05-01 & 7 & 17 & 31.7 \\  \hline
HD 186704 & 2005-05-01 & 10 & 13 & 17.3 \\  \hline
HD 201091 & 2005-04-30 & 20 & 7 & 37.33 \\ \hline
\enddata
\end{deluxetable}

\begin{deluxetable}{lccccc}
\tablecolumns{5}
\tablewidth{0pc}
\tablecaption{Limiting H mag (5$\sigma$) at 0.5''}
\tablehead{
\colhead{Object} & \colhead{$\Delta$F1} & \colhead{Separation(AU)} & \colhead{$\Delta$H (T8 SpT)} & \colhead{m$_H$} & \colhead{M$_H$} }
\startdata
$\epsilon\,$Eri & 9.4 $\pm$ 0.12 & 1.61 & 10.3 & 12.2 & 14.7 \\ 
$\epsilon\,$Ind A & 10.6 $\pm$ 0.12 & 1.81 & 11.5 & 13.8 & 16 \\ 
HD 201091 & 8.08 $\pm$ 0.52 & 1.74 & 8.95 & 11.5 & 13.8 \\ 
HD 114613 & 6.13 $\pm$ 0.26 & 10.2 & 7 & 10.3 & 8.74 \\ 
HD 17925 & 9.69 $\pm$ 0.14 & 5.19 & 10.6 & 14.8 & 14.7 \\ 
HD172555A & 9.14 $\pm$ 0.12 & 15 & 10 & 14.3 & 11.9 \\ 
$\pi_1\,$UMa & 8.04 $\pm$ 0.15 & 7.14 & 8.91 & 13.2 & 12.4 \\ 
HD 48189A & 8.54 $\pm$ 0.052 & 10.8 & 9.41 & 14.2 & 12.5 \\ 
GJ803 & 9.54 $\pm$ 0.091 & 4.97 & 10.4 & 15.2 & 15.2 \\ 
AB Dor & 9.04 $\pm$ 0.019 & 7.47 & 9.91 & 14.8 & 13.9 \\ 
HD155555AB & 5.87 $\pm$ 0.14 & 15 & 6.74 & 11.6 & 9.21 \\ 
GJ 417 & 7.79 $\pm$ 0.23 & 10.9 & 8.66 & 13.7 & 12 \\ 
HD181321 & 7.42 $\pm$ 0.13 & 10 & 8.29 & 13.3 & 11.8 \\ 
SRX1 & 9.95 $\pm$ 0.079 & 11.7 & 10.8 & 16 & 14.1 \\ 
GJ799A & 7.48 $\pm$ 0.082 & 5.11 & 8.35 & 13.6 & 13.6 \\ 
DX Leo & 8.24 $\pm$ 0.19 & 8.87 & 9.11 & 14.4 & 13.2 \\ 
GJ 862 & 9.51 $\pm$ 0.25 & 7.72 & 10.4 & 15.7 & 14.8 \\ 
V834 Tau & 9.08 $\pm$ 0.18 & 6.74 & 9.95 & 15.3 & 14.6 \\ 
HD 166435 & 8.42 $\pm$ 0.17 & 12.6 & 9.29 & 14.7 & 12.7 \\ 
LQ Hya & 9.82 $\pm$ 0.16 & 9.17 & 10.7 & 16.3 & 15 \\ 
HD 186704 & 7.13 $\pm$ 0.091 & 15.1 & 8 & 13.6 & 11.2 \\ 
HD92945 & 9.91 $\pm$ 0.0099 & 10.8 & 10.8 & 16.6 & 14.9 \\ 
UY Pic & 9.96 $\pm$ 0.11 & 11.9 & 10.8 & 16.7 & 14.8 \\ 
HD224228 & 9 $\pm$ 0.15 & 11 & 9.87 & 15.9 & 14.2 \\ 
EK Dra & 7.85 $\pm$ 0.39 & 17 & 8.72 & 14.7 & 12 \\ 
HIP   1481 & 9.22 $\pm$ 0.13 & 20.5 & 10.1 & 16.3 & 13.2 \\ 
CD-64 1208 & 9.33 $\pm$ 0.087 & 14.6 & 10.2 & 16.5 & 14.2 \\ 
HD 135363 & 7.9 $\pm$ 0.27 & 14.7 & 8.77 & 15.1 & 12.8 \\ 
HIP23309 & 8.45 $\pm$ 0.092 & 13.1 & 9.32 & 15.7 & 13.6 \\ 
GJ 182 & 8.01 $\pm$ 0.16 & 13.3 & 8.88 & 15.3 & 13.2 \\ 
V577 PerA & 8.9 $\pm$ 0.33 & 16.9 & 9.77 & 16.2 & 13.6 \\ 
HIP9141 & 8.92 $\pm$ 0.29 & 21.2 & 9.79 & 16.3 & 13.2 \\ 
HIP30030 & 6.91 $\pm$ 0.17 & 26.2 & 7.78 & 14.4 & 10.8 \\ 
KW Lup & 8.76 $\pm$ 0.091 & 20.5 & 9.63 & 16.3 & 13.2 \\ 
ERX8 & 9.4 $\pm$ 0.2 & 18.6 & 10.3 & 17.2 & 14.4 \\ 
ERX6 & 9.38 $\pm$ 0.4 & 24.6 & 10.2 & 17.1 & 13.6 \\ 
AO Men & 6.91 $\pm$ 0.33 & 19.2 & 7.78 & 14.8 & 11.9 \\ 
AB Pic & 9.65 $\pm$ 0.027 & 22.8 & 10.5 & 17.6 & 14.3 \\ 
GJ 207.1 & 7.5 $\pm$ 0.094 & 8.41 & 8.37 & 15.5 & 14.4 \\ 
HIP112312A & 9.09 $\pm$ 0.27 & 11.8 & 9.96 & 17.1 & 15.2 \\ 
BD +05 378 & 8.31 $\pm$ 0.088 & 20.3 & 9.18 & 16.4 & 13.4 \\ 
TWA25 & 9.5 $\pm$ 0.035 & 22 & 10.4 & 17.9 & 14.7 \\ 
RX J1224.8-7503 & 7.16 $\pm$ 0.024 & 12.1 & 8.03 & 15.9 & 14 \\ 
HD155555C & 10.5 $\pm$ 0.085 & 15 & 11.4 & 19.3 & 16.9 \\ 
TWA 14 & 8.38 $\pm$ 0.03 & 33.3 & 9.25 & 18 & 13.9 \\ 
\enddata
\end{deluxetable}

\begin{deluxetable}{lccccc}
\tablecolumns{5}
\tablewidth{0pc}
\tablecaption{Limiting H mag (5$\sigma$) at 1.0''}
\tablehead{
\colhead{Object} & \colhead{$\Delta$F1} & \colhead{Separation(AU)} & \colhead{$\Delta$H (T8 SpT)} & \colhead{m$_H$} & \colhead{M$_H$} }
\startdata
$\epsilon\,$Eri & 11.3 $\pm$ 0.2 & 3.22 & 12.2 & 14.1 & 16.6 \\ 
$\epsilon\,$Ind A & 12 $\pm$ 0.16 & 3.63 & 12.9 & 15.2 & 17.4 \\ 
HD 201091 & 9.42 $\pm$ 0.05 & 3.48 & 10.3 & 12.8 & 15.1 \\ 
HD 114613 & 7.24 $\pm$ 0.13 & 20.5 & 8.11 & 11.5 & 9.94 \\ 
HD 17925 & 11.3 $\pm$ 0.19 & 10.4 & 12.2 & 16.4 & 16.3 \\ 
HD172555A & 11.2 $\pm$ 0.098 & 30 & 12.1 & 16.4 & 14 \\ 
$\pi_1\,$UMa & 9.28 $\pm$ 0.14 & 14.3 & 10.1 & 14.4 & 13.6 \\ 
HD 48189A & 9.87 $\pm$ 0.24 & 21.7 & 10.7 & 15.4 & 13.7 \\ 
GJ803 & 10.7 $\pm$ 0.03 & 9.94 & 11.6 & 16.4 & 16.4 \\ 
AB Dor & 11 $\pm$ 0.17 & 14.9 & 11.9 & 16.7 & 15.8 \\ 
HD155555AB & 7.3 $\pm$ 0.046 & 30 & 8.17 & 13.1 & 10.7 \\ 
GJ 417 & 8.44 $\pm$ 0.05 & 21.7 & 9.31 & 14.3 & 12.6 \\ 
HD181321 & 8.63 $\pm$ 0.048 & 20 & 9.5 & 14.6 & 13.1 \\ 
SRX1 & 11.2 $\pm$ 0.13 & 23.5 & 12.1 & 17.3 & 15.4 \\ 
GJ799A & 9.55 $\pm$ 0.14 & 10.2 & 10.4 & 15.6 & 15.6 \\ 
DX Leo & 9.98 $\pm$ 0.039 & 17.7 & 10.8 & 16 & 14.8 \\ 
GJ 862 & 10.7 $\pm$ 0.12 & 15.4 & 11.6 & 16.9 & 16 \\ 
V834 Tau & 10.2 $\pm$ 0.18 & 13.5 & 11.1 & 16.4 & 15.7 \\ 
HD 166435 & 9.98 $\pm$ 0.061 & 25.2 & 10.8 & 16.2 & 14.2 \\ 
LQ Hya & 11 $\pm$ 0.035 & 18.3 & 11.9 & 17.5 & 16.2 \\ 
HD 186704 & 7.35 $\pm$ 0.052 & 30.3 & 8.22 & 13.8 & 11.4 \\ 
HD92945 & 10.8 $\pm$ 0.062 & 21.6 & 11.7 & 17.5 & 15.8 \\ 
UY Pic & 11.5 $\pm$ 0.033 & 23.9 & 12.4 & 18.3 & 16.4 \\ 
HD224228 & 10.8 $\pm$ 0.11 & 22.1 & 11.7 & 17.7 & 16 \\ 
EK Dra & 8.86 $\pm$ 0.14 & 33.9 & 9.73 & 15.7 & 13 \\ 
HIP   1481 & 10.8 $\pm$ 0.046 & 41 & 11.7 & 17.9 & 14.8 \\ 
CD-64 1208 & 9.88 $\pm$ 0.54 & 29.2 & 10.8 & 17.1 & 14.8 \\ 
HD 135363 & 8.65 $\pm$ 0.025 & 29.4 & 9.52 & 15.8 & 13.5 \\ 
HIP23309 & 10 $\pm$ 0.051 & 26.3 & 10.9 & 17.3 & 15.2 \\ 
GJ 182 & 10.2 $\pm$ 0.15 & 26.7 & 11.1 & 17.6 & 15.5 \\ 
V577 PerA & 10 $\pm$ 0.062 & 33.8 & 10.9 & 17.4 & 14.8 \\ 
HIP9141 & 10.5 $\pm$ 0.028 & 42.4 & 11.4 & 18 & 14.9 \\ 
HIP30030 & 8.3 $\pm$ 0.09 & 52.4 & 9.17 & 15.8 & 12.2 \\ 
KW Lup & 9.86 $\pm$ 0.17 & 40.9 & 10.7 & 17.3 & 14.2 \\ 
ERX8 & 10.7 $\pm$ 0.12 & 37.1 & 11.6 & 18.5 & 15.7 \\ 
ERX6 & 10.6 $\pm$ 0.12 & 49.3 & 11.5 & 18.4 & 14.9 \\ 
AO Men & 7.9 $\pm$ 0.015 & 38.5 & 8.77 & 15.8 & 12.9 \\ 
AB Pic & 10.8 $\pm$ 0.013 & 45.5 & 11.7 & 18.8 & 15.5 \\ 
GJ 207.1 & 8.74 $\pm$ 0.089 & 16.8 & 9.61 & 16.8 & 15.7 \\ 
HIP112312A & 10.6 $\pm$ 0.068 & 23.6 & 11.5 & 18.7 & 16.8 \\ 
BD +05 378 & 9.52 $\pm$ 0.074 & 40.5 & 10.4 & 17.6 & 14.6 \\ 
TWA25 & 10.5 $\pm$ 0.18 & 44.1 & 11.4 & 18.9 & 15.7 \\ 
RX J1224.8-7503 & 8.04 $\pm$ 0.16 & 24.2 & 8.91 & 16.8 & 14.9 \\ 
HD155555C & 10.8 $\pm$ 0.043 & 30 & 11.7 & 19.6 & 17.2 \\ 
TWA 14 & 8.74 $\pm$ 0.047 & 66.7 & 9.61 & 18.3 & 14.2 \\ 
\enddata
\end{deluxetable}

\begin{deluxetable}{lcccccc}
\tablecolumns{7}
\tablewidth{0pc}
\rotate
\tablecaption{Star/Planet Projected Minimum Detectable Separations for 5 and 
10 M$_{Jup}$ Planets}
\tablehead{
\colhead{Object} & \colhead{Age (Myr)} & \colhead{Distance (pc)} & \colhead{Separation 5 M$_{Jup}$} & \colhead{(AU)} & \colhead{Separation 10 M$_{Jup}$} & \colhead{(AU)}}
\startdata
 & & & Burrows et al. & Baraffe et al. & Burrows et al. & Baraffe et al. \\
AB Dor            & 70         & 14.94      & 13.45      & 20.92      & 6.28       & 10.31    \\ 
AO Men            & 12         & 38.48      & 30.01      & ---\tablenotemark{*}      & 11.54      & 20.01    \\ 
BD+05 378         & 12         & 40.54      & 16.21      & 28.37      & 7.70       & 9.32     \\ 
CD -64 1208       & 12         & 34.21      & 11.29      & 17.45      & 5.82       & 8.55     \\ 
DX Leo            & 115        & 17.75      & ---      & ---      & 16.68      & 33.54    \\ 
EK Dra            & 70         & 33.94      & ---      & ---      & ---      & ---    \\ 
$\epsilon\,$Eri           & 800        & 3.22       & ---      & ---      & ---      & ---    \\ 
$\epsilon\,$Ind A        & 1300       & 3.63       & ---      & ---      & ---      & 3.81    \\ 
GJ 174            & 160        & 13.49      & ---      & ---      & 12.01      & 17.54    \\ 
GJ 182            & 12         & 26.67      & 6.67       & 16.27      & 4.80       & 5.60    \\ 
GJ 207.1          & 100        & 16.82      & ---      & ---      & 10.77      & 15.48    \\ 
GJ 417            & 115        & 21.72      & ---      & ---      & ---      & ---    \\ 
GJ 799A           & 12         & 10.22      & 1.94       & 4.60       & 1.12       & 1.64     \\ 
GJ 799B           & 12         & 10.22      & 1.43       & 1.94       & 1.12       & 1.12     \\ 
GJ 803            & 12         & 9.94       & 1.39       & 2.39       & 1.09       & 1.09     \\ 
GJ 862            & 6300       & 15.45      & ---      & ---      & ---      & ---    \\ 
HD 114613         & 4200       & 20.48      & ---      & ---      & ---      & ---    \\ 
HD 128311         & 630        & 16.57      & ---      & ---      & ---      & ---    \\ 
HD 135363         & 3          & 29.44      & 8.54       & 11.78      & 3.24       & 6.77     \\ 
HD 155555 AB      & 12         & 30.03      & ---      & ---      & 25.23      & ---    \\ 
HD 155555 C       & 12         & 30.03      & 3.30       & 3.30       & 3.30       & 3.30     \\ 
HD 166435         & 100        & 25.24      & ---      & ---      & 29.03      & ---    \\
HD 172555 A       & 12         & 29.23      & 20.17      & 28.65      & 6.43       & 14.62    \\ 
HD 17925          & 115        & 10.38      & 14.43      & ---      & 5.40       & 8.62     \\ 
HD 181321         & 160        & 20.86      & ---      & ---      & ---      & ---    \\ 
HD 201091         & 2000       & 3.48       & ---      & ---       & ---      & ---     \\ 
HD 224228         & 70         & 22.08      & 22.31      & ---      & 9.50       & 13.69    \\ 
HD 45270          & 70         & 23.50      & 30.78      & ---      & 10.57      & 15.98    \\ 
HD 48189 A        & 70         & 21.67      & ---      & ---      & 19.93      & 27.95    \\ 
HD 81040          & 2500       & 32.56      & ---      & ---       & ---      & ---     \\ 
HD 8558           & 30         & 49.29      & 27.11      & 67.52      & 11.34      & 22.18    \\ 
HD 186704          & 200        & 30.26      & ---      & ---      & ---      & ---    \\ 
HD 9054           & 30         & 37.15      & 19.32      & 31.20      & 7.06       & 9.29     \\ 
HD 92945          & 70         & 21.57      & 24.59      & 44.00      & 6.04       & 10.14    \\ 
HIP 112312 A      & 12         & 23.61      & 3.54       & 5.19       & 2.60       & 2.83     \\
HIP 1481          & 30         & 40.95      & 30.71      & 59.38      & 12.69      & 20.88    \\ 
HIP 23309         & 12         & 26.26      & 6.83       & 16.02      & 2.89       & 4.99     \\ 
HIP 30030         & 30         & 52.36      & ---      & ---      & 53.93      & ---    \\ 
AB Pic            & 30         & 45.52      & 19.12      & 40.05      & 9.56       & 13.65    \\ 
HIP 9141          & 30         & 42.35      & 37.70      & 78.78      & 15.25      & 22.45    \\ 
KW Lup            & 2          & 40.92      & 6.14       & 7.36       & 4.50       & 4.50     \\ 
LQ Hya            & 13         & 18.34      & 3.30       & 4.95       & 2.02       & 2.75     \\ 
RXJ1224.8-7503    & 16         & 24.17      & 6.53       & 17.16      & 3.63       & 4.59     \\ 
TWA 14            & 10         & 66.67      & 14.00      & 27.33      & 7.33       & 10.67    \\ 
TWA 25            & 10         & 44.05      & 9.25       & 14.10      & 4.85       & 7.93     \\ 
UY Pic            & 70         & 23.87      & 19.81      & 29.12      & 5.73       & 11.46    \\ 
V577 Per A        & 70         & 33.77      & 81.73      & ---      & 17.90      & 27.02    \\ 
$\pi_1\,$UMa          & 210        & 14.27      & ---      & ---      & ---      & ---    \\ \hline
\enddata
\tablenotetext{*}{--- means that such an object is too low in mass to be detected with
our current survey contrast level for that star}
\end{deluxetable}

\begin{deluxetable}{lcc}
\tablecolumns{3}
\tablewidth{0pc}
\tablecaption{Binary Properties}
\tablehead{
\colhead{Object} & \colhead{Separation} & \colhead{Position Angle} }
\startdata \hline
SDI survey discoveries \\ \hline
AB Dor AC\tablenotemark{a} & 0.16'' & 127$^{\circ}$ \\
HIP 9141 & 0.15'' & 355$^{\circ}$ \\
HD 48189AC & 0.14'' & 143$^{\circ}$ \\
HD 135363 & 0.26'' & 132$^{\circ}$ \\
CD -64 1208 & 0.18'' & 95$^{\circ}$ \\ \hline
SDI survey confirmations \\ \hline
RXJ 1243.6-7834\tablenotemark{b} & 0.068'' & 171$^{\circ}$/351$^{\circ}$ \\
LH 98 062 & 2.4'' & 354$^{\circ}$ \\
TWA 4 & 0.78'' & 3$^{\circ}$ \\
EK Dra & 0.67'' & 176$^{\circ}$ \\
\enddata
\tablenotetext{a}{Separation and position angle from Close et al. 2005b.  
For updated photometry and astrometry see Close et al. 2007b.}
\tablenotetext{b}{As RXJ 1243.6-7834 is nearly an equal-magnitude binary, 
we were unable to determine which star was the primary (as selected by 
Brandner et al. (2000)) and thus present two values for the position
angle (assuming each star is the primary in turn).} 
\end{deluxetable}

\end{document}